\documentclass[12pt,preprint]{aastex}

\usepackage{natbib}

\shorttitle{Bolocam Atmospheric Noise}
\shortauthors{Sayers and Golwala}

\begin{document}

\title{Studies of Millimeter-Wave Atmospheric Noise Above Mauna Kea}

\author{J.~Sayers\altaffilmark{1,6,7}, S.~R.~Golwala\altaffilmark{2},
  P.~A.~R.~Ade\altaffilmark{3},
  J.~E.~Aguirre\altaffilmark{3}, J.~J.~Bock\altaffilmark{1},
  S.~F.~Edgington\altaffilmark{2}, J.~Glenn\altaffilmark{5},
  A.~Goldin\altaffilmark{1}, D.~Haig\altaffilmark{3},
  A.~E.~Lange\altaffilmark{2}, G.~T.~Laurent\altaffilmark{5},  
  P.~D.~Mauskopf\altaffilmark{3}, H.~T.~Nguyen\altaffilmark{1}, 
  P.~Rossinot\altaffilmark{2}, and~J.~Schlaerth\altaffilmark{5}}
%  \\ 
%  \vspace{12pt}
%  \copyright 2009. All rights reserved.}
\altaffiltext{1}
  {Jet Propulsion Laboratory, California Institute of Technology, 
  4800 Oak Grove Drive, Pasadena, CA 91109}
\altaffiltext{2}
  {Division of Physics, Mathematics, \& Astronomy,
  California Institute of Technology, 
  Mail Code 59-33, Pasadena, CA 91125}
\altaffiltext{3}
  {Physics and Astronomy, Cardiff University, 5 The Parade,
  P. O. Box 913, Cardiff CF24 3YB, Wales, UK}
\altaffiltext{4}
  {University of Pennsylvania, 209 South 33rd St, Philadelphia, PA 19104}
\altaffiltext{5}
  {Center for Astrophysics and Space Astronomy \& Department of 
  Astrophysical and Planetary Sciences, 
  University of Colorado, 389 UCB, Boulder, CO 80309}
\altaffiltext{6}
  {NASA Postdoctoral Program Fellow}
\altaffiltext{7}
  {jack@caltech.edu}

\begin{abstract}

  We report measurements of the fluctuations in atmospheric emission
  (atmospheric noise) above Mauna Kea recorded with Bolocam
  at 143 and 268~GHz from the Caltech Submillimeter Observatory (CSO).
  The 143~GHz data were collected during a 40 night observing run
  in late 2003, and the 268~GHz observations were made in early 2004
  and early 2005 over a total of 60 nights.
  Below $\simeq 0.5$~Hz, the data time-streams are dominated
  by atmospheric noise 
%  with an $f^{-\delta}$ spectrum
  in all observing conditions.
  The atmospheric noise data are consistent with
  a Kolmogorov-Taylor (K-T) turbulence model for a thin wind-driven
  screen, and the median
  amplitude of the fluctuations is 280~mK$^2$ rad$^{-5/3}$
  at 143~GHz and 4000~mK$^2$ rad$^{-5/3}$ at 268~GHz.
  Comparing our results with previous ACBAR data,
  we find that the normalization of the power spectrum
  of the atmospheric noise
  fluctuations is a factor of $\simeq 80$
  larger above Mauna Kea than above the South Pole
  at millimeter wavelengths.
  Most of this difference is due to the fact that the atmosphere
  above the South Pole is much drier than the atmosphere above
  Mauna Kea.
  However, the atmosphere above the South Pole is 
  slightly more stable as well:  
  the fractional fluctuations in the column depth of precipitable
  water vapor are a 
  factor of $\simeq \sqrt{2}$ smaller at the South Pole compared
  to Mauna Kea.
  Based on our atmospheric modeling, we developed several algorithms
  to remove the atmospheric noise, and the best results 
  were achieved when we described the fluctuations using 
  a low-order polynomial in detector position over the 
  8~arcmin field of view (FOV).
  However, even with these algorithms, we were not able to
  reach photon-background-limited instrument photometer (BLIP) 
  performance at 
  frequencies below $\simeq 0.5$~Hz in any observing 
  conditions. 
  We also observed an excess low-frequency noise that
  is highly correlated between detectors separated by
  $\lesssim (f/\#)\lambda$; 
  this noise appears to be caused by atmospheric fluctuations,
  but we do not have an adequate model to explain its source.
  We hypothesize that the correlations arise from
%  We also observed excess correlations in the atmospheric noise 
%  between pixels separated by $\lesssim (f/\#)\lambda$, which 
%  are 
%  explained by 
  the classical coherence of 
  the EM field across a distance of $\simeq (f/\#)\lambda$ 
  on the focal plane. 

\end{abstract}  
\keywords{atmospheric effects: instrumentation}

\section{Introduction}

  A number of wide-field ground-based mm/submm imaging arrays
  have been commissioned during the past 15 years, 
  including SCUBA \citep{holland99}, 
  MAMBO \citep{kreysa98}, 
  Bolocam \citep{glenn98}, 
  SHARC II \citep{dowell03},
  APEX-SZ \citep{dobbs06},  
  LABOCA \citep{kreysa03},
  ACT \citep{kosowsky03},
  and SPT \citep{ruhl04}.
  Since these cameras are operated at ground-based telescopes,
  they all see emission from water vapor in the atmosphere.
  In almost all cases, the raw data from these cameras is 
  dominated by atmospheric noise caused by fluctuations
  in this emission.\footnote{
    The column depth of oxygen in the atmosphere
    also produces a non-negligible amount of emission,
    a factor of a few less than the emission from water
    vapor under typical conditions at Mauna Kea.
    However, the oxygen in the atmosphere is well
    mixed, and therefore fluctuations in the emission
    are minimal.
    In contrast, the temperature of the atmosphere tends to
    be close to the condensation point of the water vapor,
    and causes the water vapor to be poorly mixed
    in the atmosphere.
    Therefore, there are in general significant fluctuations
    in the emission from water vapor~\citep{masson94}.}
  All of these cameras make use of the fact that the
  atmospheric water vapor is in the near field,
  and therefore most of the fluctuations in the atmospheric
  emission are recorded as a common-mode signal among
  all of the detectors
  \citep{jenness98, borys98, 
  reichertz01, weferling01, archibald02}.
  Most of the atmospheric noise can be removed from the data
  by subtracting this common-mode signal,
  and this method has been shown to be at least
  as effective as the
  traditional beam-switching or chopping techniques
  \citep{conway63, weferling01, archibald02}.

  However, this subtraction does not allow recovery of BLIP
  performance on scales where the atmospheric signal
  is largest (\emph{i.e.}, at low frequencies in the time-stream data).
  In the case of Bolocam, the majority of the atmospheric
  fluctuations can be removed by subtraction of the common mode signal;
  but the residual atmospheric noise still limited the sensitivity
  of our data, thus motivating further study of these
  atmospheric fluctuations.
  This study focused on two main topics:
  1) determining the 
  phenomenology of the atmospheric noise
  (\emph{i.e.}, could it be modeled in a simple and robust way) and
  2) finding more effective ways to remove the
  atmospheric noise based on this modeling.

  \subsection{Instrument Description}
   \label{sec:inst}

    Bolocam is a large format, millimeter-wave camera designed
    to be operated at the CSO,
    and $\simeq 115$ optical detectors were used 
    for the observations described in this paper.
    Cylindrical waveguides and a metal-mesh filter are used 
    to define the passbands for the detectors,
    which can be centered at either 143 or 268~GHz with
    a $\simeq 15$\% fractional bandwidth.
    Note that, for either configuration, 
    the entire focal plane uses the same passband.
    A cold (4~K) Lyot stop is used to define
    the illumination of the 10.4~m primary mirror
    with a diameter of $\simeq 8$~meters, 
    and the resulting far-field beams
    have full-width half-maximums (FWHMs)
    of 60 or 30 arcsec (143 or 268 GHz).
    The detector array,
    which utilizes silicon nitride micromesh (spider-web)
    bolometers \citep{mauskopf97}, has a hexagonal geometry
    with nearby detectors separated by 40~arcsec, and the
    FOV is approximately 8~arcmin.

    The optical efficiency from the cryostat window to the
    detectors is 8\% at 143~GHz and 19\% at 268~GHz; 
    at each frequency approximately half of the loss
    in efficiency is due to coupling to the Lyot
    stop and half is due to 
    inefficiencies (reflection, standing waves, or loss) in the 
    metal-mesh filter stack.
    At 143~GHz the typical optical load from the 
    atmosphere is relatively small ($\simeq 0.5$~pW or
    10~K), but the total optical load is 
    $\simeq 4$~pW (80~K), most of which is
    sourced by warm surfaces inside the relay optics box.
    The atmosphere contributes an optical load of
    $5-15$~pW (20-60~K) per detector at 268~GHz,
    and there is an additional load of 
    $\simeq 10$~pW (40~K) due to the warm and 
    cold optics.
    Optical shot and Bose noise contribute in roughly
    equal amounts to the total photon noise at
    each observing frequency,
    with the BLIP $NEP_{\gamma} \simeq 1.5$~mK/$\sqrt{\textrm{Hz}}$ 
%2009/10/30
    (2.3~mK$_{CMB}$/$\sqrt{\textrm{Hz}}$) at 143~GHz and the BLIP
    $NEP_{\gamma} \simeq 0.8$~mK/$\sqrt{\textrm{Hz}}$
    (4.5~mK$_{CMB}$/$\sqrt{\textrm{Hz}}$)
    at 268~GHz.\footnote{
      The subscript $CMB$ is used throughout this paper to
      denote CMB temperatures;
      all temperatures given without a subscript refer to
      Rayleigh-Jeans temperatures.}
%      Note that these NEP values are given in Rayleigh-Jeans
%      temperatures, the subscript $CMB$ is used throughout
%      this paper when referrring to CMB temperatures.}
    More details of the Bolocam instrument can be found 
    in \cite{glenn98}, \cite{glenn03}, 
    \cite{haig04}, and \cite{golwala08}.

    The data we describe in this paper were collected during three 
    separate observing runs at the CSO:
    a 40 night run at 143~GHz in late 2003, 
    a 10 night run at 268~GHz in early 2004, and
    a 50 night run at 268~GHz in early 2005.
    For the 143~GHz observations, we focused on two science fields, one
    centered on the Lynx field at 08h49m12s, +44d50m24s (J2000) and
    one coinciding with the Subaru/XMM Deep Survey
    (SXDS or SDS1) centered at 02h18m00s, -5d00m00s (J2000).
    The 268~GHz observations were all focused on the COSMOS
    field at 10h00m29s, +2d12m21s (J2000).
    All three of these fields are blank, which means they contain
    very little astronomical signal.
    Therefore, our data are well suited to measure
    the signal caused by emission from the atmosphere.
    To map these fields, we raster-scanned the telescope
    parallel to the RA or dec axis at 4 arcmin/sec for
    the 143~GHz observations and 2 arcmin/sec for the 
    268~GHz observations.\footnote{
    Slower scan speeds improve our observing efficiency
    by reducing the fractional amount of time spent
    turning the telescope around between scans
    (the CSO turnaround time is approximately 10 seconds
    regardless of scan speed),
    but faster scan speeds improve the instantaneous 
    sensitivity of the camera by moving the signal
    band to higher frequencies where there is less
    atmospheric noise.
    Several scan speeds were tried at each observing
    frequency to find the best combination of
    observing efficiency and instantaneous sensitivity,
    and we found that 4 arcmin/sec is optimal for
    143~GHz observations and 2 arcmin/sec is 
    optimal for 268~GHz observations.
    Note that it may be possible to optimize the CSO
    telescope drive servo to improve the turnaround time,
    but this has not been attempted.
    \label{fn:CSO}}
    Throughout this paper, we will refer to single scans
    and single observations;
    a scan is one raster across the field and is $\simeq 15$
    seconds ($ \simeq 30-60$ arcmin) in length and
    an observation is a set of $\simeq 15-20$ scans that completely
    map the science field, which takes $\simeq 10$~minutes.
    Our total data set contains approximately 1000 
    observations at each observing frequency, 
    with the 143~GHz data split evenly among Lynx and SDS1.
    Flux calibration was determined from
    observations of Uranus, Neptune, and Mars, and 
    nearby quasars were used for pointing reconstruction.
    A more detailed description of the data is given
    in \cite{sayers09} and \cite{aguirre09}.

  \subsection{Typical Observing Conditions}

\begin{figure}
  \plotone{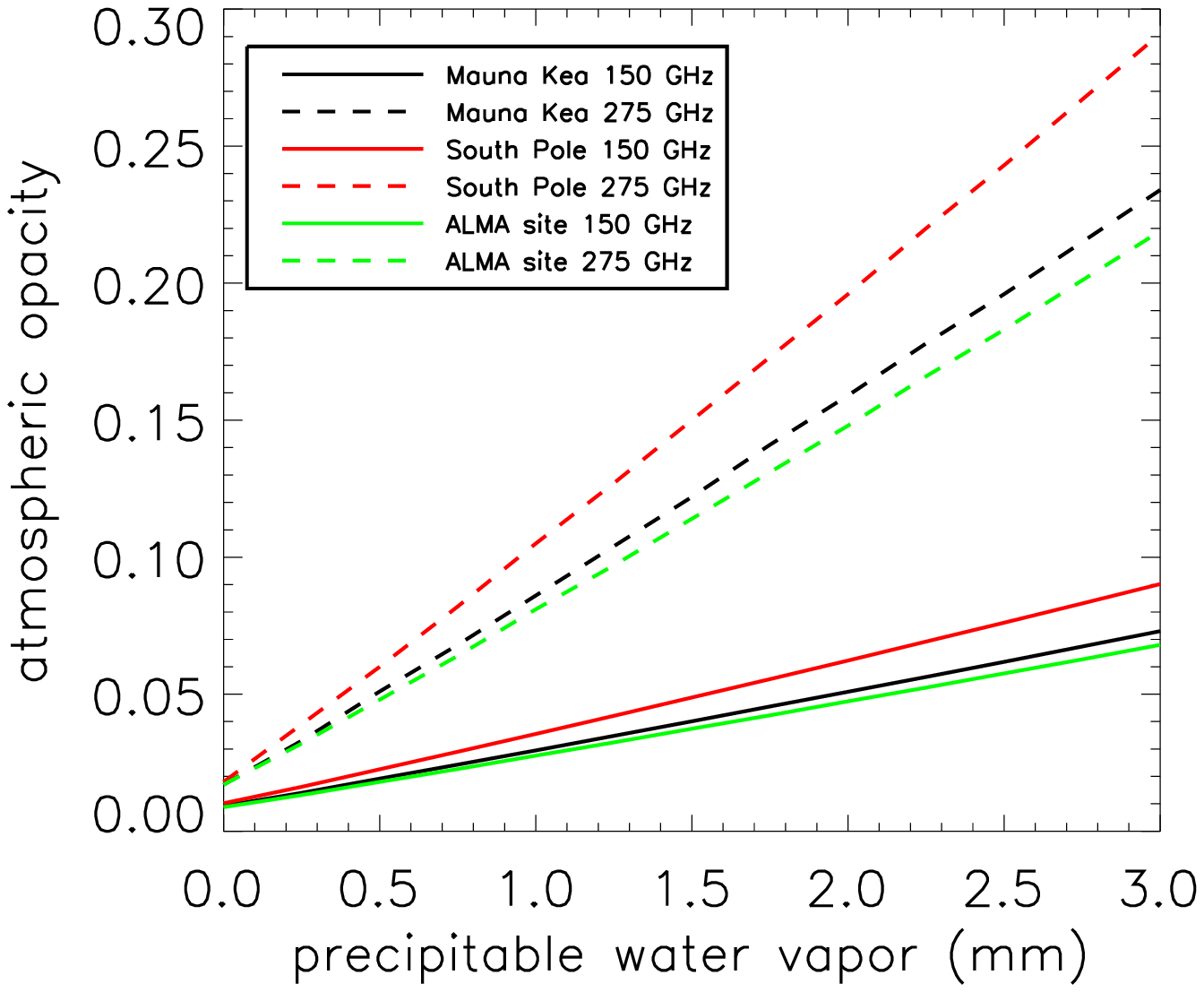}
  \caption{Atmospheric opacity as a function of $\mathcal{C}_{PW}$
	for Mauna Kea, the South Pole, and the ALMA site.
	The opacity is shown at 150~GHz and 275~GHz, the approximate
	centers of the Bolocam/ACBAR observing bands.
	All of the scalings were derived using the Pardo ATM
	algorithm \citep{pardo05, pardo01, pardo01_2}.}
  \label{fig:tau_vs_pwv}
\end{figure}

    Since atmospheric noise from water vapor is generally the
    limiting factor in the sensitivity of broadband, ground-based, 
    millimeter-wave observations, 
    the premier sites for these observations, 
    which include Mauna Kea, Atacama, and the South Pole,
    are extremely dry.
    On Mauna Kea, the CSO continuously monitors the
    atmospheric opacity
    with a narrow-band, heterodyne $\tau$-meter that
    measures the optical depth at 225~GHz 
    ($\tau_{225}$)~\citep{chamberlin99}.
    Since $\tau_{225}$ is a monotonically increasing function
    of the column depth of precipitable water vapor in the atmosphere,
    these $\tau_{225}$ measurements can be used to quantify
    the dryness of the atmosphere above Mauna Kea.
    Historically, the median value of $\tau_{225}$ is 0.091
    during winter nights, 
    which corresponds to a column depth of 
    precipitable water vapor of $\mathcal{C}_{PW} = 1.68$~mm
    \citep{pardo05, pardo01, pardo01_2}.
    The 25th and 75th centiles at Mauna Kea are 1.00 and 2.92~mm.
    Note that the 25th, 50th, and 75th centiles of our data
    sets closely match these historical averages, so our
    data are a fair representation of the average conditions on 
    Mauna Kea.
    For comparison, the median value of $\mathcal{C}_{PW}$
    at the ALMA site in Atacama is $\simeq 1.00$~mm
    during winter nights, while
    the median value at the South Pole
    is around 0.25~mm during the
    winter~\citep{radford00, lane98, peterson03, stark01}.\footnote{
	Note that the scaling between $\mathcal{C}_{PW}$ and 
	opacity is different at the three sites due to the
	different atmospheric conditions at each location.
	See Figure~\ref{fig:tau_vs_pwv}.}  

\section{Kolmogorov-Taylor/Thin-Screen Atmospheric Model}

  The K-T model of turbulence provides a good
  description of air movement in the 
  atmosphere~\citep{kolmogorov41,taylor38, tatarskii61}.
  According to the model, processes such as convection
  and wind shear inject energy into the atmosphere on
  large length scales, of order several 
  kilometers \citep{kolmogorov41,wright96}.
  This energy is transferred to smaller scales by
  eddy currents, until it is dissipated by viscous
  forces at Kolmogorov microscales, 
  corresponding to the smallest scales in
  turbulent flow and of order several millimeters
  for the atmosphere \citep{kolmogorov41}.
  For a three-dimensional volume, 
  the model predicts a power spectrum for the fluctuations
  from this turbulence that is
  proportional to $|\vec{q}|^{-11/3}$, where $\vec{q}$ is a 
  three-dimensional spatial
  frequency with units of 1/length.
  The same spectrum holds for particulates that
  are passively entrained in the atmosphere, such as 
  water vapor~\citep{tatarskii61}.

\begin{figure}
  \plotone{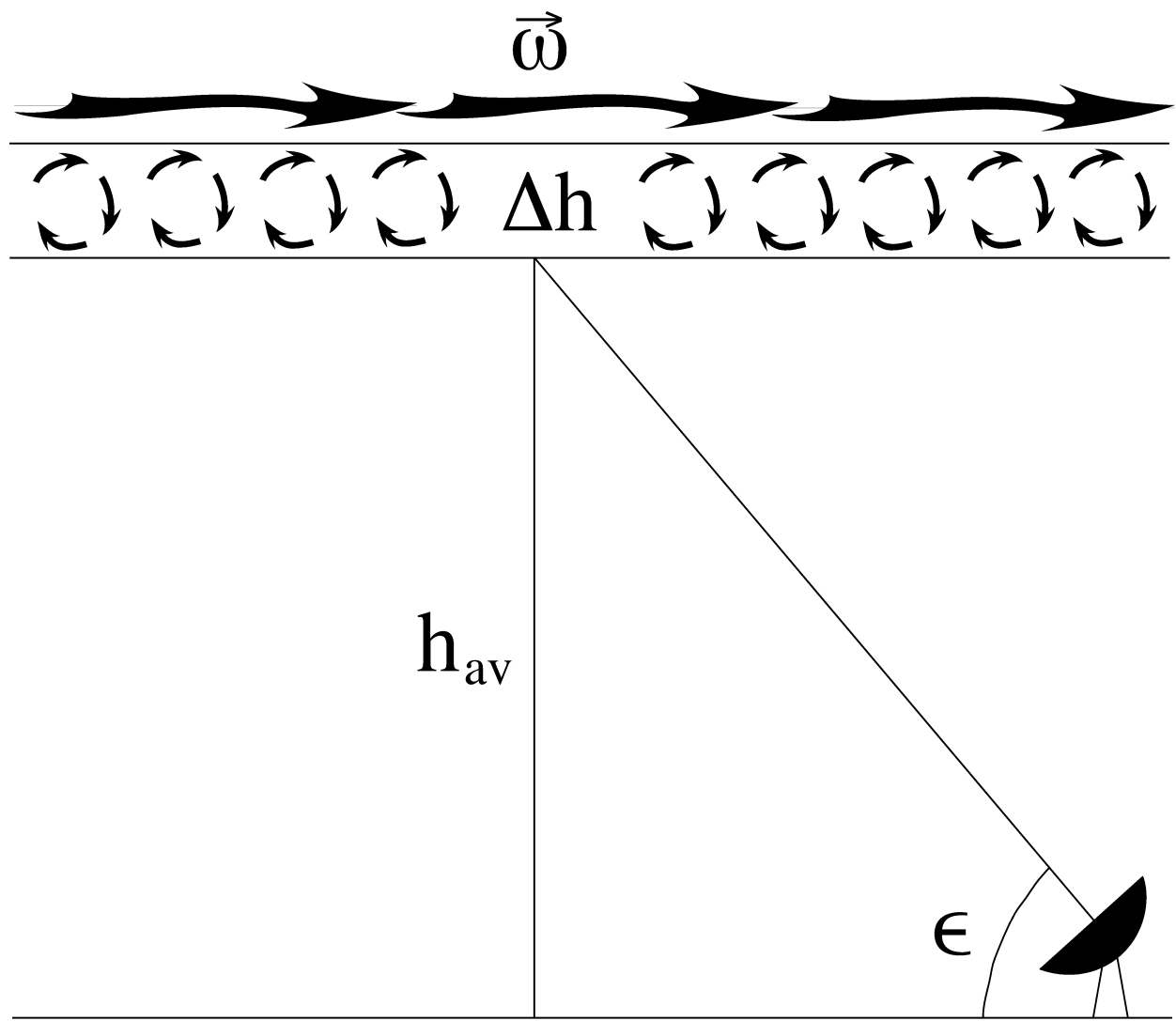}
  \caption{A diagram of the thin-screen turbulence model 
    described by
    \citet{Lay00} that is used throughout this paper.}
  \label{fig:diagram}
\end{figure}

  For our analysis, we adopted the two-dimensional
  thin-screen model described by \citet{Lay00},
  and a schematic of this thin-screen model is given in
  Figure~\ref{fig:diagram}.
  This model assumes that the fluctuations in water vapor occur
  in a turbulent layer at a height h$_{\rm{av}}$ with a 
  thickness $\Delta$h, where h$_{\rm{av}} \gg \Delta$h.
  This layer is moved horizontally across the sky by
  wind at an angular velocity $\vec{w}$.
  Given these assumptions and following the notation
  of \citet{Bussmann05}, the three-dimensional
  Kolmogorov-Taylor power spectra reduces to
  \begin{equation}
    P(\vec{\alpha}) = B_{\nu}^2 (\sin{\epsilon})^{(1-b)}
    |\vec{\alpha}|^{-b},
    \label{eqn:2_d_atm_psd}
  \end{equation}
  where $B_{\nu}^2$ is the amplitude of the power
  spectrum at zenith, 
  $\epsilon$ is the elevation angle of the telescope, 
  $\vec{\alpha}$ is the two-dimensional angular frequency
  with units of 1/radians,
  and $b$ is the power law of the model (equal to 11/3 for the 
  K-T model).
  Note that $B_{\nu}^2$ has units of mK$^2$ rad$^{-5/3}$ for
  $b = 11/3$.

\section{Fitting Bolocam Data to the K-T Theory}

  \subsection{Calculating the Wind Velocity}
    \label{sec:time_lag}

    If the angular wind velocity, $\vec{w}$,
    is assumed to be constant and the 
    spatial structure of the turbulent layer is static
    on the time scales required for the wind to move the
    layer past our beams \citep{taylor38},
    then detectors aligned with the angular wind velocity will
    see the same atmospheric emission, but at different 
    times~\citep{church95}.
    Making reasonable assumptions for the wind speed (10~m/s)
    and height of the turbulent layer (1~km) yields
    an angular speed of approximately
    30~arcmin/sec for the layer.
    Note that this is much faster than our maximum scan speed of 
    4~arcmin/sec.
    Since the diameter of the Bolocam focal plane is 8~arcmin,
    the angular wind velocity and spatial structures only need to
    be stable for a fraction of a second to make 
    our assumption valid.
    To look for these time-lagged correlations, we computed
    the relative cross power spectrum between every pair
    of bolometers, described by
    \begin{displaymath}
      xPSD_{i,j}(f_m) = \frac{D_{i}(f_m)^{*} D_{j}(f_m)}
      {\sqrt{|D_i(f_m)|^2} \sqrt{|D_{j}(f_m)|^2}},
    \end{displaymath}
    where $xPSD_{i,j}(f_m)$ is the relative cross PSD
    between bolometers $i$ and $j$, 
    $D_i(f_m)$ is the Fourier transform of the 
    data time-stream for bolometer $i$
    at Fourier space sample $m$,
    and $f_m$ is the frequency (in Hz) of
    sample $m$.

\begin{figure}
  \plottwo{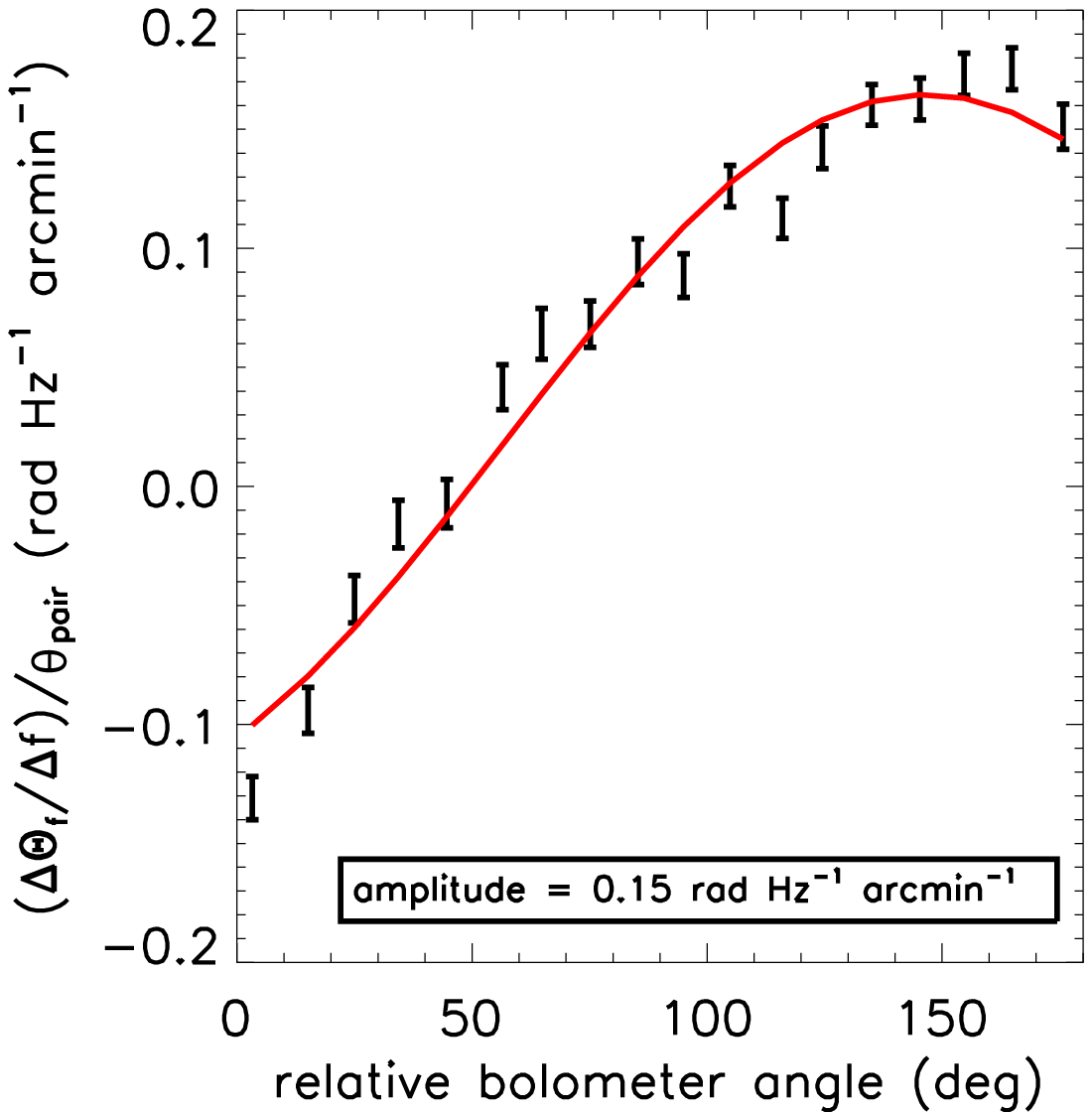}{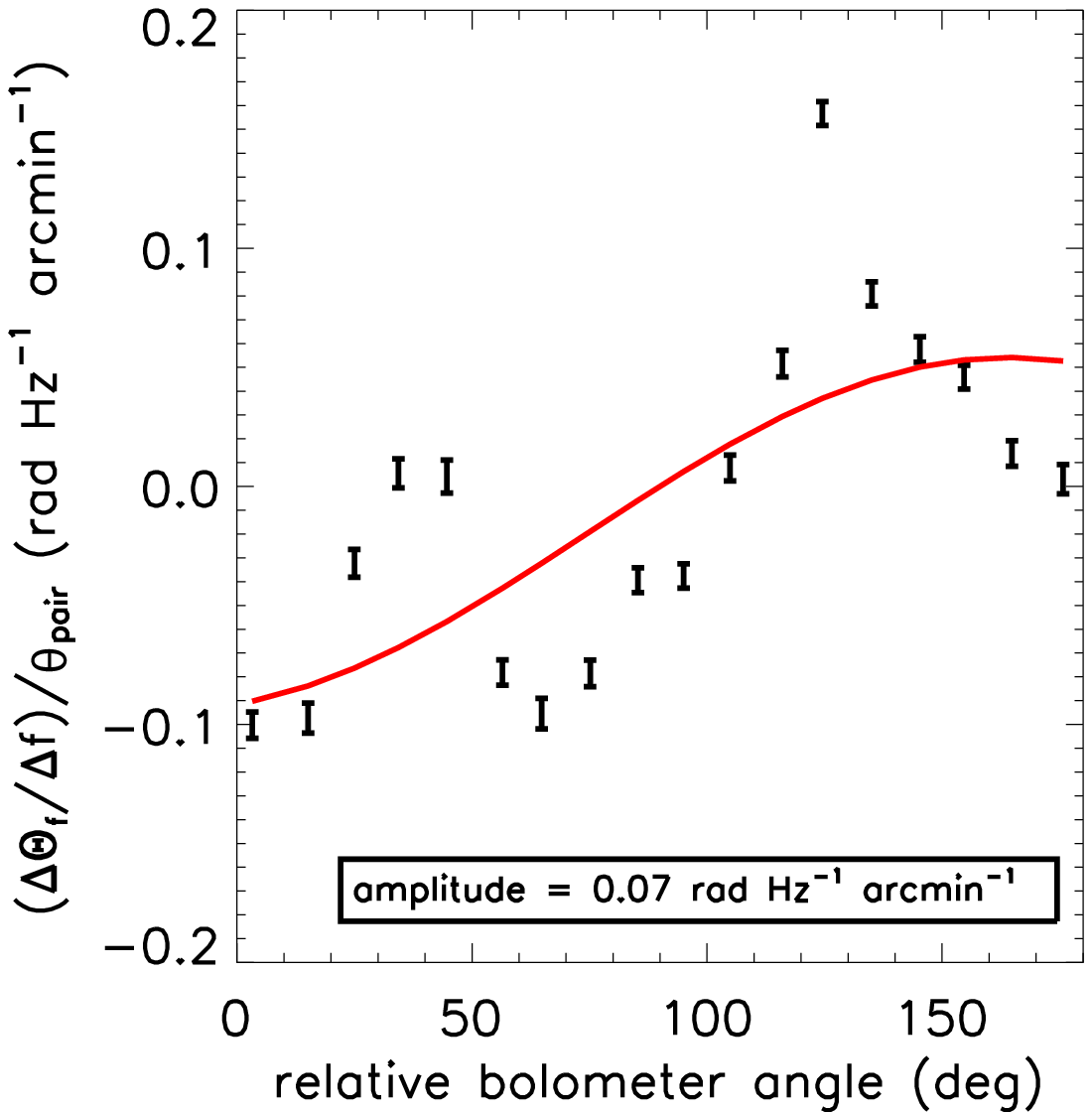}
  \caption{
    Plots of  
    $\frac{\left( \Delta \Theta_f/\Delta f \right)}
    {\theta_{pair}}$ averaged over all bolometer
    pairs and all scans for a single observation.
    This slope is binned according 
    to $\phi_{pair}$, and the sinusoidal fit predicted from
    the thin-screen K-T model	
    is overlaid in red.
    In general, roughly half our data are well described by
    this model, with a typical example shown in the left-hand plot.
    The other half of the data tend to contain outliers and/or
    additional features; the right-hand plot shows an example
    of one of these data sets.}
  \label{fig:gamma_hist}
\end{figure}

    If two bolometers see the same signal at different times, then the 
    cross PSD of these bolometers will have a phase angle described by
    \begin{displaymath}
      \tan^{-1}(xPSD) = \Theta_f = 2\pi f \Delta t
    \end{displaymath}
    where $f$ is the frequency in 
    Hz and $\Delta t$ is the time difference (in sec)
    between the signal recorded by the two bolometers.
    Therefore, the slope of a linear fit to $\Theta_f$ versus $f$ will be 
    proportional to $\Delta t$.
    If the simple atmospheric model we have assumed is correct, then 
    $\Delta t / \theta_{pair}$ should be a sinusoidally varying function
    of the relative angle on the focal plane
    between the bolometer pair, $\phi_{pair}$,
    where $\theta_{pair}$ is the angular
    separation of the two bolometers
    (\emph{i.e.}, if one bolometer is located at position $(x_1,y_1)$
    and another bolometer is located at position $(x_2,y_2)$,
    then $\phi_{pair} = \tan^{-1}(\frac{y_2-y_1}{x_2-x_1})$ and 
    $\theta_{pair} = \sqrt{(x_2-x_1)^2 + (y_2-y_1)^2}$).
    Some examples of 
    $2\pi\Delta t / \theta_{pair}$ versus $\phi_{pair}$ are given in
    Figure~\ref{fig:gamma_hist}.
    In general, the model provides an excellent fit for roughly
    half of our data (typically the data collected in better 
    weather as quantified by the time-stream RMS).
    The remaining data tend to contain several outliers
    and/or features in addition to the underlying sinusoid
    given by the model.

\begin{figure}
  \plotone{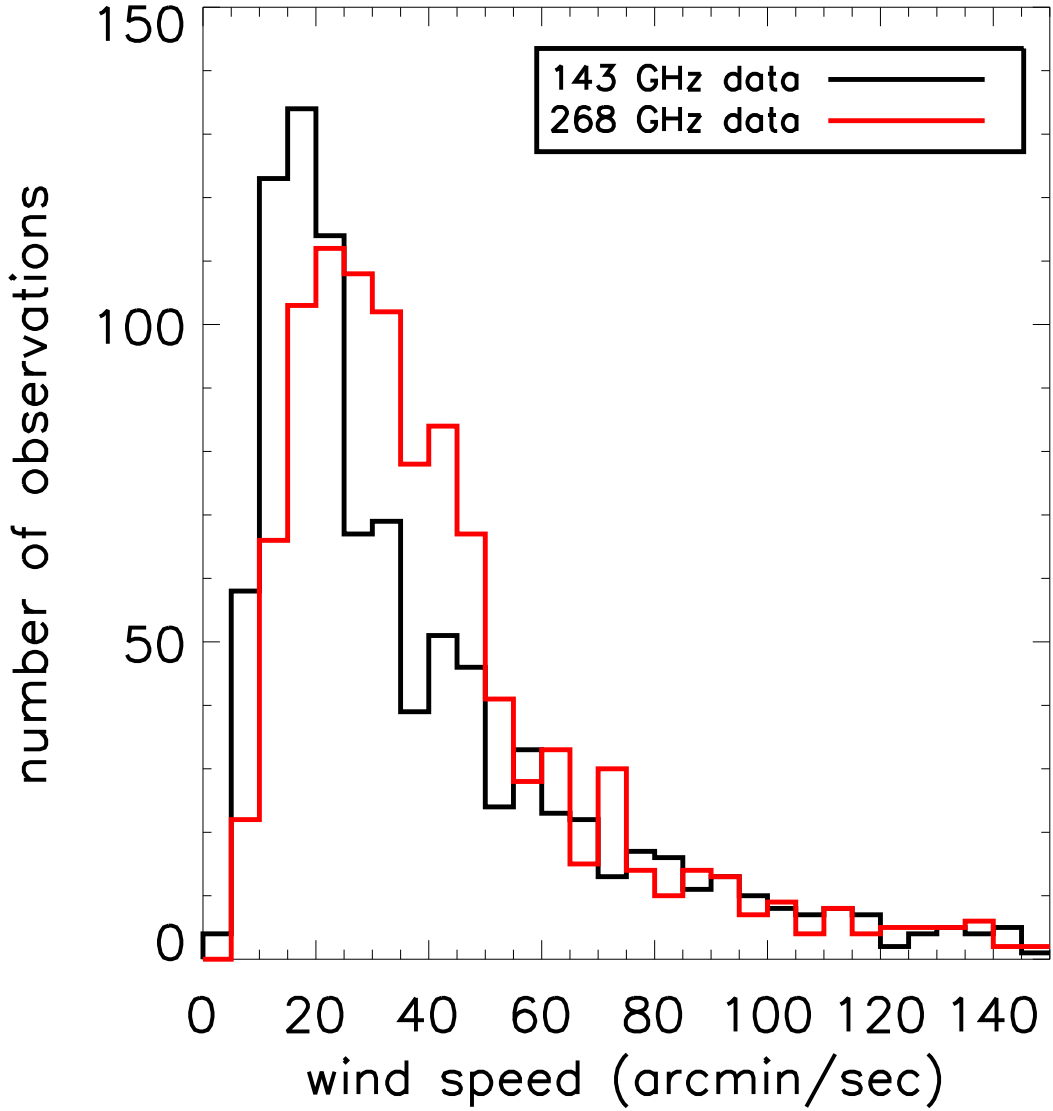}
  \caption{The angular wind speed of the turbulent layer for
    every observation at both 143 GHz and 268 GHz.
    Note that the median value of the distributions is 31 and
    35 arcmin/sec, respectively.
    This corresponds to a linear speed of 10 m/s if the layer is
    at a height of 1 km, which is physically reasonable.}
  \label{fig:wind_speed}
\end{figure}

    The model fits also provide an estimate of the angular wind
    speed, with
    \begin{displaymath}
      |\vec{w}| = \theta_{pair} / \Delta t,
    \end{displaymath}
    where $\theta_{pair} \simeq 40$ arcsec for adjacent detectors
    on the Bolocam focal plane.
    Histograms showing the angular wind speed for all of our
    observations at both 143 and 268~GHz are given in
    Figure~\ref{fig:wind_speed}.
    Note that the median
    angular wind speed is 31 arcmin/sec for the 143~GHz data
    and 35 arcmin/sec for the 268~GHz data,
    which is approximately what we
    expected for a physically reasonable model of the
    atmosphere.

  \subsection{Instantaneous Correlations}
    \label{sec:time_inst_model}

    Equation~\ref{eqn:2_d_atm_psd} can be converted from a 
    power spectrum in angular frequency space to a correlation
    function as a function of angular separation.
    Since the power spectrum is azimuthally symmetric,
    we can write $P(\vec{\alpha})$ as $P(\alpha)$, where
    $\alpha = | \vec{\alpha} |$.
    This power spectrum will produce a correlation function according to
    \begin{equation}
      C(\theta) = 2\pi \int_{\alpha_{min}}^{\infty} d\alpha \
      \alpha \ P(\alpha) \ J_0(2\pi\alpha\theta),
      \label{eqn:atm_corr}
    \end{equation}
    where $\theta$ is the angular separation in radians,	
    $\alpha_{min}$ is the maximum length scale of the 
    turbulence,
    and $J_0$ is the 0$^{\rm{th}}$-order Bessel function
    of the first kind.
	
    To compare our data to this model, we calculated the 
    correlation between the time-streams of every bolometer pair
    according to
    \begin{displaymath}
      C_{ij} = \frac{1}{N} \sum_n d_{in} d_{jn},
    \end{displaymath}
    where $C_{ij}$ is the correlation
    between bolometer $i$ and bolometer $j$ in mK$^2$,
    $N$ is the number of time-stream samples, 
    and $d_{in}$ is the time-stream data for
    bolometer $i$ at time sample $n$.
    A single correlation value for each pair
    was calculated for each 
    $\simeq 15$-second-long
    scan made while observing one of the science fields, and then averaged
    over the twenty scans in one complete observation of 
    the field.
    Therefore, we have assumed that the atmospheric
    noise conditions do not change over the 
    $\simeq 10$-minute-long observation
    and are independent of the scan direction, which
    is reasonable given that the typical angular wind speed
    is much larger than our scan speed.
    The $C_{ij}$ were then binned as a function of angular
    separation between bolometer $i$ and bolometer $j$
    to give correlation as a function of $\theta$.

    Ideally, we would like to compare our data directly to the theoretical
    model using Equation~\ref{eqn:atm_corr}.
    However, evaluating the integral in Equation~\ref{eqn:atm_corr}
    is non-trivial, especially when
    the effects of Bolocam's finite beams, data processing, etc.
    are included.
    Therefore, we have determined the theoretical correlation
    function based on the K-T model via simulation.
    First, we generate 50 two-dimensional projections 
    (\emph{i.e.}, maps) of the
    atmospheric fluctuation signal according to the 
    power spectrum given in Equation~\ref{eqn:2_d_atm_psd}.
    In each of these realizations, the phases of the different
    spatial frequency components are taken to be random.
    Next, we convolve each map with the profile of a 
    Bolocam beam.\footnote{
    Since the far field distance for Bolocam is tens of
    kilometers, we assume that the atmospheric fluctuations
    occur in the near field.
    Therefore, the Bolocam beams can be well approximated
    by the primary illumination pattern, which is approximately a top
    hat with a diameter of 8~m.
    This means that the angular size of the beam will
    depend on the height of the turbulent layer.}
    Then, we generate time-stream data by moving
    the atmospheric fluctuation map across our detector
    array at a rate given by
    the angular wind speed we calculated in Section~\ref{sec:time_lag}.
    These simulated time-streams are then processed in the 
    same way as our real data, including removing the mean
    signal level from each $\simeq 15$-second-long scan.
    Finally, we determine the values of $C_{ij}$ for the 
    simulated data, averaging over all 50 realizations,
    and bin these $C_{ij}$ as a function of bolometer
    separation.

\begin{figure}
  \plottwo{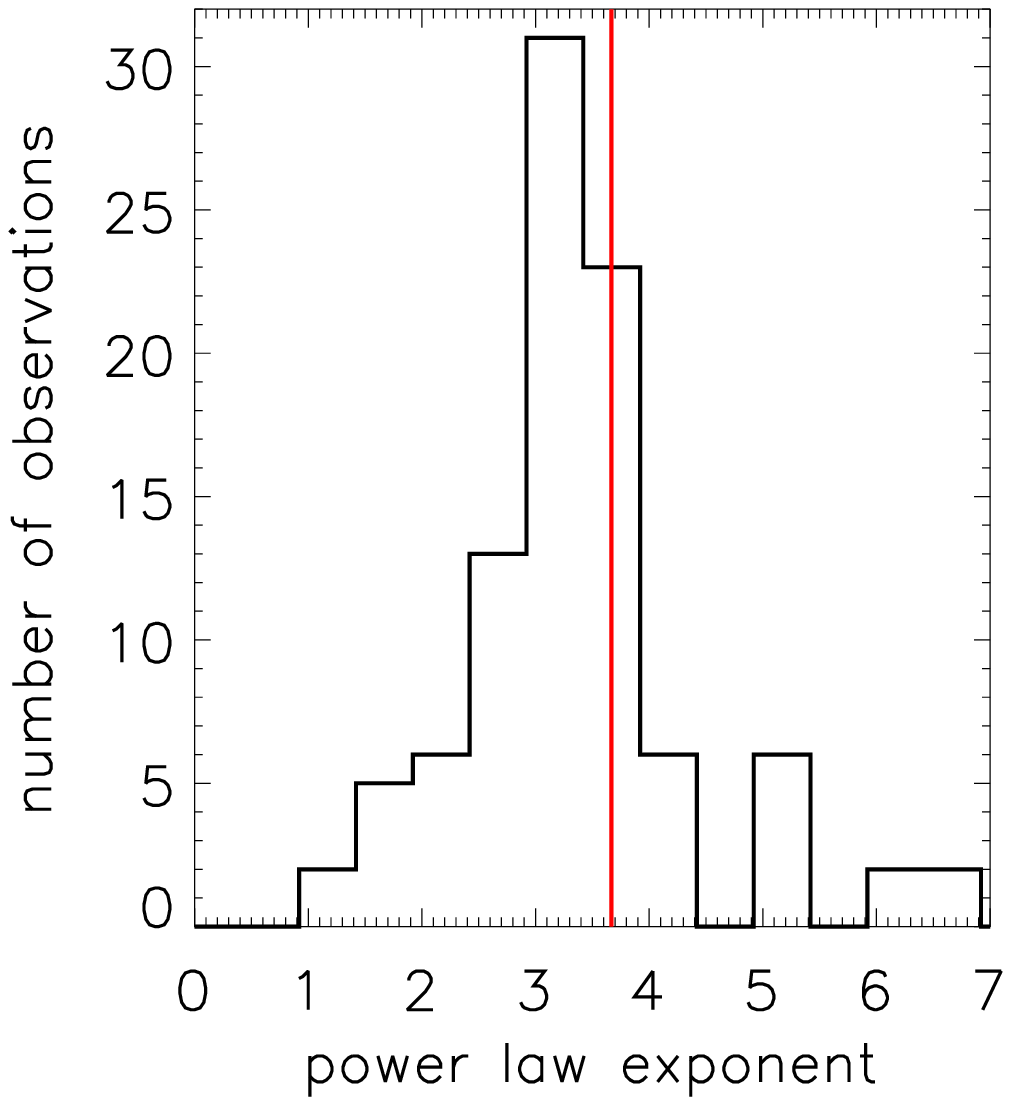}{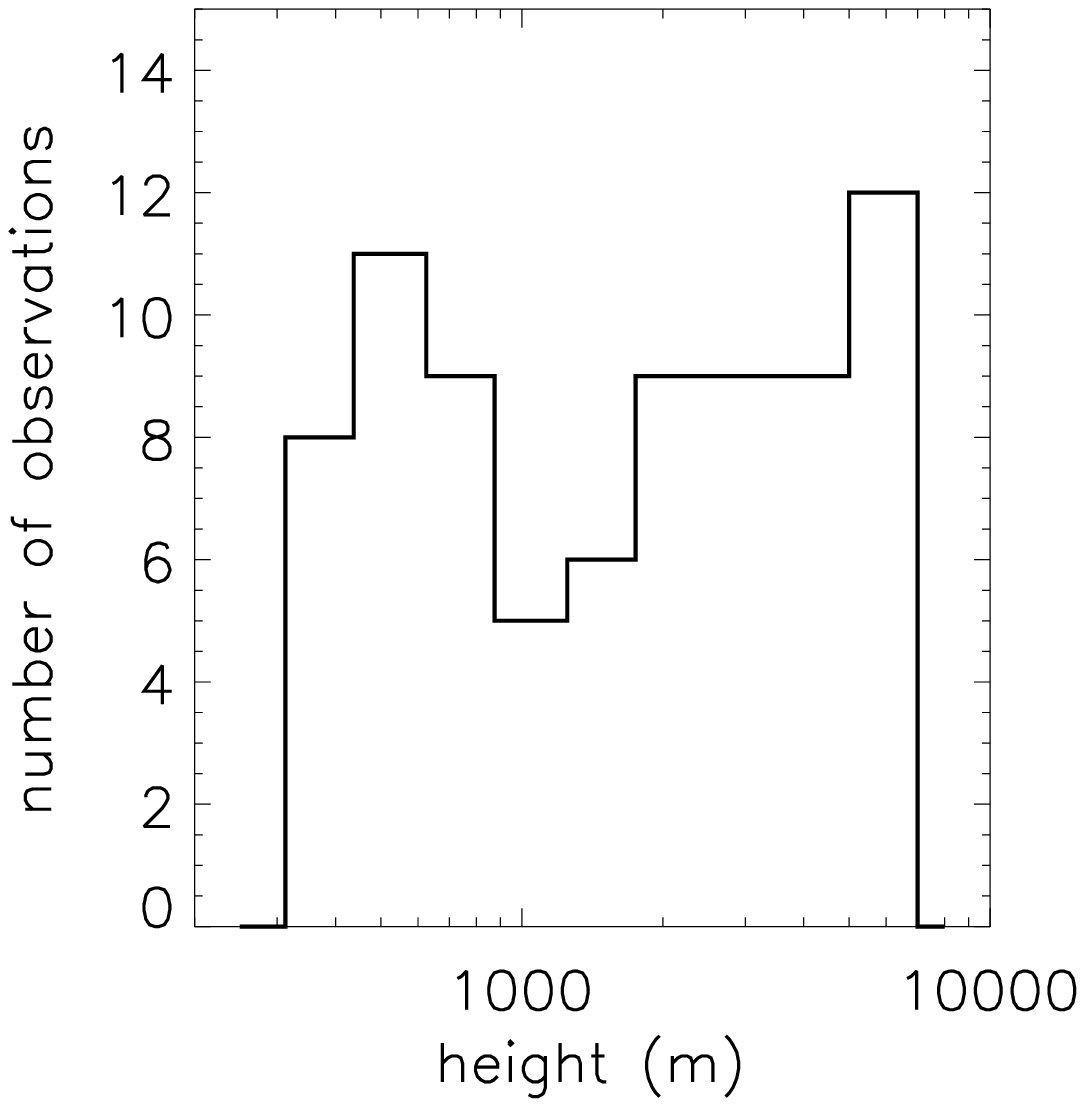}
  \caption{The histogram on the left shows the best fit value
    of the power law exponent $b$ for the K-T model of the atmosphere
    for a randomly selected subset of 96 143~GHz observations.
    The mean is 3.3 and the standard deviation is 1.1, indicating
    that our data are consistent with the K-T model
    prediction of $b=11/3$, which is shown as a red vertical line.
    The histogram on the right shows the best fit value for the
    height of the turbulent layer.
%2009/10/30
    The uniform distribution of $h$ over the allowed range 
    indicates that we do not meaningfully constrain $h$.
%    Note that the bin widths are approximately logarithmic in height.
%    Our data are sensitive to the quantity $1/h$; the best fit
%    value of $1/h$ corresponds to a height of $1000^{+3000}_{-500}$~m.
%    This large spread in the value of $h$ is likely not physical,
%    but rather is caused by our relative inability to constrain the
%    height of the turbulent layer.
    }
  \label{fig:power_law_hist}
\end{figure}

    The shape of the theoretical $C(\theta)$ determined from
    these simulations will depend not only on the 
    value of the power law index, $b$, but also on the 
    height of the turbulent layer, $h$.
    Any reasonable value of $h$ will be in the near field
    for Bolocam, so the physical size of the beam profiles
    (in meters) will
    be approximately independent of $h$, which means that the
    angular size of the beams in the
    turbulent layer will be a function of $h$.
    Therefore, a change in the height of the turbulent layer
    will cause a change in the way that the angular
    emission profile of the atmosphere is smoothed by
    the Bolocam beams, which will result in a 
    different profile for $C(\theta)$.
%    Consequently, our measured correlation profiles
%    as a function of separation are sensitive to both $b$ 
%    and $h$ (along with $B_{\nu}^2$), although the constraints
%    on $b$ and $h$ are not precise.
%2009/10/30
    Thus, in principle, our measured correlation profiles as a function
    of separation are sensitive to both $b$ and $h$	
    (along with $B_{\nu}^2$).
    However, as we explain below and show in Figure~\ref{fig:power_law_hist},
    we obtain no meaningful constraint on $h$
    because our measurement uncertainty on $C(\theta)$ is large compared to
    the 
    variations in $C(\theta)$ with $h$.
%    In practice, the variations in $C(\theta)$ over a reasonable range
%    of $h$ are too mild for us to place a constraint on the height of 
%    the turbulent layer.
	
\begin{figure}
  \plottwo{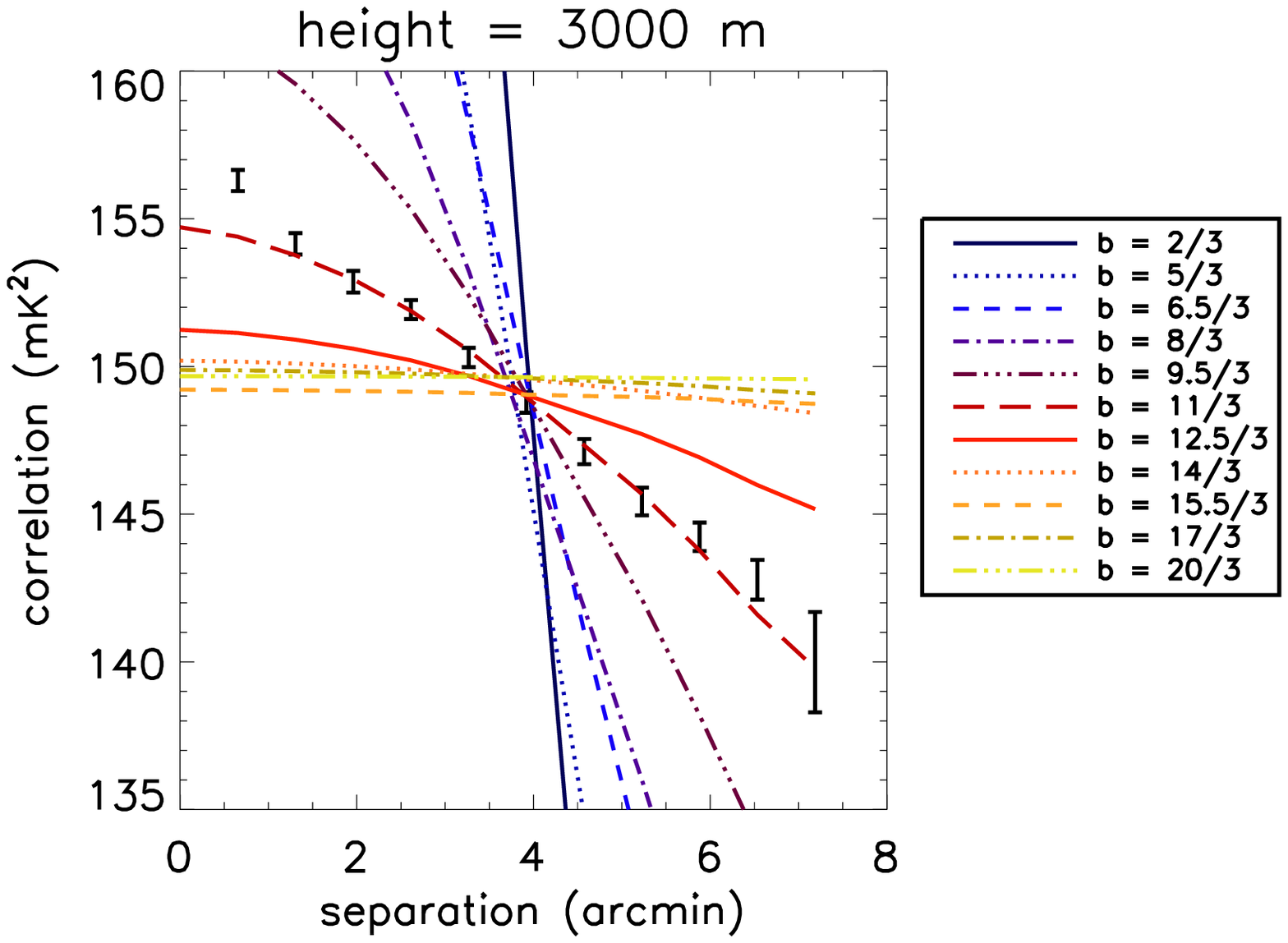}{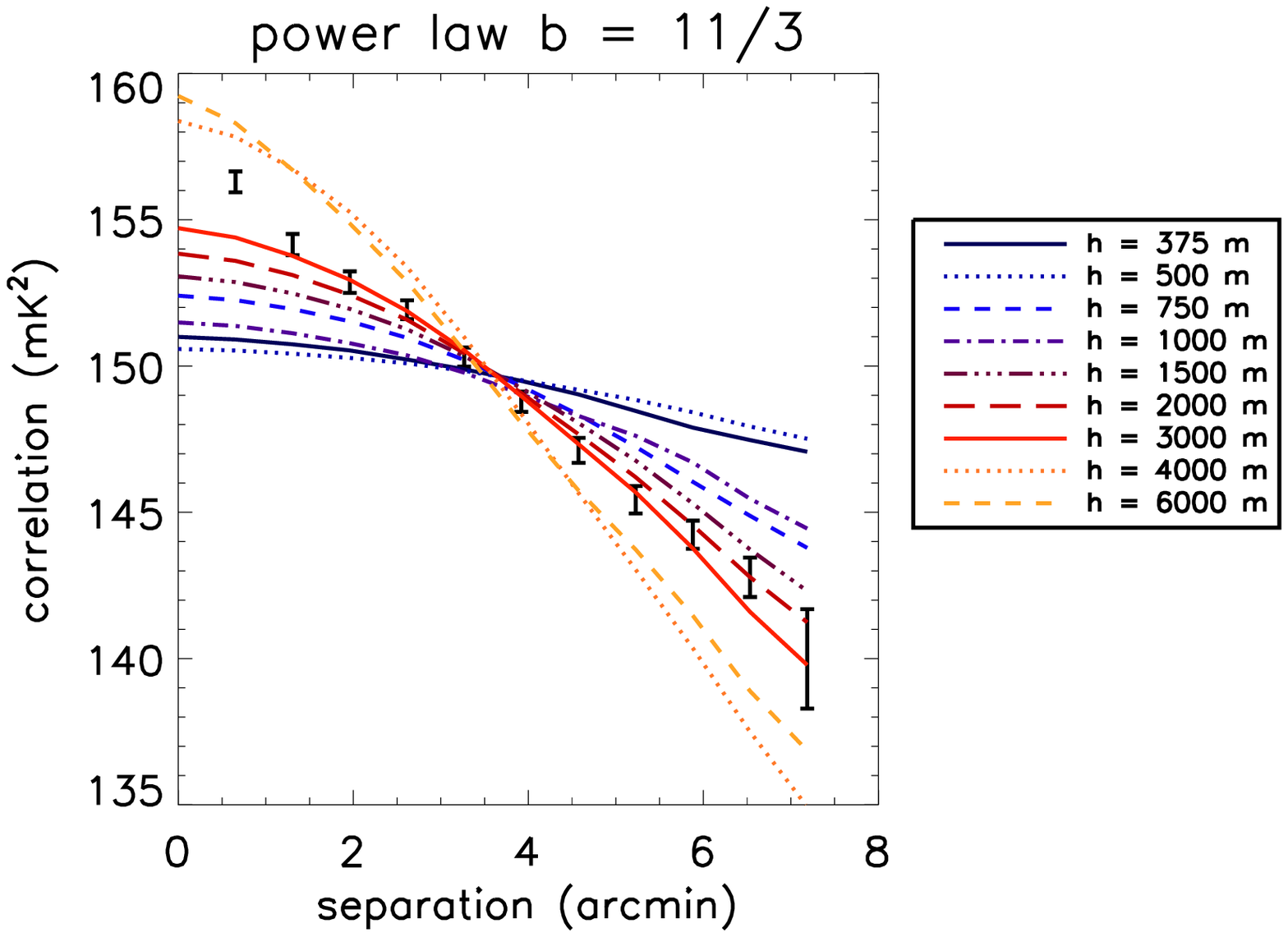} \\
  \plottwo{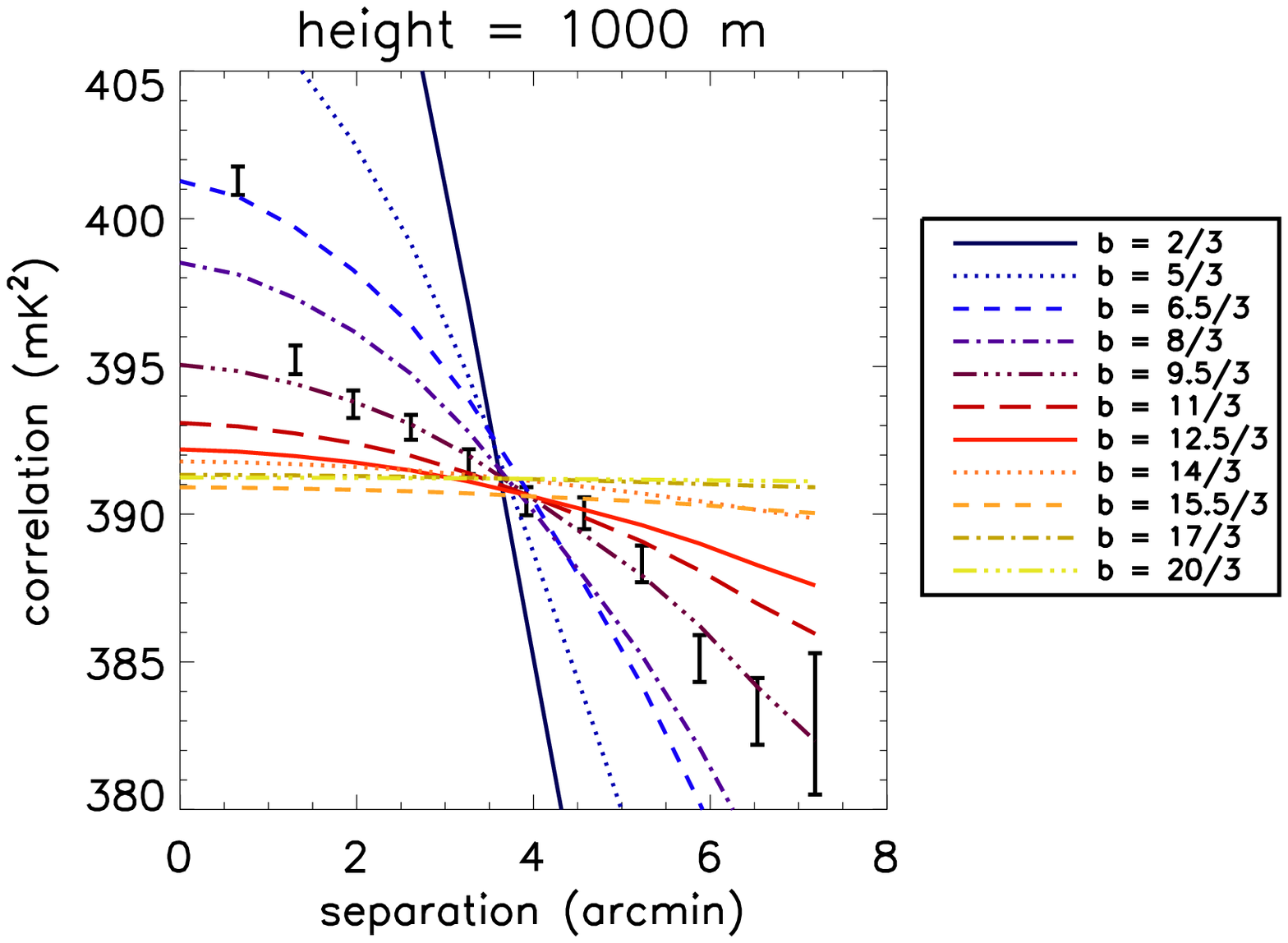}{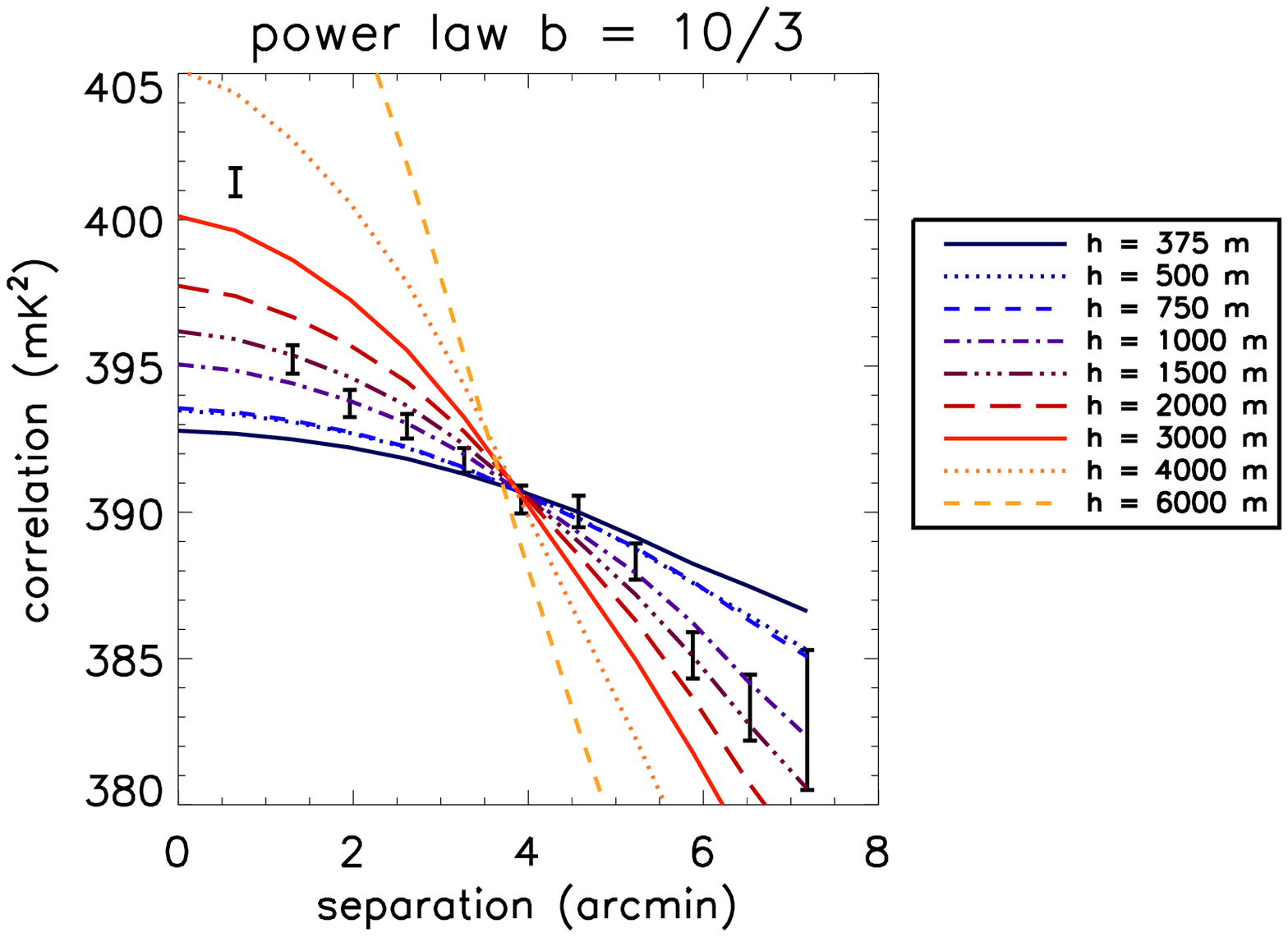}
  \caption{Plots of the average correlation between bolometer pairs
    as a function of separation between the bolometers.
    The top row shows data from a 143~GHz observation taken when the 
    amplitude of the atmospheric noise is better than average, and the 
    bottom row shows data from a 143~GHz observation taken when the 
    amplitude of the atmospheric noise is worse than average.
    The model fits overlaid on the left plots show a range of 
    power law indices, $b$, at the best fit value of $h$ for the 
    data set.
    The model fits overlaid on the right plots show a range
    of heights, $h$, at the best fit value of $b$ for the
    data set.
    The $\chi^2$ value of the model fit 
    for these two observations is similar, and is at
    roughly the 30th centile of our complete set of data
    (\emph{i.e.}, 1/3 of our observations produce a better fit to the
    K-T model, and 2/3 of our observations
    produce a worse fit to the K-T model).
    Therefore, the quality of the model fit for these observations is 
    fairly typical.
    Note the degeneracy between $b$ and $h$ in the 
    general shape of the model fits,
    which makes it difficult to constrain either value
    precisely for a single observation, especially $h$.
    These plots clearly show the excess correlation among adjacent
    bolometers, and note that the adjacent bolometer correlations
    are discarded when fitting the K-T model to the data.}
  \label{fig:corr_plots}
\end{figure}

    Initially, we assumed that both the height $h$ and the
    power law index $b$ were unknown, and ran simulations over
    a grid of values for each parameter.
    In our grid the values of $b$ ran from 2/3 to 20/3 in steps of 1/2, 
    and the values of $h$ were 375, 500, 750, 1000, 1500, 
    2000, 3000, 4000, and 6000~m.
    Note that we used an irregular step size for $h$ because the
    beam size is proportional to $1/h$.
    Since the computation time required for our simulation
    is substantial, we were only able to run the full grid
    of 121 different parameter values over a randomly selected
    subset of 96 143~GHz observations (approximately 10\% of
    our 143~GHz data).
    After computing the best fit value of $B_{\nu}^2$ for
    each observation and each grid point, we determined
    what values of $h$ and $b$ provided the best fit
    to the data.
    Note that the data from adjacent bolometer pairs is
    discarded before fitting a model, due to the 
    excess correlations between these pairs 
    (see Section~\ref{sec:bolo_corr}).
    Additionally, the constraints on $b$ or $h$ for a single
    observation are not very precise
    because there
    is a wide
    range of combinations of $b$ and $h$ that will produce
    very similar model profiles.
    Some examples of data with model fits overlaid are
    given in Figure~\ref{fig:corr_plots}.
    We found the 
    average best fit value of the power law
    $b$ is 3.3 with a standard deviation
    of 1.1, indicating that our data are consistent
    with the K-T model prediction of $b = 11/3$.	
    Note that \cite{Bussmann05} previously found the
    atmosphere above the South Pole to be consistent
    with the K-T model ($b = 3.9 \pm 0.6$ when only
    high signal to noise scans are included, $b = 4.1 \pm 0.8$
    when all scans are included) using 
    ACBAR data that was sensitive to much different
    physical scales in the atmosphere 
    ($\simeq 1.5$~m beams and a $\simeq 1$ deg FOV).\footnote{
      For ACBAR, the primary mirror is $\simeq 1.5$ m in
      diameter and adjacent detectors are separated by
      16 arcmin.
      As a result, the typical separation between ACBAR
      beams is larger than the diameter of a single
      beam as they pass through the water vapor in
      atmosphere (\emph{i.e.}, each ACBAR beam passes
      through a different column of atmosphere).
      In contrast, the $\simeq 10$ m primary at the CSO
      and 40 arcsec separation between adjacent 
      Bolocam detectors means that there is significant
      overlap between the beams as they pass through
      the water vapor in the atmosphere.} 
%    For the height of the water vapor we found that
%    the average best fit value of $1/h$ corresponds to
%    $1000^{+3000}_{-500}$~m.
%    Again, this is consistent with the 1.3~km scale
%    height that \cite{Bussmann05} found at the South
%    Pole using ACBAR data.
%    See Figure~\ref{fig:power_law_hist}.
%2009/10/23
    Figure~\ref{fig:power_law_hist} shows that 
    the best-fit values of $h$ were uniformly distributed over 
    the allowed range, indicating our data do not 
    meaningfully constrain $h$.

    We have so far assumed that the beams have a tophat profile
    while passing through the atmosphere.
%    To the extent that this assumption is false, and that the	
    If the profile is not a tophat and/or 
    varies among pixels, 
    then our simulation will predict 
    a $C(\theta)$ that is too flat.
%as $\theta$ increases.
%the amount of
%    correlation between pixels (\emph{i.e.}, 
%    our predicted $C(\theta)$ will 
%an 
%    additional decorrelation as a function of $\theta$
%    will be introduced.
    However, given that the data are consistent with the 
    K-T model prediction of $b = 11/3$, 
%    with a large fractional uncertainty, 
    there is no
    indication that such an effect is significant.

  \subsection{Atmospheric Noise Amplitude}

\begin{figure}
  \plotone{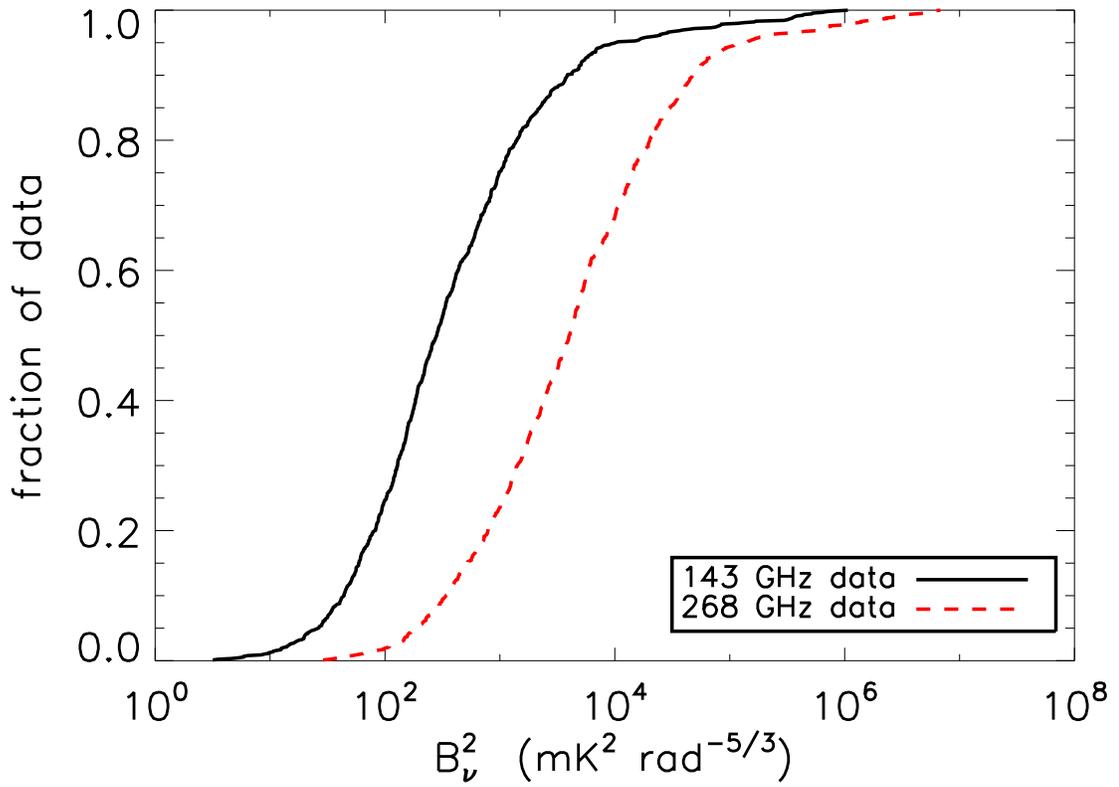}
  \caption{Plots of the cumulative distribution function
    of $B_{\nu}^2$ at both 143 and 268~GHz.}
  \label{fig:CDF}
\end{figure}

    After showing that our data are consistent with the 
    K-T model, we repeated the analysis of 
    Section~\ref{sec:time_inst_model} for all of our
    data.
    For each observation we generated 50 simulated
    atmospheric noise maps with the value of $b$
    fixed at 11/3 and the value of $h$ fixed at 1000~m.
%2009/10/30 
    We set $b = 11/3$ because this is the power
    law predicted by the theory and is consistent with our data.
    The value of $h$ was chosen based on
    independent measurements of the water vapor profile
    above Mauna Kea (\emph{e.g.},~\cite{pardo01_2}, estimated
    from Hilo radiosonde data).
    Note that 
    our primary result, a measurement of the distribution
    of $B_{\nu}^2$, does not depend strongly on 
    the choice of $h$ because 
    the best-fit value of $B_{\nu}^2$ is fairly
    insensitive to $h$.\footnote{
%    The value of $h$ was chosen to be roughly equal
%    to the best fit value of $h$ we determined
%    in Section~\ref{sec:time_inst_model}.
%    This choice of $h$ was somewhat arbitrary, 
%    but there is no need to be more precise
%    because the the value of $B_{\nu}^2$ is fairly
%    insensitive to the exact value of $h$.
%2009/10/30
      Varying $h$ 
      over the physically reasonable range that we allowed in
      Section~\ref{sec:time_inst_model} ($375-6000$~m)
      causes $B_{\nu}^2$ to vary by $\pm 15$\%
      compared to the value of $B_{\nu}^2$ at $h=1000$~m.
%      the average fractional difference in $B_{\nu}^2$
%      between these two heights is $\simeq 25$\%, indicating
%      that our choice of $h=1000$~m will add
%      $\simeq 12$\% to our uncertainty in 
%      determining $B_{\nu}^2$, which 
      This variation is comparable
      to the uncertainty in $B_{\nu}^2$ due to
      our flux calibration uncertainty.}
    For the 143~GHz data, the quartile values of $B_{\nu}^2$
    are 100, 280, and 980~mK$^2$ rad$^{-5/3}$, and for the 268~GHz
    data the quartile values are 1100, 4000, and 14000~mK$^2$ rad$^{-5/3}$.
    Note that the uncertainty in these values due to our	  
    flux calibration is approximately 12\%.
    Plots of the cumulative distribution function of 
    $B_{\nu}^2$ at each frequency are given in 
    Figure~\ref{fig:CDF}.

\begin{figure}
  \epsscale{0.75}
  \plotone{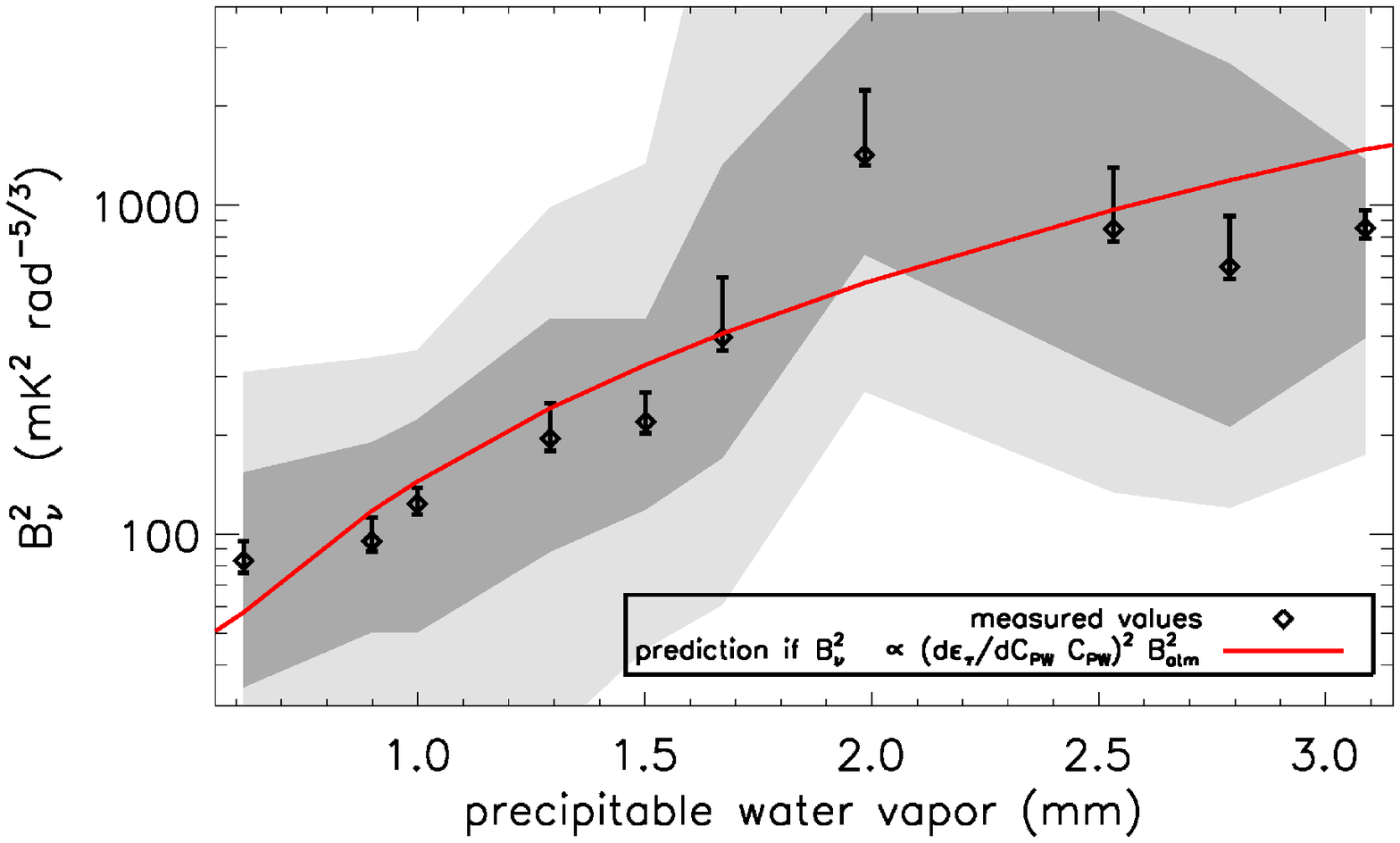} \\
  \plotone{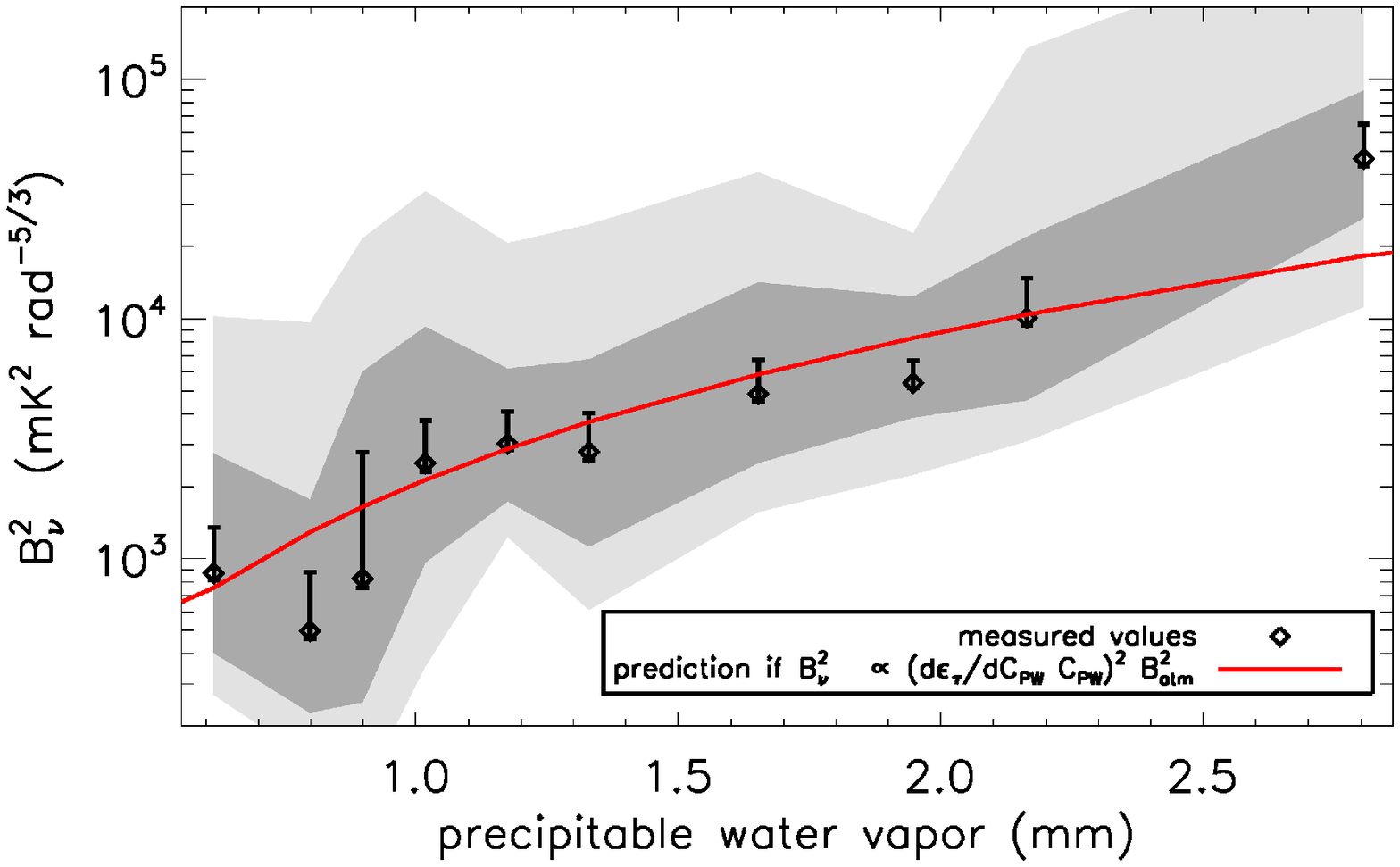}
  \caption{Plots of the amplitude of the atmospheric noise, 
    $B_{\nu}^2$, as a function of column depth of precipitable water
    vapor, $\mathcal{C}_{PW}$.
    The data points show the median value of $B_{\nu}^2$ and the
    error bars give the uncertainty on this median value;
    the light shaded region spans the $10-90$th centile values of
    $B_{\nu}^2$, and the darker shaded region spans the 
    $25-75$th centile values of $B_{\nu}^2$.
    The top plot shows Bolocam data collected at 143~GHz and 
    the bottom plot shows Bolocam data collected at 268~GHz.
    Overlaid on the plots is a fit to the data assuming that the
    fractional fluctuations in the column depth of precipitable
    water are constant (\emph{i.e.}, that $B_{\nu}^2$ is 
    proportional to
    $\left( \frac{d \epsilon_{\tau}} {d \mathcal{C}_{PW}} 
    \mathcal{C}_{PW} \right)^2 \mathsf{B}_{atm}^2$).}
  \label{fig:B_nu_vs_PW}
\end{figure}

    A reasonable phenomenological expectation is that
    the fractional 
    fluctuations in the column depth of
    water vapor are independent of the amount of water vapor
    (\emph{i.e.}, 
    $\delta\mathcal{C}_{PW} \propto \mathcal{C}_{PW}$).
    Since 
    \begin{displaymath}
      B_{\nu}^2 \propto (\delta \epsilon_{\tau})^2
      \mathsf{B}_{atm}^2,
    \end{displaymath}
    where $\epsilon_{\tau} = 1-e^{\tau_{\nu}}$ is the 
    emissivity of the atmosphere and
    $\mathsf{B}_{atm} = \frac{2 \nu^2}{c^2} k_B T_{atm}$
    is the brightness of the atmosphere in the
    Rayleigh-Jeans limit,
    this means that 
    \begin{displaymath}
      B_{\nu}^2 
      \propto \left( \frac{d \epsilon_{\tau}} {d \mathcal{C}_{PW}} 
      \delta\mathcal{C}_{PW} \right)^2 \mathsf{B}_{atm}^2 \propto
      \left( \frac{d \epsilon_{\tau}} {d \mathcal{C}_{PW}} 
      \mathcal{C}_{PW} \right)^2 \mathsf{B}_{atm}^2.
    \end{displaymath}
    Note that $\tau_{\nu}$
    is the total opacity of the atmosphere
    at observing frequency $\nu$.
    To test the validity of this expectation,
    we first considered the
    data in each observing band separately.
    The data sets for each observing band spanned a 
    wide range of weather conditions, and 
    in general our predicted scaling fit the data fairly
    well over the entire range.\footnote{
    During the course of our observations 
    $0.5 \lesssim \mathcal{C}_{PW} \lesssim 3.5$
    mm, and the value of
    $\left( \frac{d \epsilon_{\tau}} {d \mathcal{C}_{PW}} 
    \mathcal{C}_{PW} \right)^2 \mathsf{B}_{atm}^2$ varies
    by almost two orders of magnitude over this range of 
    $\mathcal{C}_{PW}$.}
    See Figure~\ref{fig:B_nu_vs_PW}.
    Additionally, we can test our 
    assumption that $\delta\mathcal{C}_{PW} \propto \mathcal{C}_{PW}$ 
    by comparing the values of $B_{\nu}^2$ at
    143~GHz to the values at 268~GHz.
    For our bands, the median value of 
    $\left( \frac{d \epsilon_{\tau}} {d \mathcal{C}_{PW}} 
    \mathcal{C}_{PW} \right)^2 \mathsf{B}_{atm}^2$, based on the
    Pardo ATM model~\citep{pardo05, pardo01, pardo01_2}, is
    approximately 16 times larger for the 268~GHz data compared
    to the 143~GHz data.
    The ratio of the values of 
    $B_{\nu}^2$ for the two frequencies is 11, 14, and 14
    for the three quartiles, 
    indicating that most of the observed difference in $B_{\nu}^2$
    between the two observing bands can be accounted for
    by assuming that $\delta\mathcal{C}_{PW} \propto \mathcal{C}_{PW}$. 

  \subsection{Comparing Mauna Kea to the South Pole and Atacama}

    At the South Pole, the median column depth of precipitable water
    vapor is $\simeq 0.25$~mm, roughly $6-7$ times lower
    than the median value at Mauna Kea.
    Therefore, the amplitude of the atmospheric noise at the 
    the South Pole is expected to be much lower than the 
    amplitude at Mauna Kea.
    Using our data, along with ACBAR data collected at the South	
    Pole, we can make a direct comparison of the amplitude
    of the atmospheric noise between the two locations.
    ACBAR had observing bands centered at 151 and 282 GHz,
    very close to the Bolocam bands, along with a third band centered
    at 222~GHz.
    For the 2002 observing season, \citet{Bussmann05} determined that
    the quartile values of $B_{\nu}^2$ for the 151~GHz band are
    3.7, 10, and 37~mK$^2$ rad$^{-5/3}$, and the quartile values
    of $B_{\nu}^2$ for the 282~GHz band are
    28, 74, and 230~mK$^2$ rad$^{-5/3}$.
    Therefore, the amplitude of the atmospheric noise is a factor
    of $\simeq 25$ 
    different for the Bolocam and ACBAR bands at $\simeq 150$~GHz,
    and a factor of $\simeq 50$ different for the bands at $\simeq 275$~GHz.
    Additionally, the ratio of $B_{\nu}^2$ between Bolocam and ACBAR
    is similar for all three quartiles in
    both observing bands, indicating that the relative variations
    in $B_{\nu}^2$ are comparable at both locations.
    See Table~\ref{tab:noise_amp}.

    Our phenomenological expectation of constant fractional
    fluctuations in $\mathcal{C}_{PW}$
    (\emph{i.e.}, $\frac{\delta\mathcal{C}_{PW}}{\mathcal{C}_{PW}}$ 
    is on average the same at both locations)    
    implies that
    the ratio of
    $\left( \frac{d \epsilon_{\tau}} {d \mathcal{C}_{PW}} 
    \mathcal{C}_{PW} \right)^2 \mathsf{B}_{atm}^2$
    should predict the ratio of $B_{\nu}^2$.
    This prediction, again based on the Pardo ATM 
    model~\citep{pardo05, pardo01, pardo01_2},\footnote{
      Note that by adjusting the input parameters, the Pardo ATM model
      can be matched to the conditions at the South Pole.}
    is that the ratio of $B_{\nu}^2$ should be
    12 for the $\simeq 150$~GHz  bands and 21 for the 
    $\simeq 275$~GHz bands.\footnote{
      We have used the measured Bolocam and ACBAR 
      bandpasses, along with the Pardo ATM model
      \citep{pardo05, pardo01, pardo01_2},
      to determine the value of
      $\left( \frac{d \epsilon_{\tau}} {d \mathcal{C}_{PW}} 
      \mathcal{C}_{PW} \right)^2 \mathsf{B}_{atm}^2$
      for each instrument for the median observing
      conditions at their respective sites.
      Although the Bolocam and ACBAR bands are similar, there
      are important differences;
      not only are the Bolocam bands centered at lower frequencies 
      than the ACBAR bands,
      but the $\simeq 150$~GHz Bolocam band is significantly
      narrower as well.
      Since the value of  
      $\left( \frac{d \epsilon_{\tau}} {d \mathcal{C}_{PW}} 
      \right)^2$ 
      is in general a strong function of observing
      frequency, these subtle differences in the observing bands
      produce noticeable differences in the predicted value
      of $B_{\nu}^2$.
      Additionally, differences in the atmosphere above each
      location can cause significant differences in the value of
      $\left( \frac{d \epsilon_{\tau}} {d \mathcal{C}_{PW}} 
      \right)^2$ 
      for a given value of $\mathcal{C}_{PW}$.
      Specifically, the ratio of 
      $\left( \frac{d \epsilon_{\tau}} {d \mathcal{C}_{PW}} 
      \right)^2$ between Bolocam and ACBAR is
      $\simeq 0.30$ for the $\simeq 150$~GHz bands and 
      $\simeq 0.45$ for the $\simeq 275$~GHz bands.}
    These predicted scalings are much lower than the observed
    scalings of 25 and 50, indicating that 
    the value of $\left(\frac{\delta\mathcal{C}_{PW}}{\mathcal{C}_{PW}} \right)^2$ 
    is a factor of $\simeq 2$ lower at the 
    South Pole compared to Mauna Kea.
    Consequently, in addition to the South Pole being on average
    much drier than Mauna Kea, 
    we conclude that 
    the fractional fluctuations in the column
    depth of water vapor are also lower by a factor of
    $\simeq \sqrt{2}$.

    Thus, because 
    $\left( \frac{\delta\mathcal{C}_{PW}}{\mathcal{C}_{PW}} \right)^2$ is a 
    factor of $\simeq 2$ larger 
    at Mauna Kea compared to the South Pole,
    and because the median
    value of $\mathcal{C}_{PW}^2$ is a factor of $\simeq 40$
    larger at Mauna Kea compared to the South Pole,
    we find that $B_{\nu}^2$
    is a factor of $\simeq 80$ larger at Mauna Kea compared to the 
    South Pole for mm-wave observations.
    Additionally, 
    \citet{Bussmann05}, 
    using the results in \citet{Lay00}, 
    found that the value of $B_{\nu}^2$ is a factor of $\simeq 30$
    lower at the South Pole compared to the ALMA site in Atacama.
    Therefore, we can infer that
    $B_{\nu}^2$ is a factor of $\simeq 3$ lower at the ALMA site
    compared to Mauna Kea.
    Since the value of 
    $\left( \frac{d \epsilon_{\tau}} {d \mathcal{C}_{PW}} 
    \mathcal{C}_{PW} \right)^2 \mathsf{B}_{atm}^2$
    is a factor of $\simeq 3.5$ lower at the ALMA site than 
    Mauna Kea for the median observing conditions at each location,
    we find that the value of 
    $\left( \frac{\delta\mathcal{C}_{PW}}{\mathcal{C}_{PW}} \right)^2$ is
    similar for Mauna Kea and the ALMA site.\footnote{
      The median value of $\mathcal{C}_{PW}$ at the Cerro Chajnantor
      site under consideration for the Cornell-Caltech Atacama
      Telescope (CCAT) is approximately 0.83~mm, so the median
      value of $B_{\nu}^2$ should be about 30\% lower
      at the CCAT site compared to the ALMA site.}
    Therefore, the fractional fluctuations in the column depth of 
    precipitable water vapor appear to be the same at Mauna Kea
    and the ALMA site, but they are significantly lower at the
    South Pole;
    these lower fluctuations at the South Pole may be due to
    the lack of diurnal variations at that site.
    We emphasize that these are statements about the fluctuations in
    $\mathcal{C}_{PW}$, and thus relate only to atmospheric noise.
    In shorter wavelength bands with higher opacity, 
    it may be that signal attenuation and photon noise
    due to the absolute opacity are more important than atmospheric
    noise in determining the quality of a given site.

  \subsection{Map Variance as a Function of Atmospheric Conditions}

    Although it is useful to determine
    the amplitude of the fluctuations in
    atmospheric emission, the quality of our data is characterized
    by the residual noise level after removing as much
    atmospheric noise as possible.
    We will use the 
    difference between
    the measured map variance, $\sigma^2_{map}$, and the
    expected map variance in the absence of atmospheric
    noise, $\sigma_{white}^2$, 
    as a proxy
    for this residual noise level.
    Note that these maps are produced after removing most of the 
    atmospheric noise using the average subtraction algorithm
    given in Section~\ref{sec:avg_skysub},
    and $\sigma_{white}^2$ is estimated from the noise level
    of the map at high spatial frequency where the atmospheric
    noise is negligible.

\begin{figure}
  \plotone{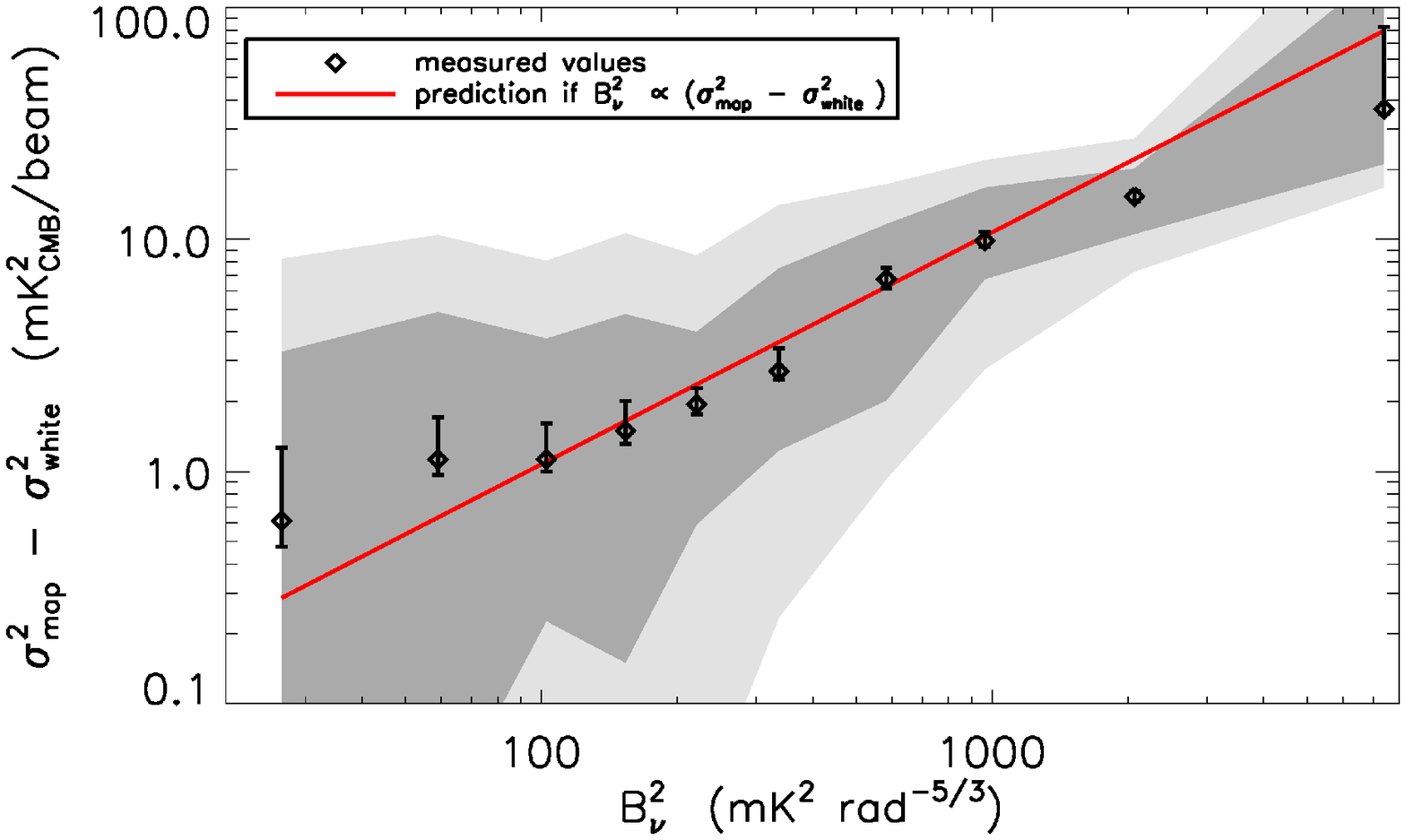} \\
  \plotone{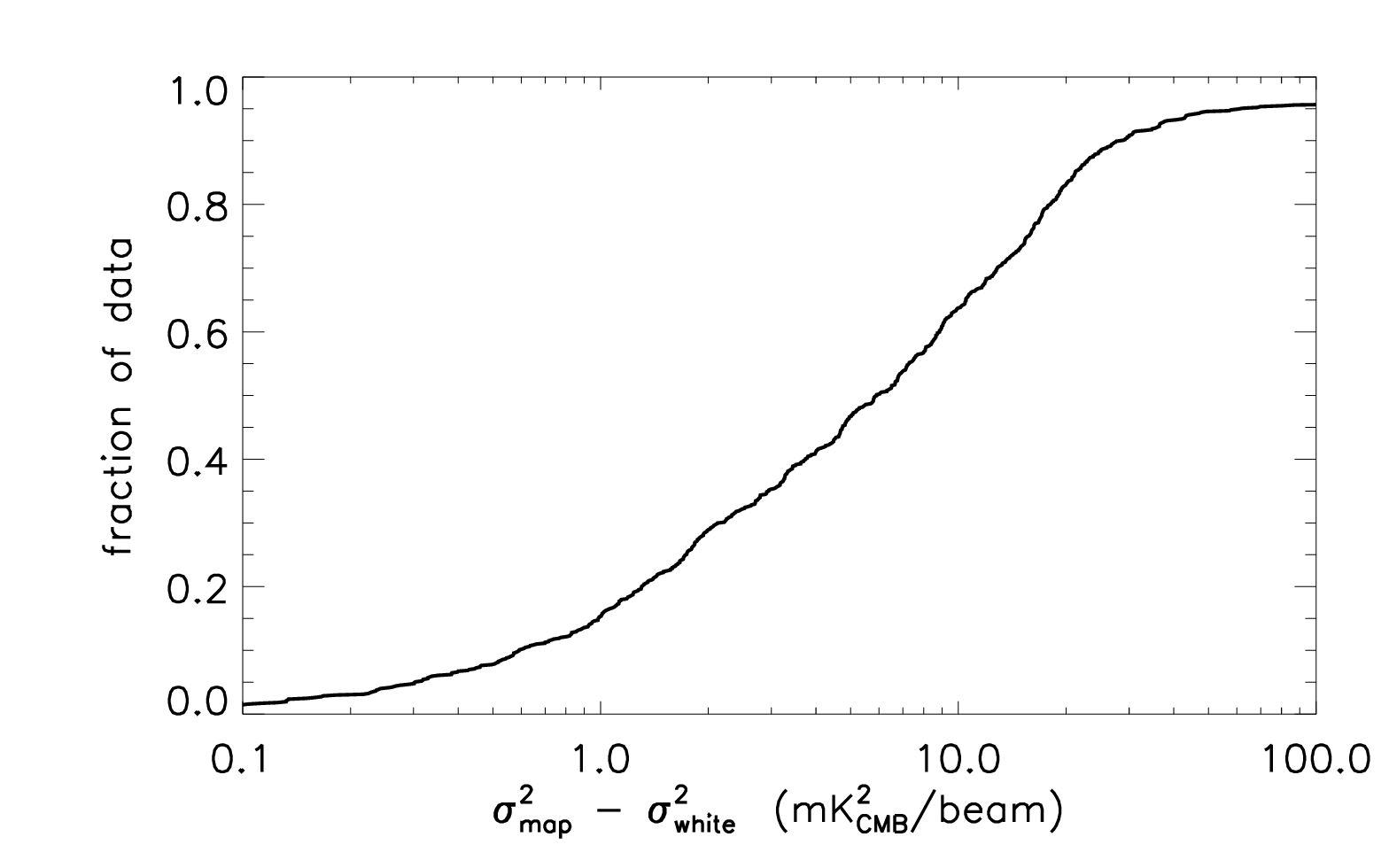}
  \caption{The top plot shows the 143~GHz single-observation
  residual map variance after subtracting the white noise level
  as a function of the amplitude of the atmospheric noise,  
  $B_{\nu}^2$.
  Note that the typical white noise level of the maps is
  $\sigma_{white}^2 \simeq 5$ mK$_{CMB}^2$.
  The error bars represent the error on the median value for
  each data point;
  the light shaded region spans the $10-90$th centile values and
  the dark shaded region spans the $25-75$th centile values.
  Note that most of the atmospheric noise has been removed from
  the data using the average subtraction algorithm described in
  section~\ref{sec:avg_skysub}.
  The red line shows the prediction assuming that the 
  residual map variance is proportional to $B_{\nu}^2$.
  The bottom plot shows cumulative distribution of 
  residual map variance for the 143~GHz data.}
  \label{fig:map_var}
\end{figure}

    As expected, we find a correlation between 
    $\sigma^2_{map} - \sigma_{white}^2$
    and $B_{\nu}^2$,
    although there is quite a bit of scatter in the 
    amount of residual atmospheric noise for a given
    value of $B_{\nu}^2$.
    See Figure~\ref{fig:map_var}.
    Most of this scatter is likely due to the fact that the 
    residual noise is inversely proportional
    to the amount of correlation in the atmospheric signal
    over our FOV;
    this correlation depends not only on the value
    of $B_{\nu}^2$, but also on the
    the height and angular wind speed of the turbulent layer.
    Since the atmosphere is in the near field for Bolocam,
    an increase in the height of the turbulent layer
    reduces the overlap of the beams from individual 
    detectors.
    Thought of in a different way, 
    a decrease in the
    height of the turbulent layer implies that
    the beam smoothing of the atmospheric signal
    is extended to larger spatial scales, making the 
    atmospheric signal more uniform over the fixed angular
    scale of our FOV.
    Therefore, for a fixed value of $B_{\nu}^2$,
    there will be less correlation in the atmospheric
    signal over the FOV as the
    height of the turbulent layer increases.
    Additionally, the angular wind speed of the turbulent layer
    will influence the amount of atmospheric noise in
    the data because our scan speed is much slower than
    the angular wind speed.
    This means that a higher angular wind speed will modulate
    the atmospheric noise to higher frequencies in
    the time-stream data; at higher frequencies more of the
    atmospheric noise will be in our signal band and less
    of the noise will be removed using the subtraction
    algorithms described in Section~\ref{sec:atm_noise_rem}.
    Also, note that, in the best conditions, our data 
    approach the white noise limit, and these conditions
    can occur over a relatively wide range of values
    for $B_{\nu}^2$.
    Thus, we find that while $B_{\nu}^2$ (and 
    also $\mathcal{C}_{PW}$ based on our assumption that
    $B_{\nu}^2 \propto
    \left( \frac{d \epsilon_{\tau}} {d \mathcal{C}_{PW}} 
    \mathcal{C}_{PW} \right)^2 \mathsf{B}_{atm}^2$)
    is not a precise predictor of $\sigma_{map}^2 - \sigma_{white}^2$,
    there is a general trend of less residual map noise at
    lower values of $B_{\nu}^2$ ($\mathcal{C}_{PW}$).

  \subsection{Summary}

    In summary, the K-T thin-screen
    model appears to provide an adequate description of 
    the atmospheric signal in our data.
    We find the angular speed of the thin-screen to be
    approximately 30 arcmin/sec, although roughly half
    of our data contain some features that cannot be
    explained with a single angular wind velocity.
    The turbulent layer has a power law exponent of $b = 3.3 \pm 1.1$,
    consistent with the K-T prediction of $b = 11/3$.
%    and the height of the turbulent layer varies between
%    $h=500$ and 4000 m.
    If we assume that $b=11/3$, then the median amplitude of the
    atmospheric fluctuations is 280 mK$^2$ rad$^{-5/3}$ at 
    143 GHz and 4000 mK$^2$ rad$^{-5/3}$ at 268 GHz.
    These amplitudes are $\simeq 80$ times larger than
    the amplitudes found at similar observing frequencies
    at the South Pole using ACBAR~\citep{Bussmann05}.
    Most of the scaling in $B_{\nu}^2$
    between observing frequencies and locations
    can be accounted for by assuming that the fractional
    fluctuations in the column depth of precipitable water 
    vapor, $\frac{\delta\mathcal{C}_{PW}}{\mathcal{C}_{PW}}$, are constant.
    However, the data indicate that 
    $\frac{\delta\mathcal{C}_{PW}}{\mathcal{C}_{PW}}$ is a factor of
    $\simeq \sqrt{2}$ smaller at the South Pole compared to Mauna Kea.
    We thus find that the bulk of the reduction in atmospheric
    noise at the South Pole is due to the consistently low
    value of $\mathcal{C}_{PW}$ at that site, and 
    the lower fractional fluctuations in the precipitable
    water vapor only reduce the RMS of the atmospheric noise
    by an additional factor of $\simeq \sqrt{2}$.
    Additionally, after removing as much atmospheric noise as 
    possible, we find a correlation between the value of 
    $B_{\nu}^2$ and the amount of residual atmospheric noise
    in our data, although it is likely that
    the height and angular speed of the
    turbulent layer also influence the amount of residual 
    atmospheric noise.
	
\section{Atmospheric Noise: Removal}
  \label{sec:atm_noise_rem}

  In this section we describe various atmospheric noise removal
  techniques, including one based on the relatively unsophisticated
  common-mode assumption and several based on
  the properties of the atmospheric noise determined
  from our fits to the K-T model.
  Additionally, we summarize the results of subtracting
  the atmospheric noise using
  adaptive principle component analysis (PCA).
  Note that in this Section, along with Section~\ref{sec:resid_corr},
  our analysis focuses entirely on the 143~GHz data.
      
  \subsection{Average Template Subtraction}
    \label{sec:avg_skysub}

\begin{figure}
  \epsscale{1}
  \plottwo{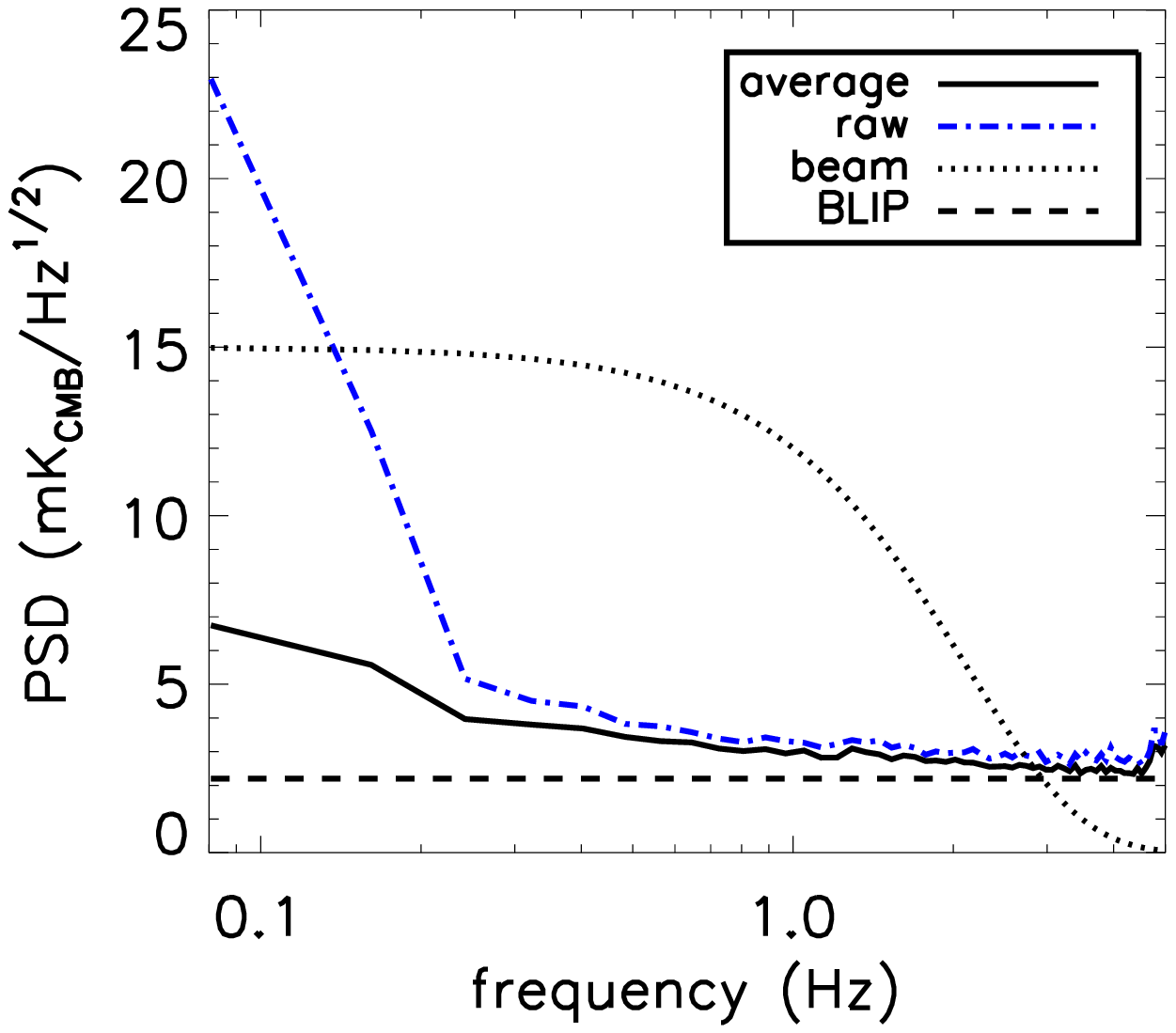}{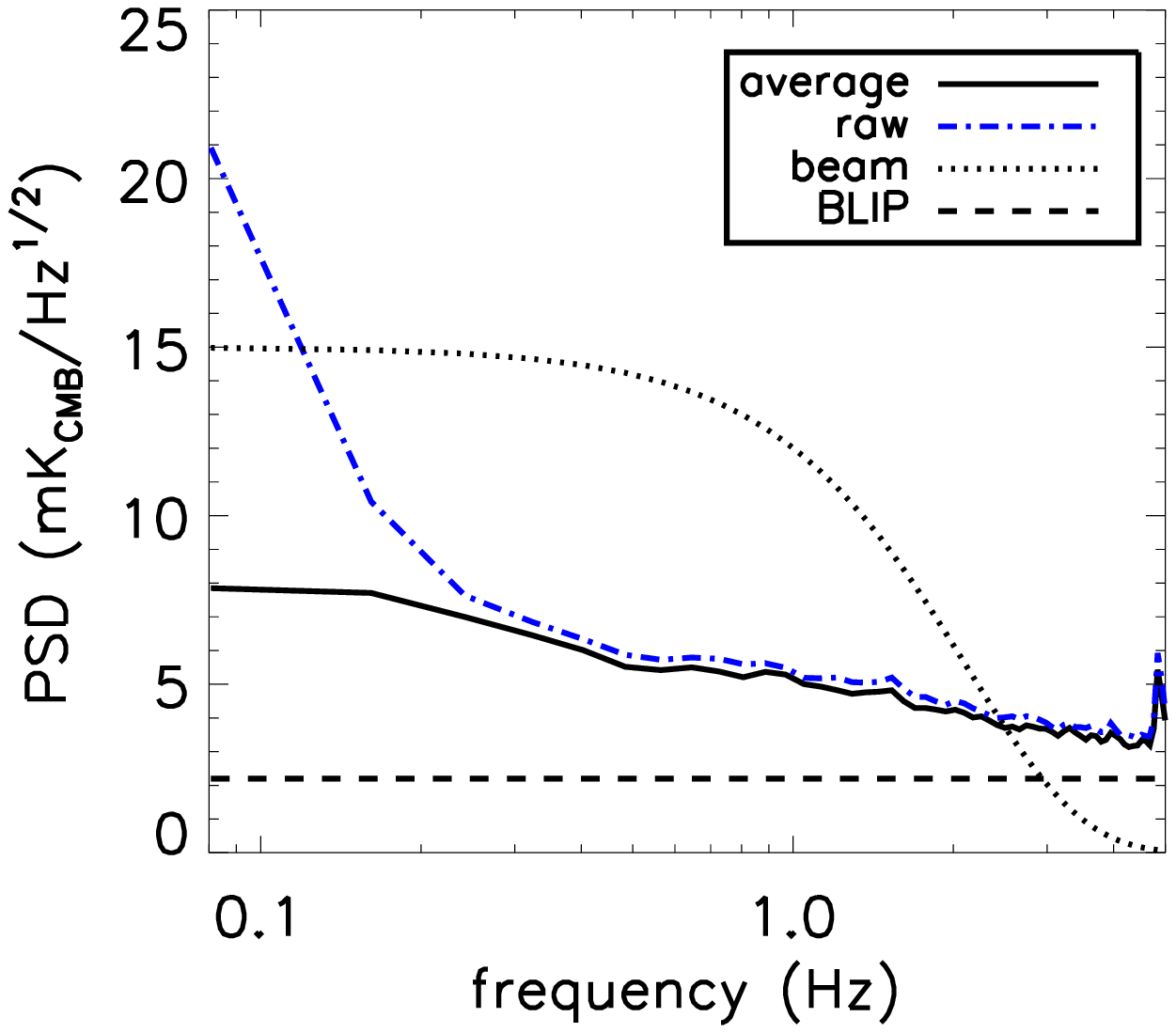} \\
  \plottwo{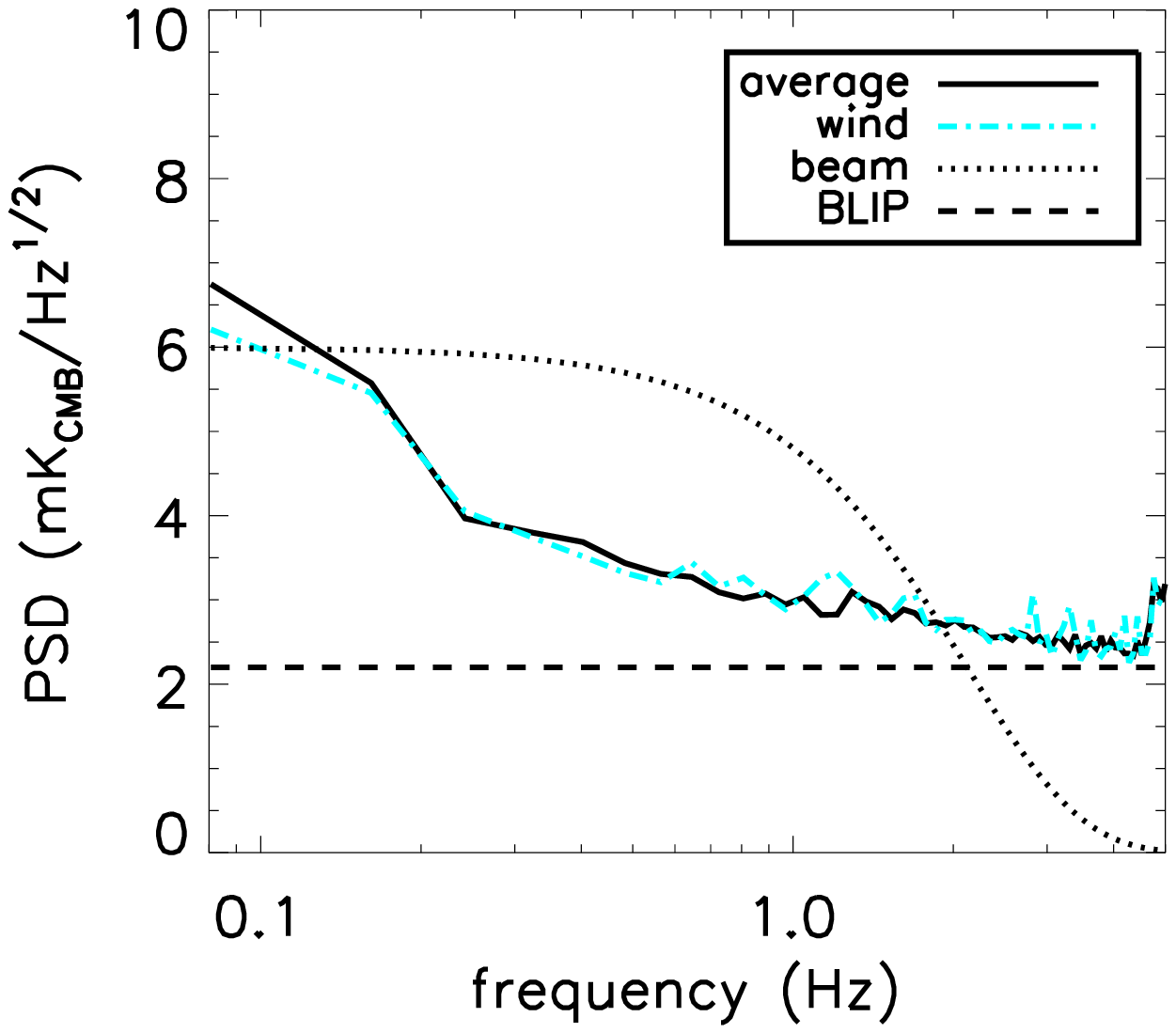}{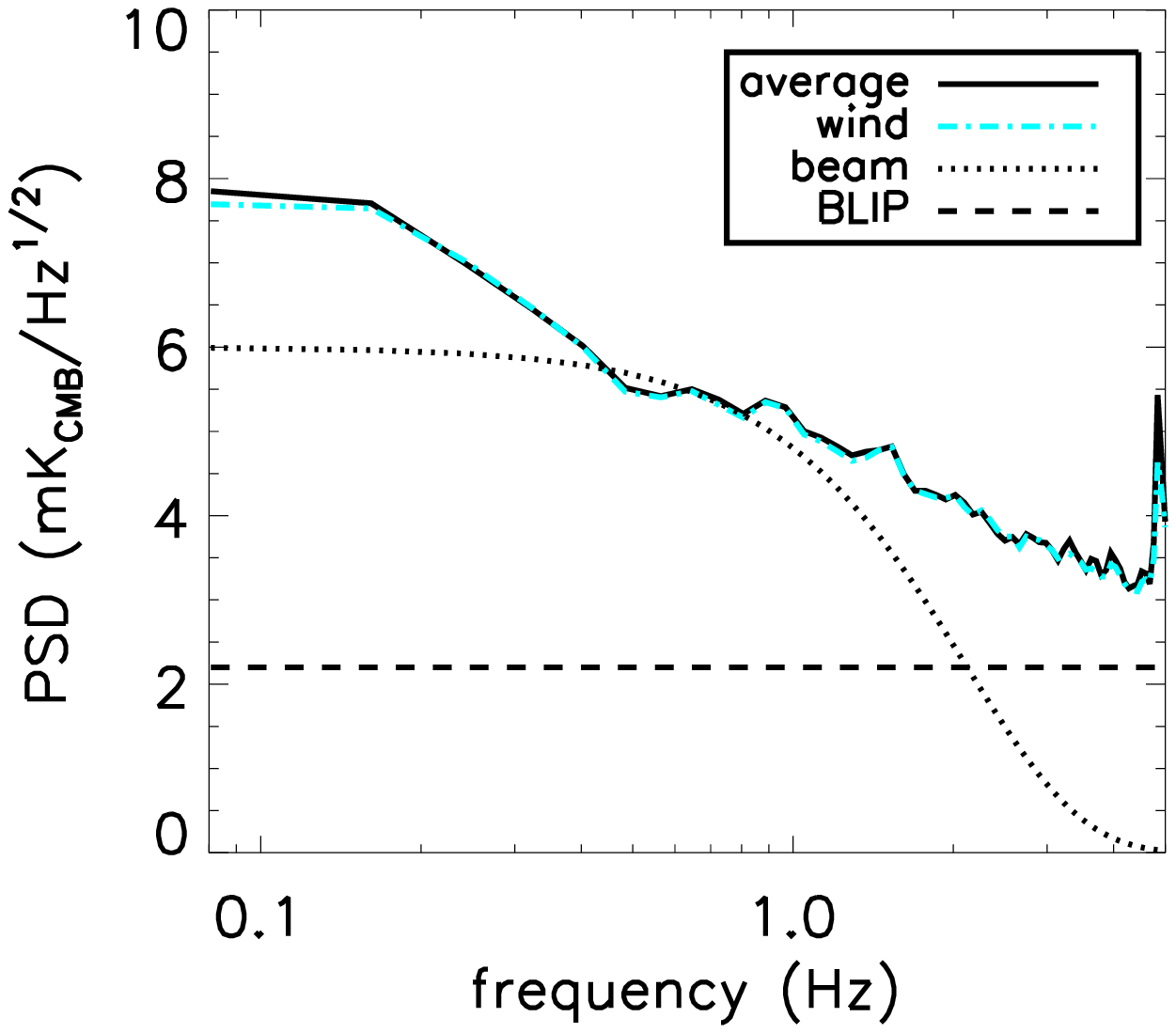} \\
  \plottwo{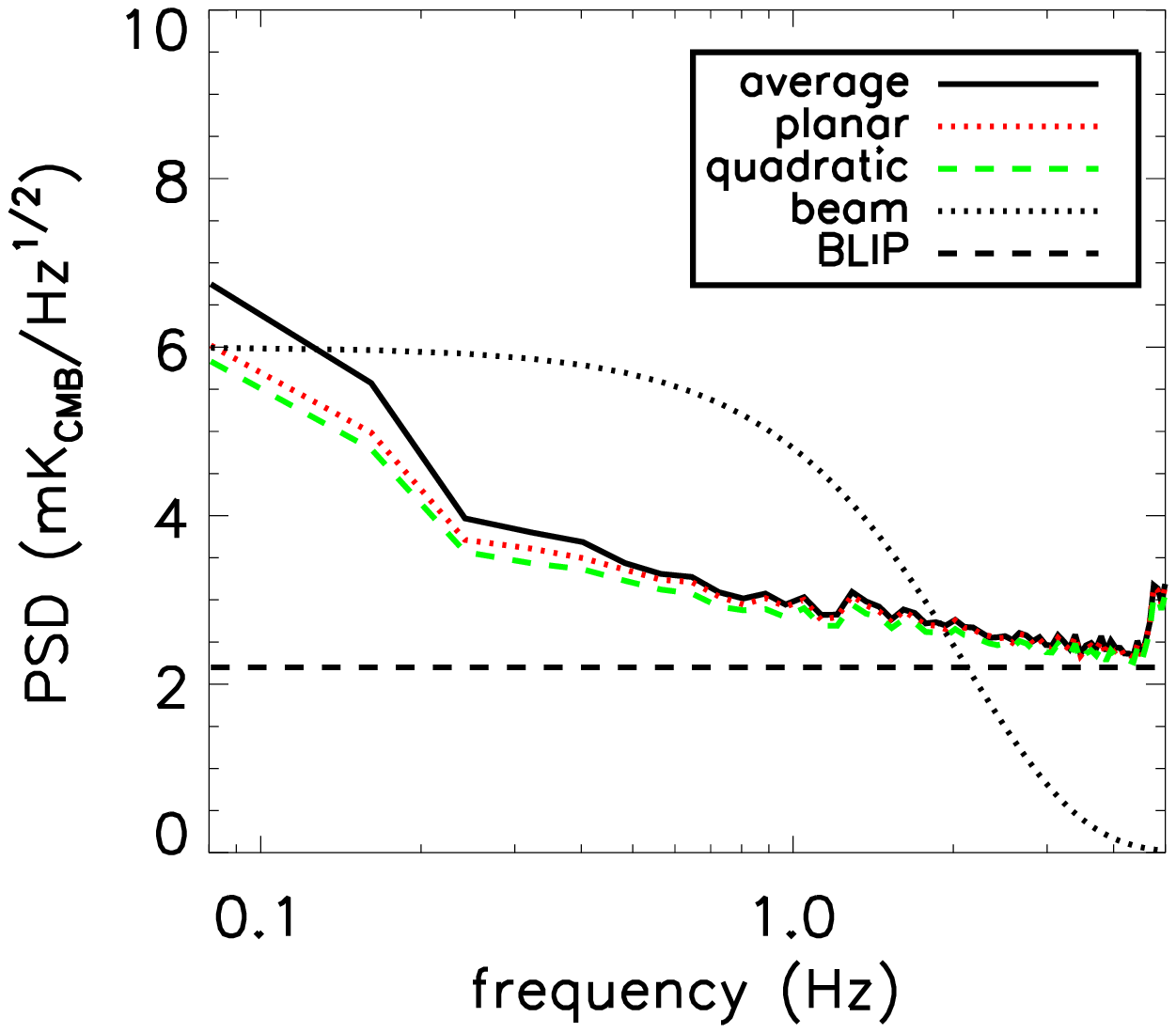}{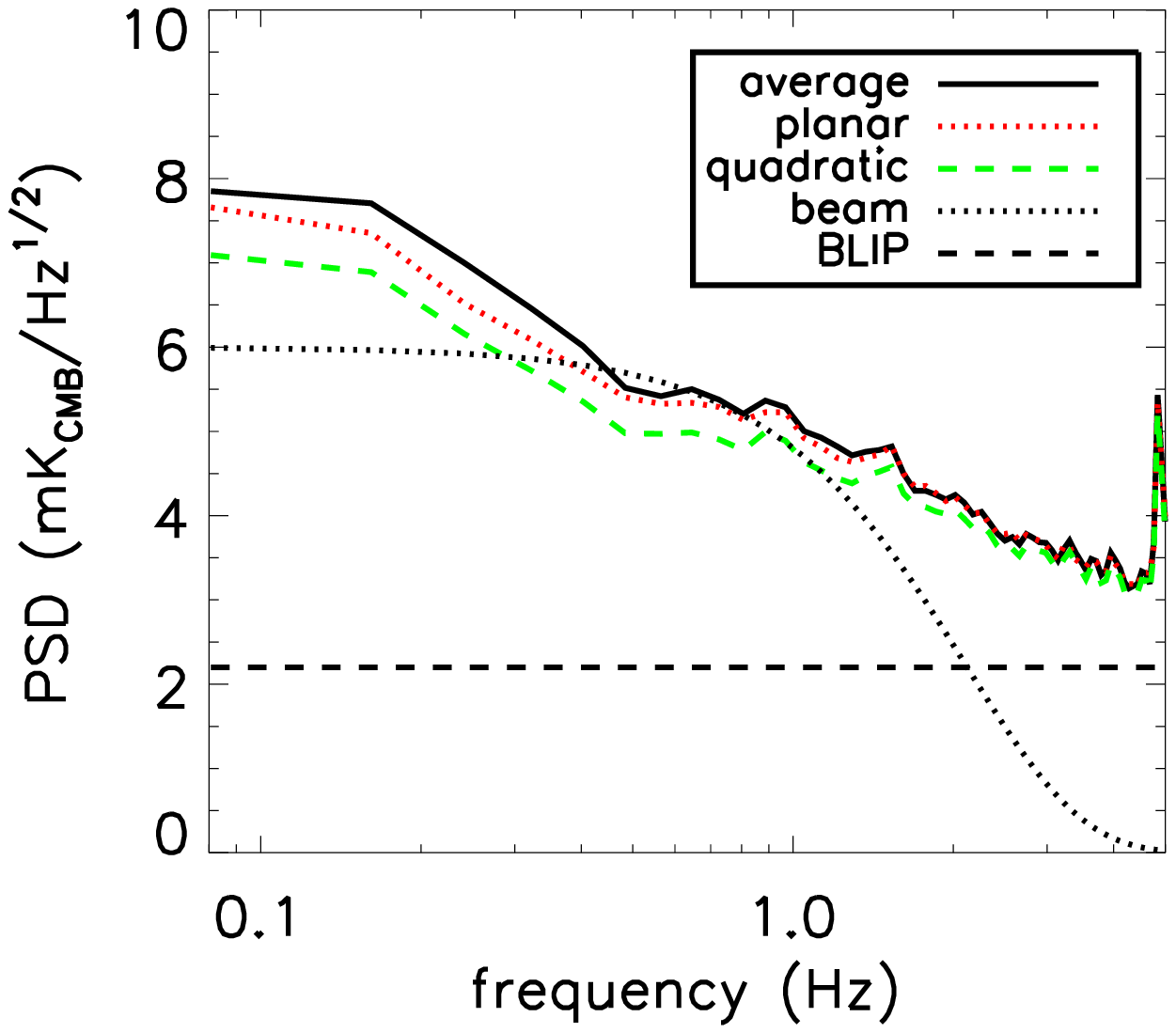}
  \caption{Plots of 143~GHz time-stream PSDs averaged over all
    scans and all bolometers for a single observation.
    The left column shows data from an observation made in relatively
    good weather, and the right column shows an observation made
    in relatively poor weather.
    For each plot, the atmospheric subtraction algorithm applied to
    the data is given in the legend.
    Overlaid as a dotted line in each plot is the profile of the
    Bolocam beam, and the approximate BLIP limit is shown as 
    a dashed line.}
  \label{fig:avg_skysub}
\end{figure}

    Our most basic method for removing atmospheric noise
    is to subtract the signal that is common to all of 
    the bolometers.
    Initially, a template is constructed according to
    \begin{equation}
      T_n = \frac{\sum_{i=1}^{i=N_{b}}
      c_i^{-1} d_{in}}{\sum_{i=1}^{i=N_{b}} c_i^{-1}}
      \label{eqn:avg_template}
    \end{equation}
    where $n$ is the sample number, $N_{b}$ is the number
    of bolometers, $c_i$ is the relative responsivity
    of bolometer $i$, $d_{in}$ is the signal recorded by
    bolometer $i$ at sample number $n$, and $T_n$ is the 
    template.
    The relative responsivity is required to account for the 
    fact that the bolometer response (in nV) to a given
    signal (in mK) is slightly different from one bolometer to 
    the next.
    A separate template is computed for each $\simeq 15$-second-long scan.
    After the template is computed, it is correlated with
    the signal from each bolometer to determine the 
    correlation coefficient, with
    \begin{equation}
      \tilde{c_i} = \frac{\sum_{j=1}^{j=N_{s}}
      T_n d_{in}}{\sum_{j=1}^{j=N_{s}} T_n^2}.
      \label{eqn:skysub_corr}
    \end{equation}
    $\tilde{c_i}$ is the correlation coefficient
    of bolometer $i$ and $N_s$ is the number of samples
    in the $\simeq 15$-second-long scan.\footnote{
      The best-fit correlation coefficients change
      from one scan to the next, typically by a 
      couple percent.}
    Next, the $c_i$ in Equation~\ref{eqn:avg_template} are
    set equal to the values of $\tilde{c_i}$ found from
    Equation~\ref{eqn:skysub_corr}, and a new template is
    computed.
    The process is repeated until the values of $c_i$ 
    stabilize.
    We generally iterate until the average fractional change
    in the $c_i$s is less than $1 \times 10^{-8}$, which
    takes five to ten iterations.
    If the $c_i$s fail to converge after 100 iterations,
    then the scan is discarded from the data.
    This algorithm generally removes the majority
    of the atmospheric noise, as shown in Figure~\ref{fig:avg_skysub}.

  \subsection{Wind Model}

    Since the moving screen atmospheric model 
    given in Section~\ref{sec:time_lag}
    provided a fairly good description of our data,
    we attempted to improve our
    atmospheric noise removal algorithm by applying
    the appropriate time delay/advance to every bolometer prior to 
    average subtraction.
    The angular wind velocity for each observation was determined using the
    formalism described in Section~\ref{sec:time_lag}, and from
    this angular wind velocity we computed the time delay/advance for
    each bolometer based on its location on the focal plane.    
    If the spatial structure of the atmospheric emission is static
    on the timescales of the delay/advance, then the 
    shifted beam centers will be pointed at the same location in
    the turbulent layer for bolometers aligned parallel to
    the angular wind velocity.
    Therefore, the atmospheric signal in these shifted
    time-streams will be identical for these bolometers, 
    modulo uncertainties
    in the angular wind velocity, 
    slight differences in the beam profiles, 
    etc.
    See Section~\ref{sec:time_inst_model} for a discussion of the 
    impact of the latter.
    For the typical angular speeds of the turbulent layer, the 
    shifts are of order 1 sample, and we used 
    a linear interpolation to account for shifts that are a 
    fraction of a sample.
    Note that this linear interpolation acts as a low-pass filter
    on our data; to
    preserve the PSDs of our time-streams, we correct for
    this attenuation in frequency space.
    See Appendix~\ref{sec:appen}.
    We applied the appropriate shift to the time-stream
    of each bolometer before performing average
    subtraction, but this did not seem to reduce the 
    post-subtraction noise PSD relative to time-instantaneous
    average subtraction. 
    See Figure~\ref{fig:avg_skysub}.
    Therefore, we abandoned this atmospheric noise subtraction
    algorithm.

  \subsection{Higher-Order Template Subtraction}
    \label{sec:planar_skysub}

    Based on the K-T model fits, we were able to
    determine which spatial Fourier modes cause the
    atmospheric emission to become uncorrelated over our
    8 arcmin FOV.
    Our time-stream PSDs show that most of the 
    atmospheric noise signal is at frequencies below
    0.1~Hz, and the atmospheric noise becomes
    negligible at frequencies above 0.5~Hz.
    Therefore, most of the atmospheric fluctuations occur
    on long time-scales, which correspond to large
    spatial scales.
    To convert these temporal frequencies to angular
    frequencies, we divide by the
    the angular wind speed we determined for the thin-screen
    model, which we found in Section~\ref{sec:time_lag}
    to be approximately 30~arcmin/sec.
    This means that most of the atmospheric noise 
    is at small angular frequencies with
    $\alpha < 300^{-1}$~arcmin$^{-1}$, and the
    atmospheric noise is negligible for 
    angular frequencies larger than $\alpha = 60^{-1}$~arcmin$^{-1}$.
    We can therefore conclude
    that very little atmospheric signal is
    sourced by spatial modes with
    wavelengths smaller than our FOV.
    Note that \cite{jenness98}, based on the atmospheric noise
    in SCUBA data and making reasonable assumptions for the height
    and angular speed of the turbulent layer, found a similar
    scale for the atmospheric fluctuations.

    Since most of the atmospheric signal is caused by
    power in spatial modes with wavelengths much larger
    than our FOV, the signal will be slowly
    varying over our focal plane.
    Therefore, we decided to model the
    atmospheric fluctuations using
    a low-order two-dimensional polynomial in detector
    position.
    This is similar to the method used by SHARC II to remove
    atmospheric noise~\citep{kovacs08}.
    Additionally, \cite{borys98} attempted a similar planar
    subtraction with SCUBA, although with limited success.

    For planar and quadratic subtraction,
    including the special case of 
    average subtraction 
    described in Section~\ref{sec:avg_skysub},
    the algorithm is implemented as follows.
    The data are modeled according to
    \begin{displaymath}
      \vec{d_n} = \mathbf{C} \mathbf{S} \vec{p_n},
    \end{displaymath}
    where $\vec{d_n}$ is a vector with $n_b$ elements representing
    the bolometer data at time sample $n$, 
    $\mathbf{C}$ is a diagonal $n_b \times n_b$ element matrix
    with the relative responsivity of each bolometer,
    $\mathbf{S}$ is an $n_b \times n_{params}$ element matrix,
    and $\vec{p_n}$ is a vector with $n_{params}$ elements.
    $n_b$ is the number of bolometers, $n$ is the sample
    number within the $\simeq 15$-second-long scan, and 
    $n_{params}$ is the number of fit parameters.
    $\mathbf{S}$ is based on the geometry of the focal plane, with
    $n_{params} = 1/3/6$ for average/planar/quadratic subtraction and
    \begin{eqnarray*}
      \mathbf{S}_{i1} = 1 & \mathbf{S}_{i2} = x_{i}  
      & \mathbf{S}_{i3} = y_i \\
      \mathbf{S}_{i4} = x_i y_i & 
      \mathbf{S}_{i5} = x_i^2 & \mathbf{S}_{i6} = y_i^2
    \end{eqnarray*}
    where $\vec{x}$ and $\vec{y}$ are vectors with $n_b$ elements that
    contain the $x$ and $y$ coordinate of each bolometer
    on the focal plane.
    The $\vec{p_n}$ are the $n_{params}$ atmospheric noise
    templates, which are obtained by 
    minimizing
    \begin{equation}
      \chi^2_n = 
      (\vec{d_n} - \mathbf{C} \mathbf{S} \vec{p_n})^T (\vec{d_n} - 
      \mathbf{C} \mathbf{S} \vec{p_n})
      \label{eqn:skysub_chi}
    \end{equation}
    with respect to $\vec{p_n}$.\footnote{
    We have assumed that the individual bolometer intrinsic 
    (\emph{i.e.}, non-atmospheric) noises at time sample $n$ are not 
    correlated with each other so that the covariance matrix is diagonal. 
    The noises of the different bolometers are sufficiently similar, 
    once corrected for relative responsivity via $\mathbf{C}$,
    that the noise covariance matrix 
    can in fact be taken to be 
    a multiple of the identity matrix. 
    The $\chi^2$ statistic is thus proportional to a statistically 
    rigorous $\chi^2$, though it is not normalized correctly. 
    The normalization is unimportant for our purposes. 
    If these assumptions are incorrect, then our estimators 
    of the atmospheric templates will not be minimum variance 
    estimators; they will, however, be unbiased. 
    We also have implicitly assumed that we should determine 
    $\vec{p_n}$ at each point in time independently, which relies 
    on the assumption that the intrinsic noise of a given bolometer 
    is uncorrelated with itself in time 
    (\emph{i.e.}, white in frequency space). 
    This is also a reasonably valid assumption, and, again, 
    if it is incorrect, then our estimators are not maximally 
    efficient but remain unbiased.}
    For a given time sample $n$, the values of $\vec{p_n}$
    give the coefficients for each term in the polynomial
    expansion of the atmospheric signal over the focal
    plane at that particular time.
    A single element in the vector $\vec{p_n}$, when
    considered over all the samples in a scan, gives
    the time dependence of that particular coefficient.
    Essentially, each element in 
    $\vec{p_n}$ can be thought of as a data time-stream that 
    gives the amplitude of the atmospheric signal with a 
    particular spatial dependence
    over the focal plane.
    Minimizing Equation~\ref{eqn:skysub_chi} yields
    \begin{equation}
      \vec{p_n} = 
      (\mathbf{S}^T \mathbf{S})^{-1} \mathbf{S}^T \mathbf{C}^{-1}
      \vec{d_n}.
      \label{eqn:template_1}
    \end{equation}
    Once $\vec{p_n}$ is known, we can construct an atmospheric
    template analogous to Equation~\ref{eqn:avg_template} for
    each bolometer according to
    \begin{equation}
      \vec{T_n} = \mathbf{S} \vec{p_n}.
      \label{eqn:template_2}
    \end{equation}
    Note that $\vec{T_n}$ varies from bolometer to bolometer as
    prescribed by the assumed two-dimensional polynomial
    form and the best-fit polynomial coefficients $\vec{p_n}$.
    A correlation coefficient is then computed for each
    bolometer according to Equation~\ref{eqn:skysub_corr},
    a new matrix $\mathbf{C}$ is computed according to
    these correlation coefficients, 
    and a new template is computed according to Equations
    \ref{eqn:template_1} and \ref{eqn:template_2}.
    The process is repeated until the fractional change in the
    values of the correlation coefficients is less
    than one part in $10^8$.

\begin{figure}
  \plottwo{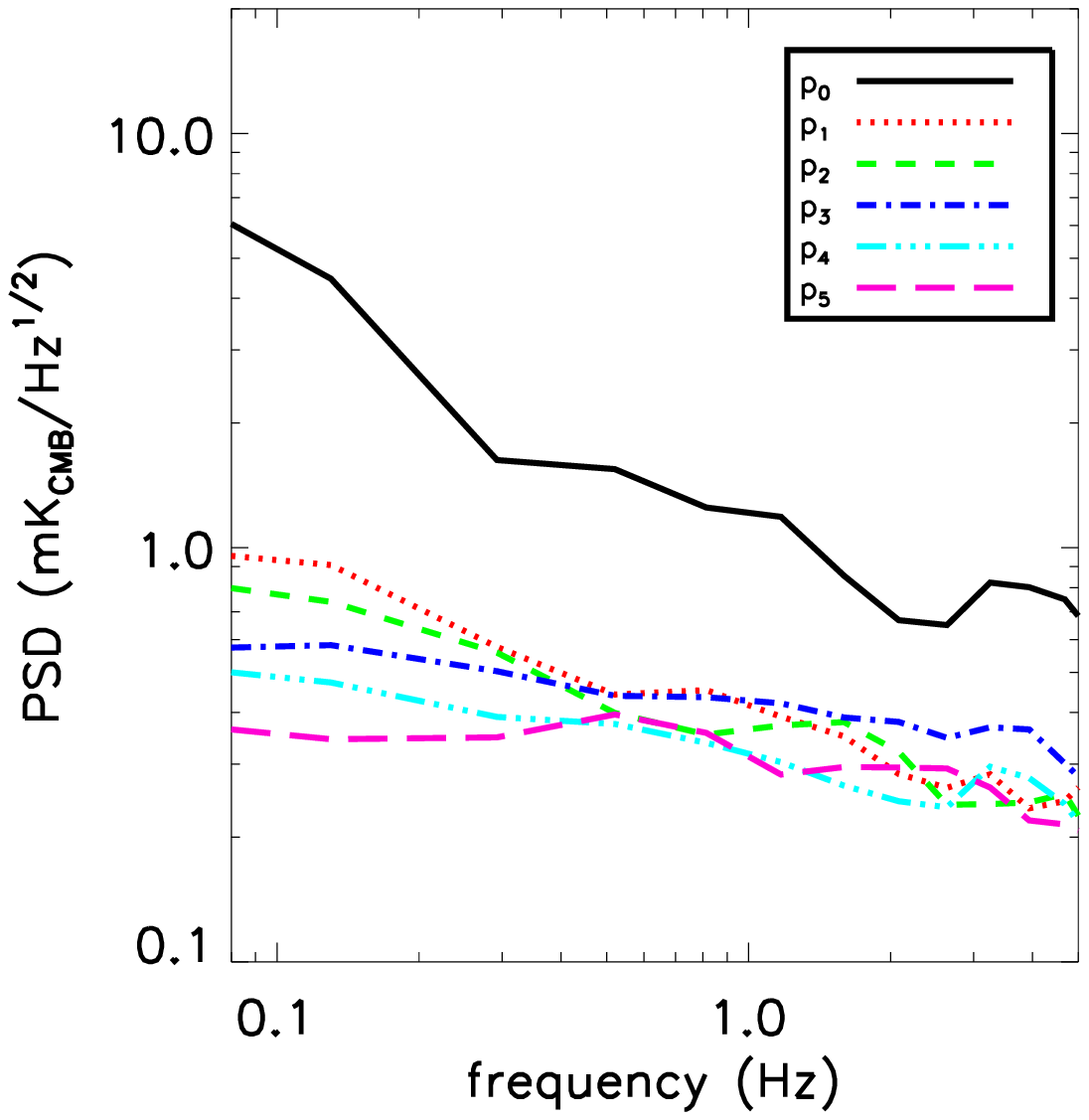}{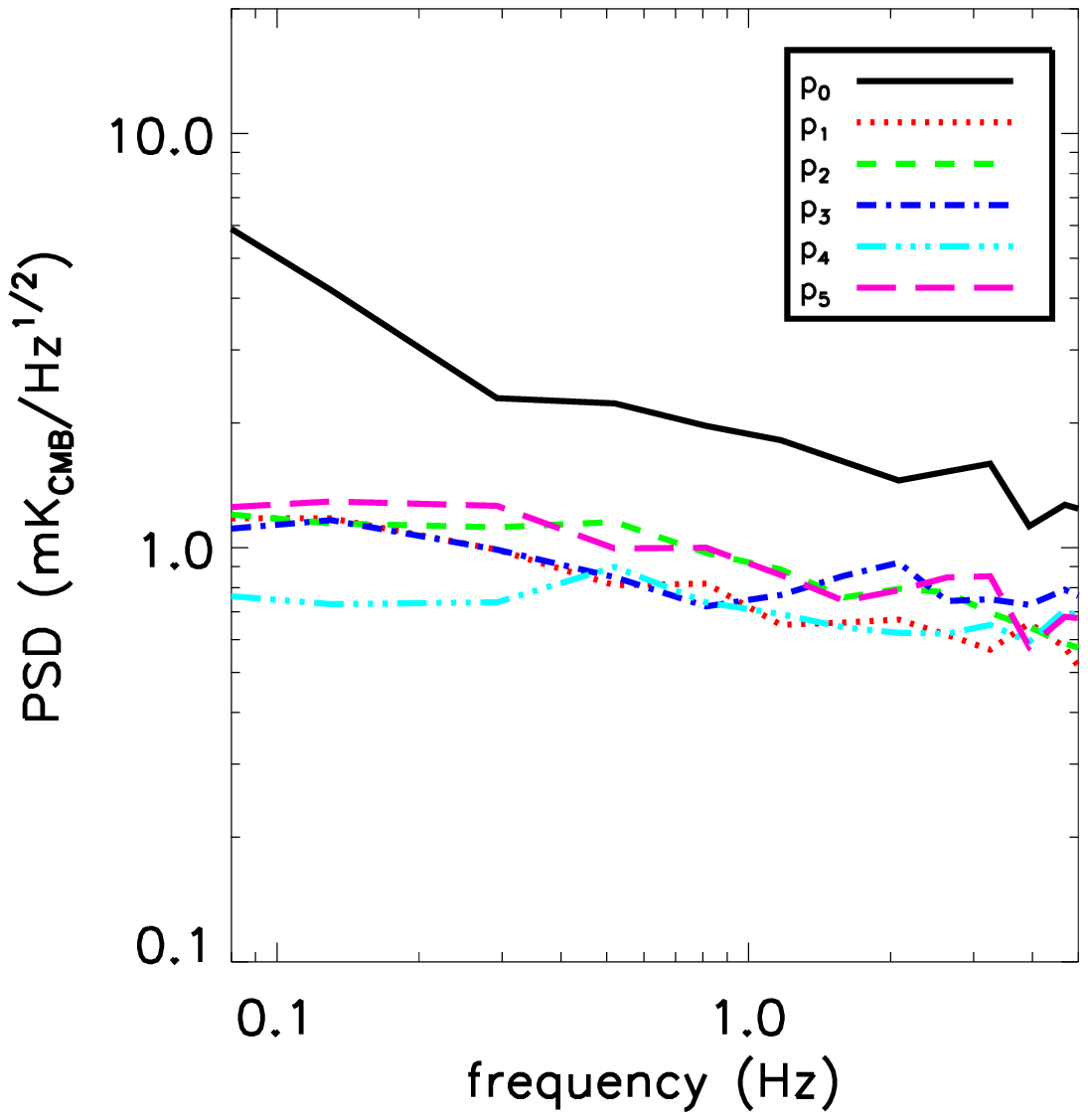}
  \caption{Power spectra for the templates generated by the
    quadratic sky subtraction algorithm for 143~GHz data.
    The plot on the left represents data collected in
    relatively good weather, and the plot on the right
    shows data collected in relatively poor weather.
    All six elements of $\vec{p_i}$ are plotted, with
    labels given in the upper right of each plot.
    The higher-order elements in $\vec{p_i}$ are shown for a
    bolometer approximately half-way between the array
    center and the edge of the array.
    Note that the magnitude of the higher-order templates
    in bad weather is a factor of $\simeq 2$ larger than the
    magnitude of the higher-order templates in good weather.}
  \label{fig:skysub_templates}
\end{figure}

    In general, the PSDs of the higher-order templates are 
    $\simeq 5$ times smaller than the PSD of the $0^{\rm{th}}$-order
    template 
    for bolometers halfway between the center
    and the edge of the focal plane.
    As expected, the ratio of the higher-order templates
    to the $0^{\rm{th}}$-order template increases as the
    weather becomes worse.
    Some typical power spectra of the $\vec{p_i}$
    are shown in Figure~\ref{fig:skysub_templates}.

\begin{figure}
  \plottwo{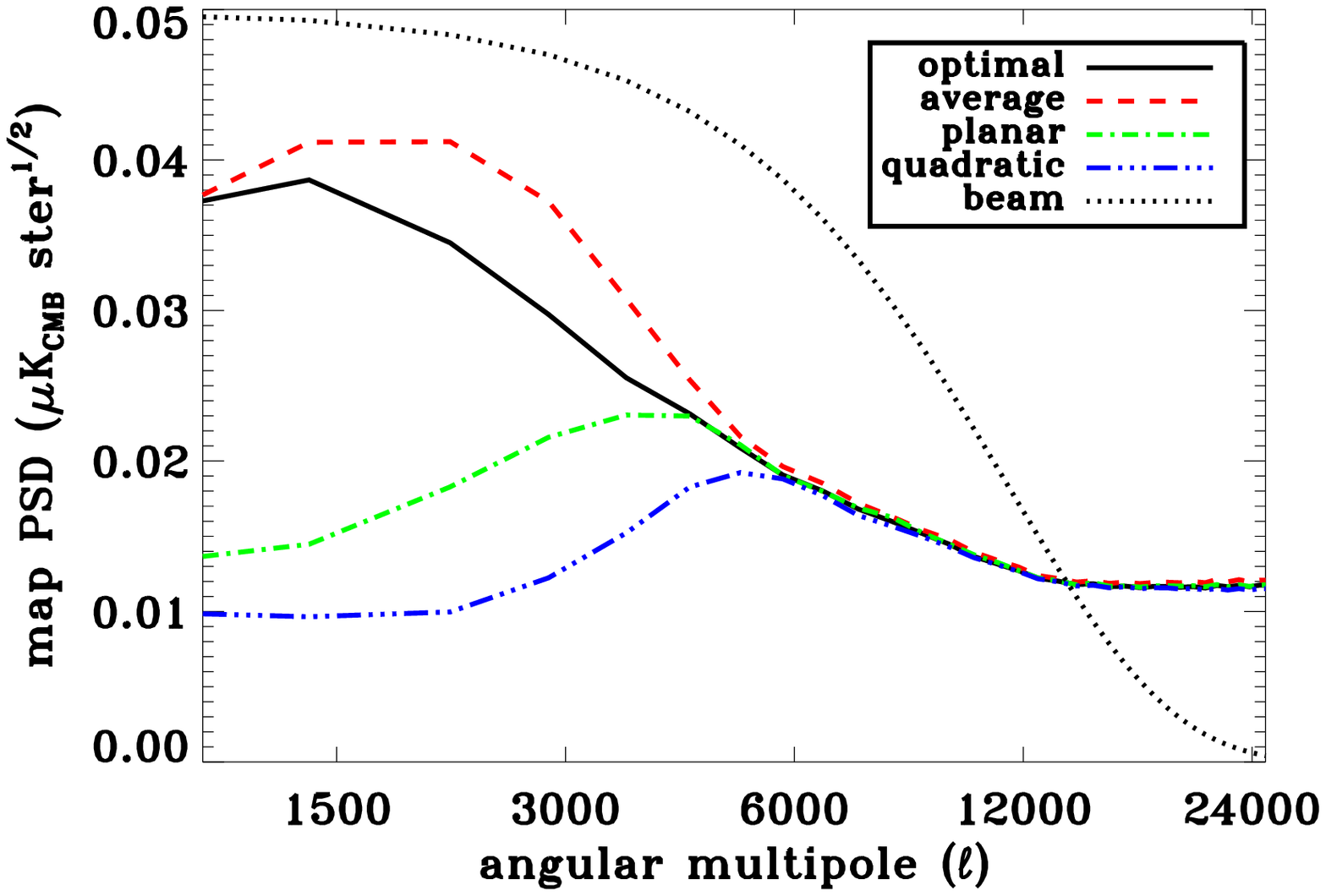}{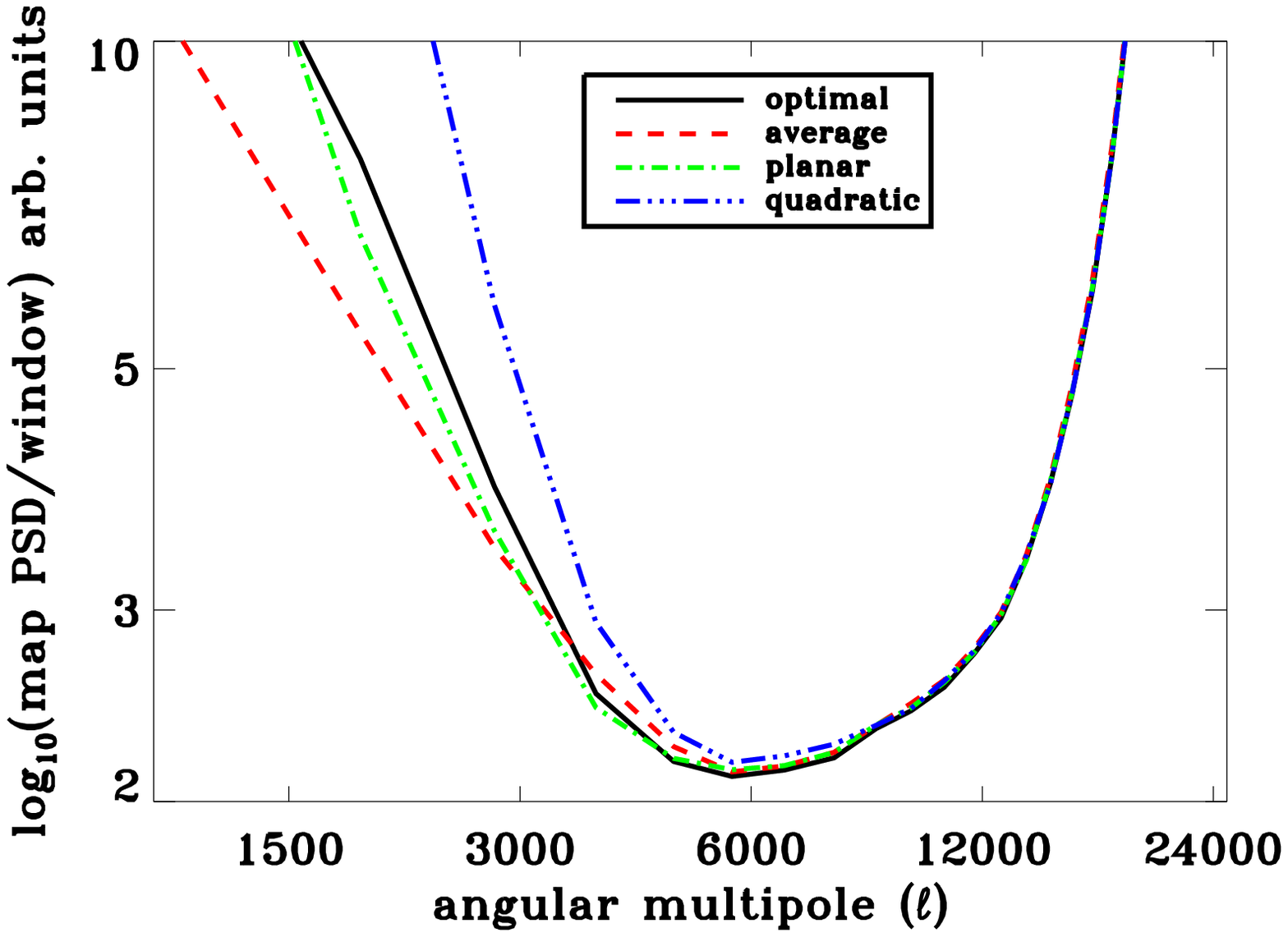} \\
  \plottwo{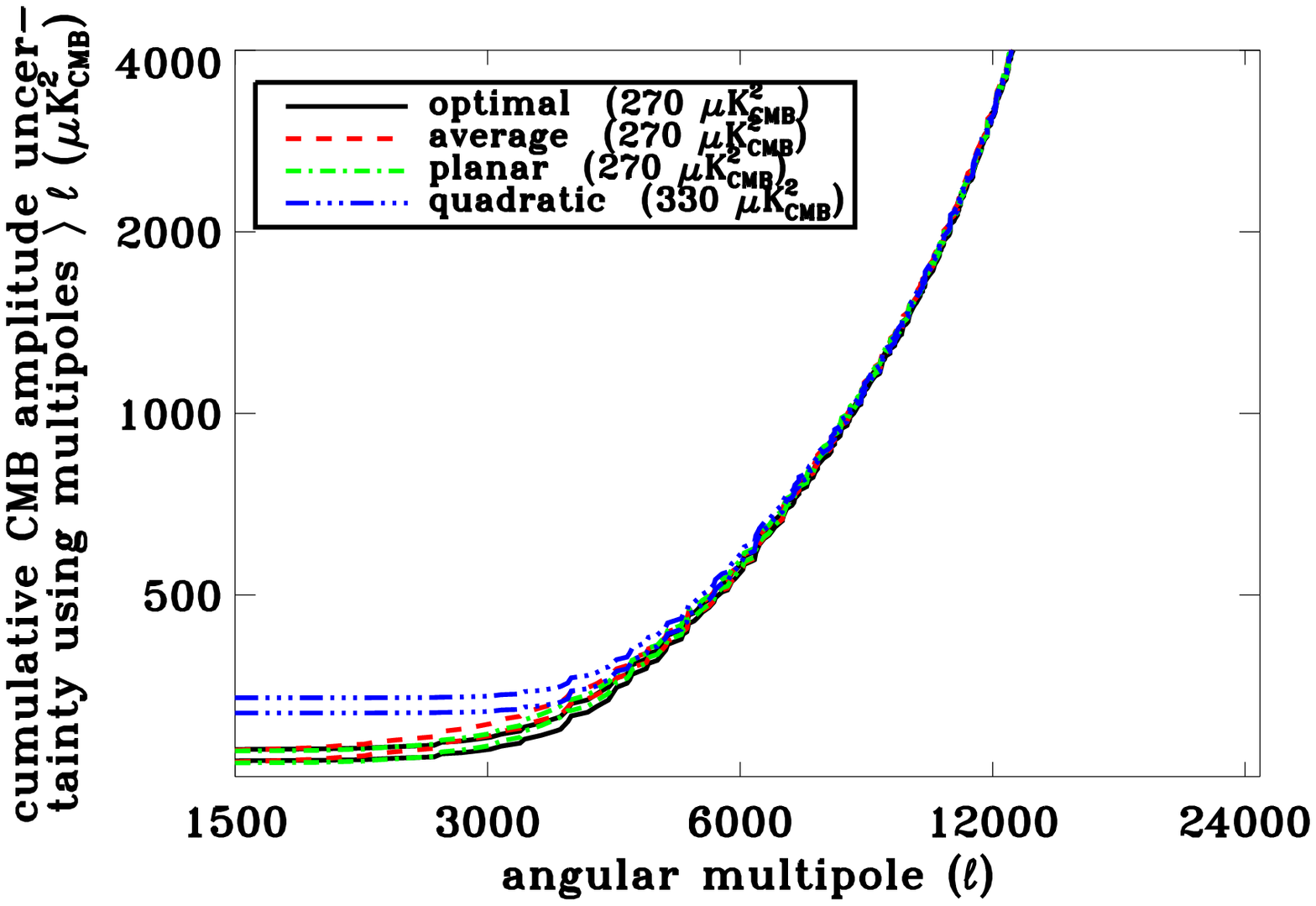}{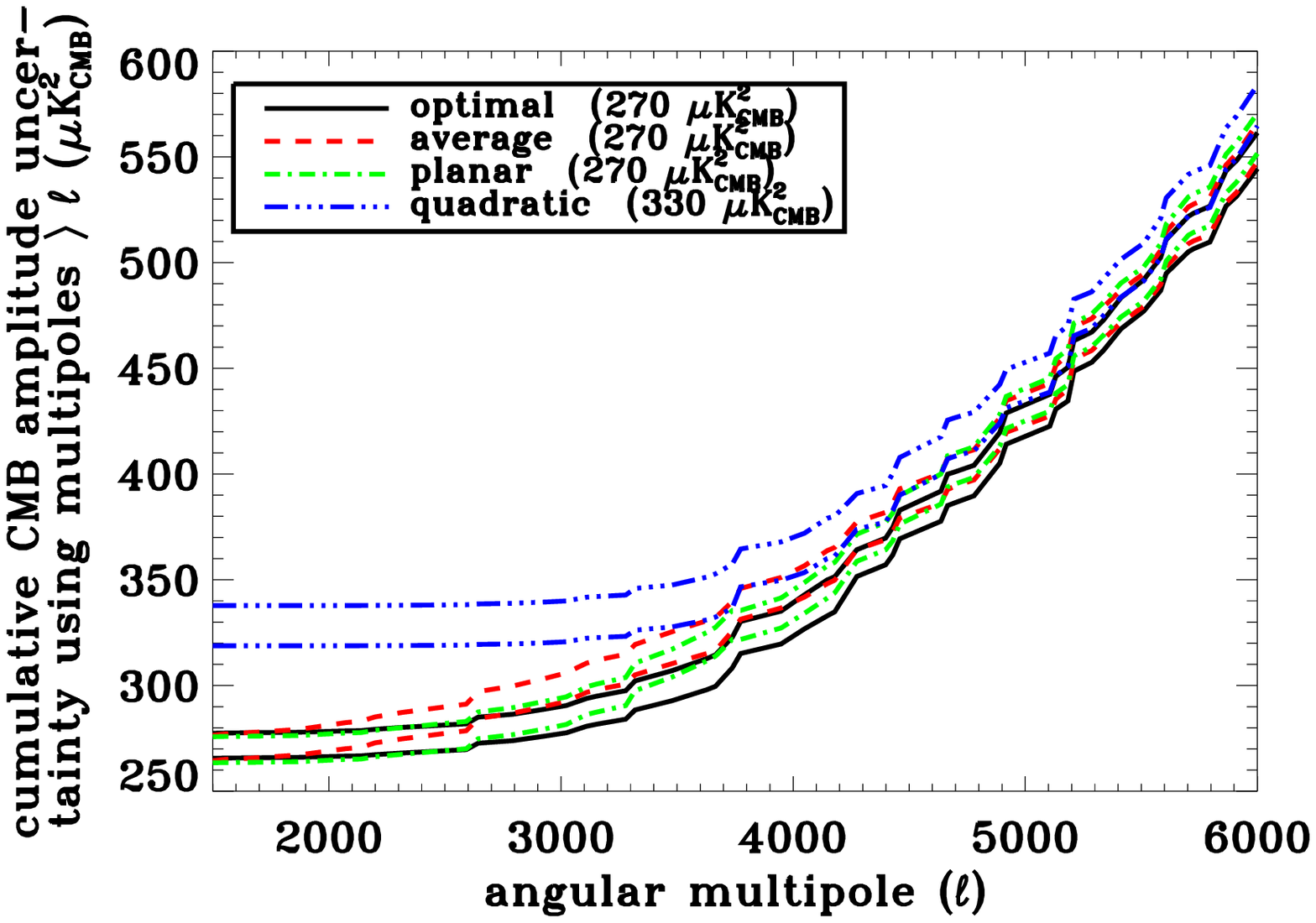}
  \caption{The plot in the top left shows the map PSD for all of the 
    143~GHz Lynx field data
    processed using average subtraction, planar subtraction,
    quadratic subtraction, or the optimal subtraction for
    each observation.
    The plot in the top right shows the same data divided by
    the window function for each subtraction algorithm and
    the window function of the beam.
    This plot shows the relative sensitivity per unit $\Delta \log(\ell)$
    to a flat band power CMB power spectrum in 
    $C_{\ell} \ell (\ell +1)/2\pi$.
    The bottom plots show the cumulative sensitivity to a flat
    band power CMB power spectrum including all of the
    data at multipoles $> \ell$.
    The two curves for each data set represent the uncertainty
    based on the RMS variations in each $\ell$-bin.
    Note that the sensitivity, including all $\ell$-bins,
    is consistent for the average, planar, and optimal data
    sets. 
    Therefore, our sensitivity to a CMB signal is largely
    independent of whether average or planar subtraction is
    used.
    This result implies that the CMB signal and the atmospheric
    noise signal are nearly indistinguishable if they are modeled as
    linearly varying over our 8 arcmin FOV.
    However, since quadratic subtraction reduces our sensitivity,
    we can infer that the CMB signal shows more correlation on
    small scales than the atmospheric noise signal,
    which is reasonable since the power spectrum of the atmosphere
    goes like $\alpha^{-11/3}$ and the power spectrum
    of the CMB goes like $\alpha^{-2}$.}
  \label{fig:map_psd_ave_plane}
\end{figure}

\begin{figure}
  \plottwo{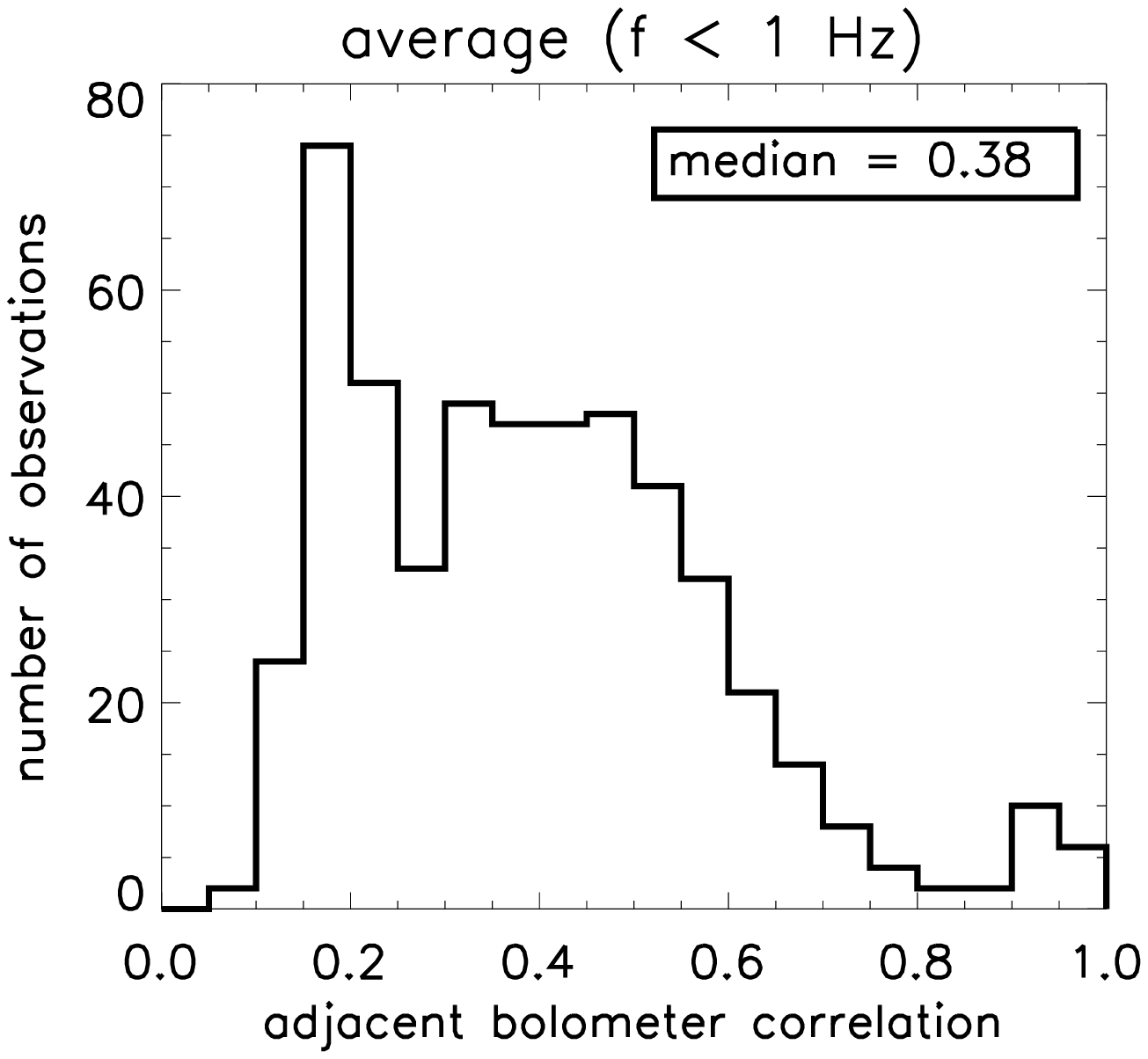}{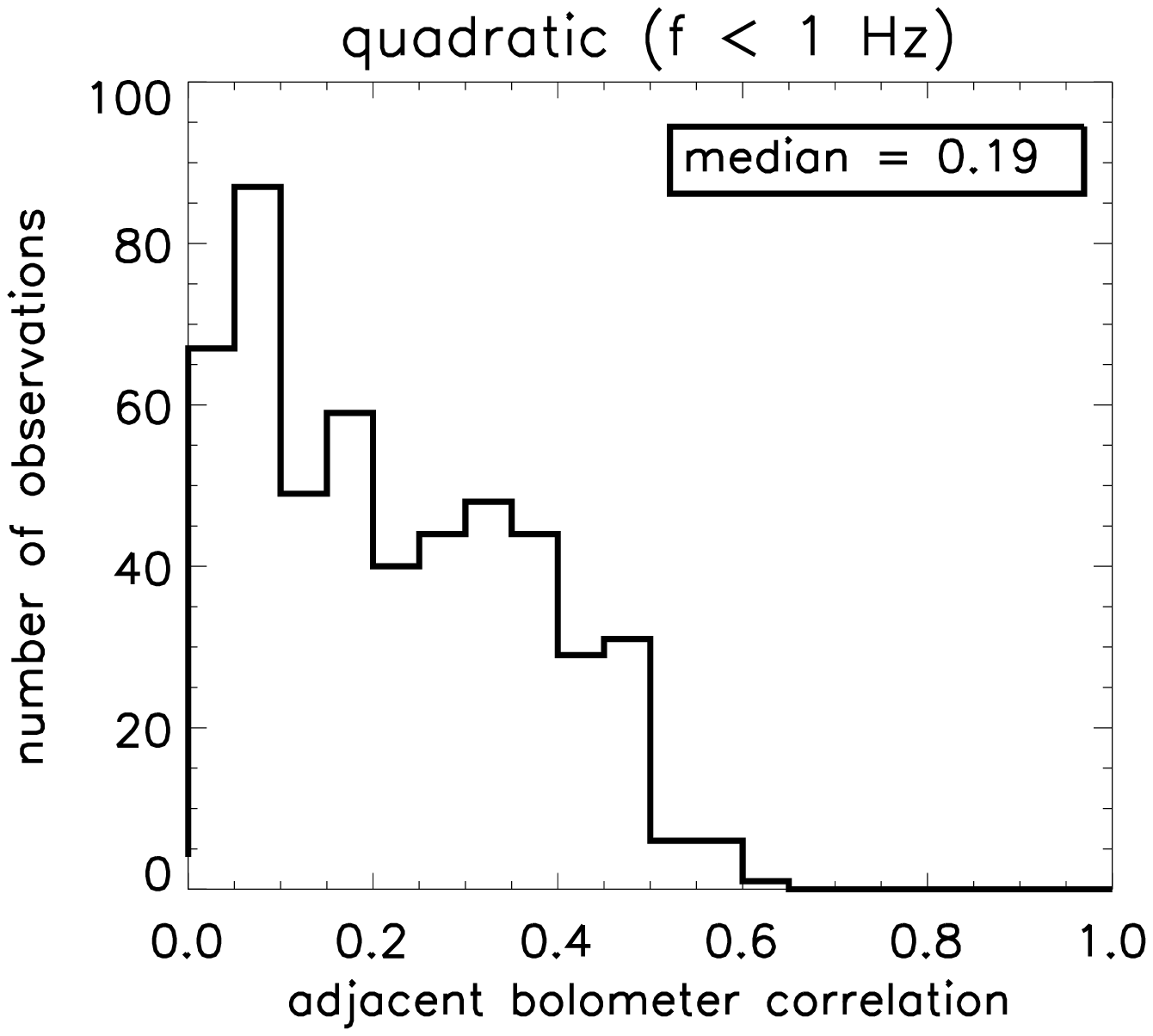} \\
  \plottwo{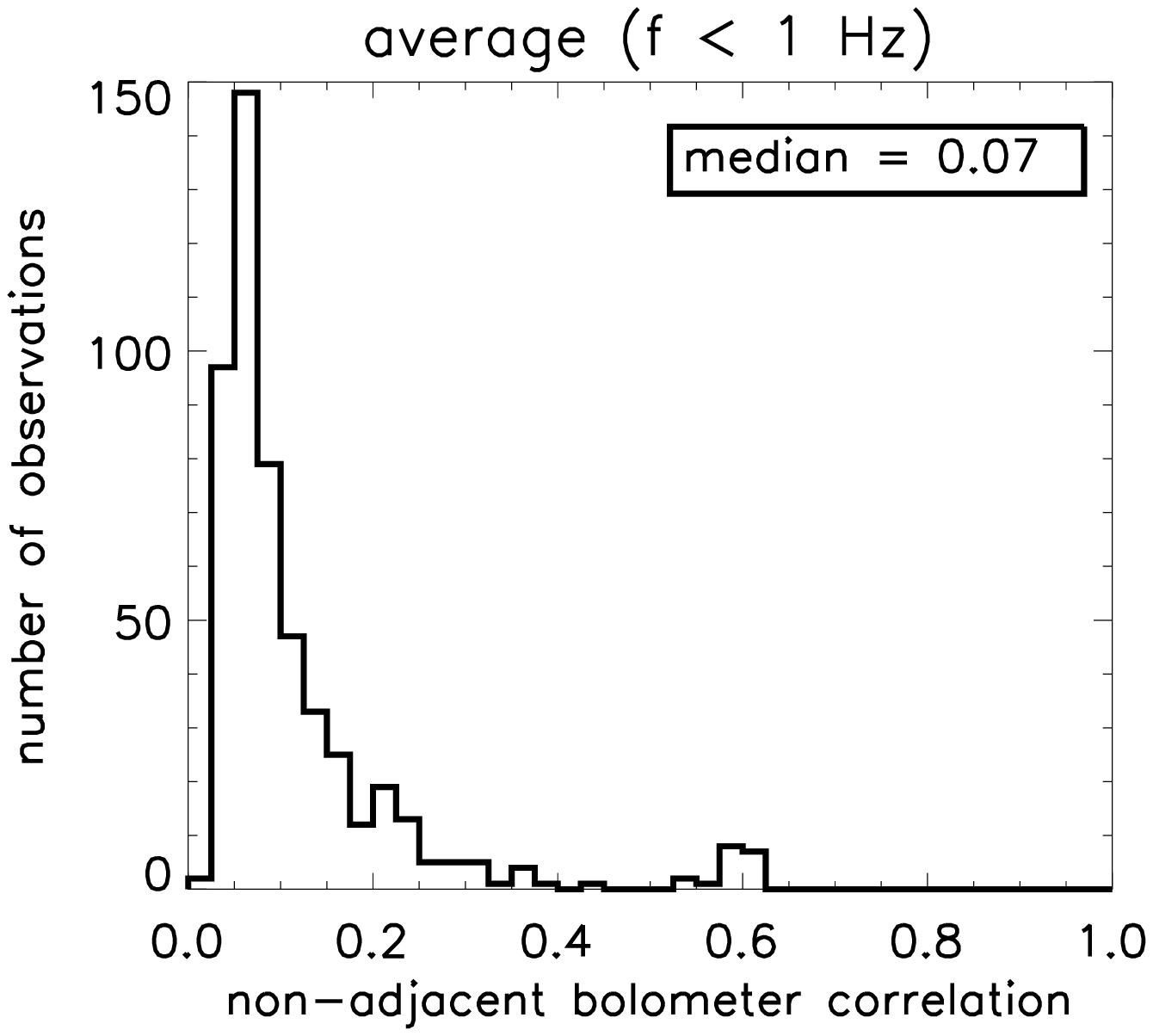}{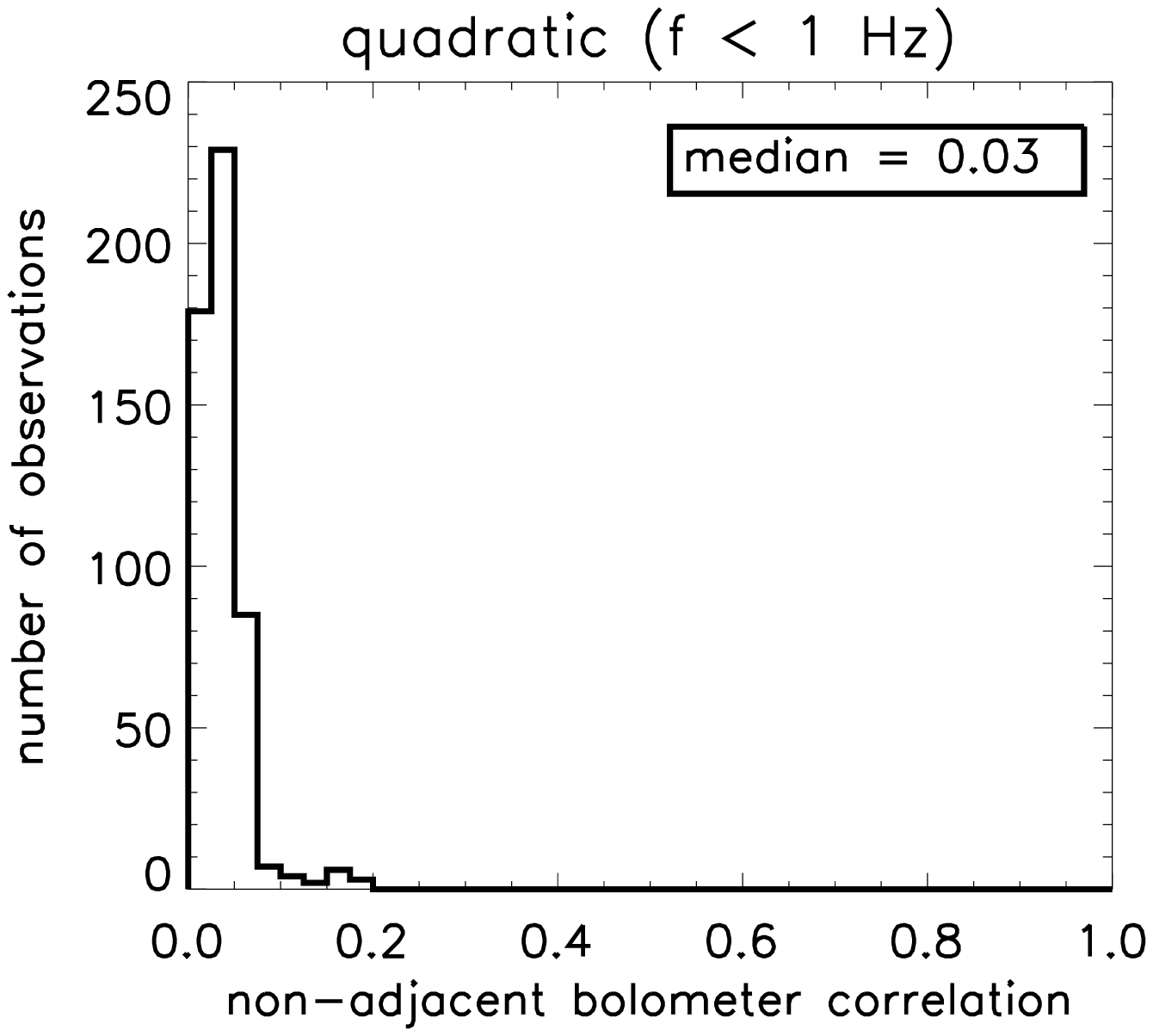}
  \caption{
%    Histograms of the bolometer-bolometer correlations, where each
%    input to the histogram represents the average correlation
%    over a single $\simeq 10$~minute long observation.
%    The top left plot contains average subtracted data at 
%    low frequency for adjacent bolometers, the top right
%    plot contains quadratic subtracted data at low
%    frequency for adjacent bolometers, the bottom left
%    plot contains high frequency data for adjacent bolometers,
%    and the bottom right plot contains quadratic subtracted data
%    at low frequency for non-adjacent bolometers.
    Histograms of the magnitude of the 
    bolometer-bolometer correlations
    at frequencies below 1~Hz
    for both adjacent and non-adjacent bolometer pairs at 143~GHz.
    The plots on the left show data processed with average subtraction,
    and the plots on the right show data processed with 
    quadratic subtraction.
    The top row shows adjacent bolometer correlations, and the 
    bottom row shows non-adjacent bolometer correlations.
    Quadratic subtraction removes almost all of the 
    atmospheric noise from the data; the residual 
    atmospheric noise in the average subtracted data is the
    reason for the much higher correlations compared to
    quadratic subtracted data.
    Adjacent bolometers are still significantly correlated 
    even after quadratic subtraction, this correlation
    is primarily due to the excess low frequency noise	
    described in Section~\ref{sec:bolo_corr}.}
  \label{fig:cross_psd_neighbor}
\end{figure}

\begin{figure}
%  \epsscale{0.7}
  \plotone{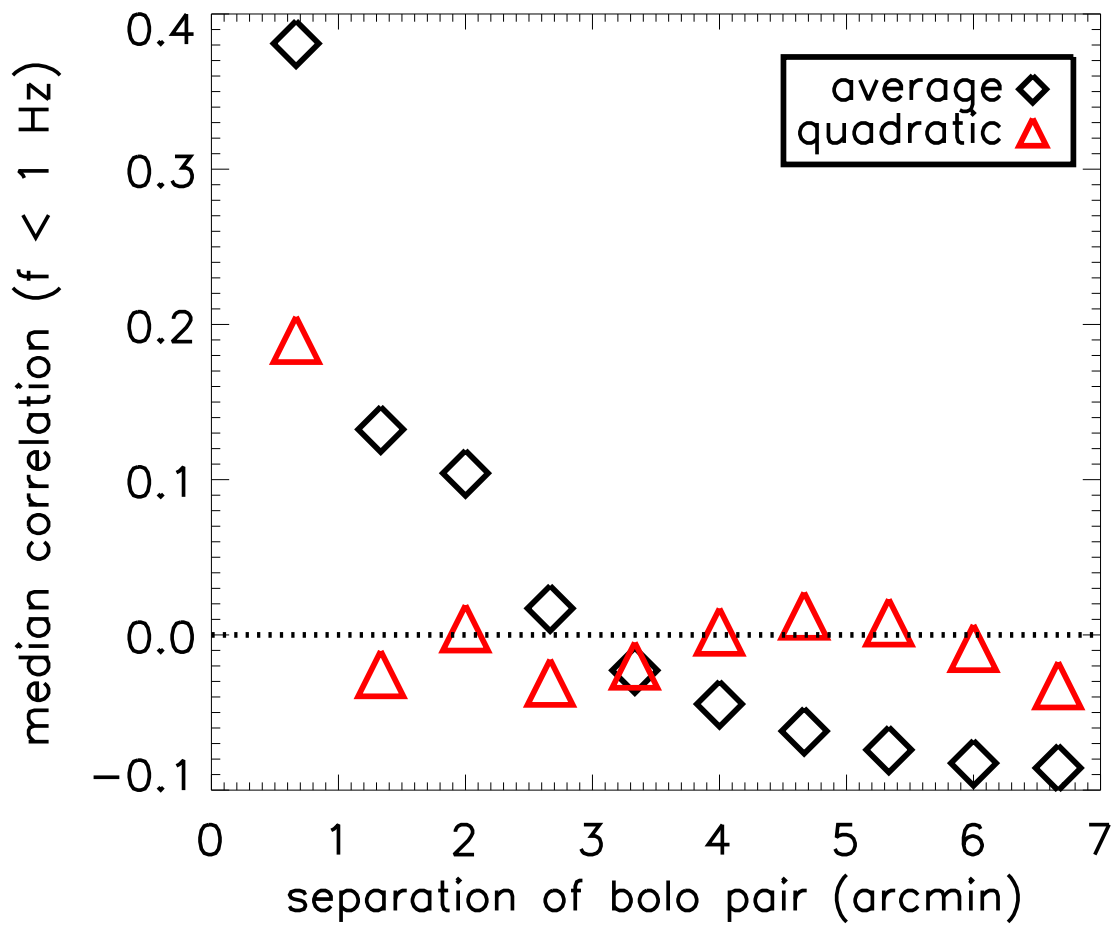}
%  \plotone{f14b.eps}
  \caption{
    Plots of median bolometer-bolometer correlation fraction
    as a function of bolometer separation for time-stream data
    below 1~Hz.
    The data has been averaged over all bolometer pairs
    and all 143~GHz observations.
    The residual atmospheric noise can be 
    easily seen in the average subtracted data as
    an excess correlation at small separations and an
    excess anti-correlation at large separations.
    In contrast, there is very little residual correlation
    in the quadratic subtracted data for non-adjacent
    bolometers, indicating that the atmospheric
    noise can be removed quite well with quadratic
    subtraction.
    The large spike in the correlation for adjacent bolometers
    is due to the excess low frequency noise described
    in Section~\ref{sec:bolo_corr}.}
%    Plots of the average 
%    $xPSD$ for frequencies below 0.25~Hz
%    for all bolometers for a single 143~GHz observation as a function
%    of bolometer separation.
%    The top plot shows data collected in relatively good weather,
%    and the bottom plot shows data collected in relatively
%    poor weather.}
  \label{fig:cross_psd_neighbor_2}
\end{figure}

    Compared to average sky subtraction,
    a slight reduction in noise, most noticeable at low frequencies,
    can be seen in the time-streams.
    See Figure~\ref{fig:avg_skysub}.
    However, the difference in the noise level of a map made from
    co-adding all $\simeq 500$ observations of the
    Lynx science field is far more dramatic.
    See Figure~\ref{fig:map_psd_ave_plane}.
    The reason such a small change in the time-stream
    PSDs produces such a large
    change in the map PSDs is because planar and quadratic subtraction
    reduce the amount of residual atmospheric-noise correlations
    remaining in 
    the time-streams of the bolometers.
    Figures~\ref{fig:cross_psd_neighbor} and \ref{fig:cross_psd_neighbor_2}
    illustrate this
    reduction in the bolometer-bolometer correlations with quadratic subtraction.
%    by showing the
%    average $xPSD$ for all of our Lynx observations.
%    Note that, to isolate the effects of the residual atmospheric
%    noise signal, these $xPSD$s were averaged over
%    frequencies below 0.25~Hz.

    However, the higher-order templates also remove more
    astronomical signal compared to average subtraction.
    Therefore, a single observation of a given astronomical source
    shape will have an optimal subtraction algorithm based
    on the noise level of the data and the amount of 
    signal attenuation.
    For an extended source, (\emph{e.g.}, a CMB anisotropy,
    which is usually modeled as flat
    in $C_{\ell} \ell (\ell+1) / 2\pi$ at
    large $\ell$, where $\ell$ is
    angular multipole)\footnote{
      A flat 
      CMB anisotropy signal profile is used throughout this 
      paper to quantify the sensitivity of our data and
      to test our subtraction algorithms.
      This signal shape was chosen because:
      1) the 143 GHz data were collected primarily to look	
      for CMB anisotropies \citep{sayers09},
      2) it has a similar power spectrum to the
      atmospheric noise, making it a good 
      indicator of the amount of atmospheric noise, and
      3) several large-format
      instruments have also been commissioned at
      mm wavelengths to study the CMB anisotropies at
      the South Pole (\emph{e.g.}, SPT
      \citep{ruhl04}) and at
      Atacama (APEX-SZ and ACT 
      \citep{dobbs06, kosowsky03}).},
    we found that
    average subtraction was optimal for $\simeq 50$\% of the
    observations, planar subtraction was optimal for
    $\simeq 42$\% of the observations, and quadratic
    subtraction was optimal for $\simeq 8$\% of
    the observations.
    Average and planar
    subtraction provide very similar sensitivity to
    a flat CMB power spectrum, likely because the CMB
    signal is nearly indistinguishable from the atmospheric noise
    signal for linear variations over our 8 arcmin FOV.
    See Figure~\ref{fig:map_psd_ave_plane}.
    For point-like sources, we found that
    average subtraction was optimal for $\simeq 37$\% of the
    observations, planar subtraction was optimal for
    $\simeq 49$\% of the observations, and quadratic
    subtraction was optimal for $\simeq 14$\% of
    the observations.
    Most observations were optimally processed
    with the same algorithm for both point-like and
    extended objects,
    indicating that weather is the primary factor in
    determining which subtraction algorithm will be optimal
    for a given observation.
    However, observations of point sources show
    a slight preference for planar and quadratic
    subtraction compared to extended sources.
    This is because the higher-order subtraction algorithms
    attenuate signal primarily on large scales, so extended
    objects are more sensitive to the signal loss caused
    by these algorithms.

  \subsection{Adaptive Principal Component Analysis (PCA)}
    \label{sec:pca_skysub}

    We have also used an adaptive PCA algorithm to
    remove atmospheric noise from Bolocam 
    data~\citep{laurent05, murtagh87}.
    The motivation for this algorithm is to produce a set
    of statistically independent modes, which hopefully
    convert the widespread spatial correlations
    into a small number of high variance modes.
    First, consider the mean-subtracted bolometer data
    for a single scan to be a matrix, $\mathbf{d}$, with
    $n_b \times n_s$ elements.
    As usual, $n_b$ denotes the number of bolometers and 
    $n_s$ denotes the number of samples in a scan.
    For our adaptive PCA algorithm,
    we first calculate a
    covariance matrix, $\mathbf{C}$, with $n_b \times n_b$ elements
    according to
    \begin{displaymath}
      \mathbf{C} = \mathbf{d}\mathbf{d}^T.
    \end{displaymath}
    Next, $\mathbf{C}$ is diagonalized in the standard way
    to produce a set of eigenvalues ($\lambda_i$) and
    eigenvectors ($\vec{\phi_i}$), where $i$ is the 
    index of the eigenvector and $\vec{\phi}$ contains $n_b$ elements.
    The $j$th element of the $i$th eigenvector, $(\phi_i)_j$, 
    indicates the contribution of the $j$th bolometer to
    the $i$th eigenvector.
    The $i$th eigenvalue gives the contribution of the $i$th
    eigenvector to the total variance of the data.
    Eigenvectors with large eigenvalues thus carry most of the
    noise in the time-stream data.
    A transformation matrix, $\mathbf{R}$, is then
    formed from the eigenvectors according to
    \begin{displaymath}
      \mathbf{R} = (\vec{\phi_1},\vec{\phi_2},...,\vec{\phi_{n_b}}).
    \end{displaymath}
    This transformation matrix is used to decompose the data
    into eigenfunctions, $\vec{\Phi_i}$, with
    \begin{displaymath}
      (\vec{\Phi_1},\vec{\Phi_2},...,\vec{\Phi_{n_b}})^T = 
      \mathbf{\Phi} =  
      \mathbf{d} \mathbf{R}^T .
    \end{displaymath}
    These eigenfunctions are the time-dependent amplitude of the 
    corresponding eigenvector in the time-stream data;
    the eigenvalue $\lambda_i$ is the variance of that
    time-dependent eigenfunction.
    At this point, we compute the logarithm for all of the 
    eigenvalues, and then determine the standard deviation
    of that distribution.
    All of the eigenvalues with a logarithm more than 
    three standard deviations from the mean are cut, and then
    a new standard deviation is calculated.
    The process is repeated until there are no more outliers
    with large eigenvalues.
    Next, all of the eigenvector columns $\vec{\phi}_i$
    in $\mathbf{R}$ that correspond to
    the cut eigenvalues are set to zero, yielding a new
    transformation matrix, $\mathbf{R'}$.
    When reconstructing the data,
    setting these columns in $\mathbf{R}$ equal to zero
    is equivalent to discarding the cut eigenvectors. 
    Finally, we transform back to the original basis, with
    the adaptive PCA cleaned data, $\mathbf{d'}$, computed
    according to
    \begin{displaymath}
      \mathbf{d'} = \mathbf{\Phi} \mathbf{R'}.
    \end{displaymath}
    In general, the eigenfunction, $\vec{\Phi_i}$, corresponding to the 
    largest eigenvalue is nearly equal to the template
    created for average sky subtraction.
    Therefore, the physical interpretation of the leading order
    eigenfunction is fairly well understood.
    However, it is not obvious what signal(s) the lower-order
    eigenfunctions correspond to.   

\begin{figure}
  \plottwo{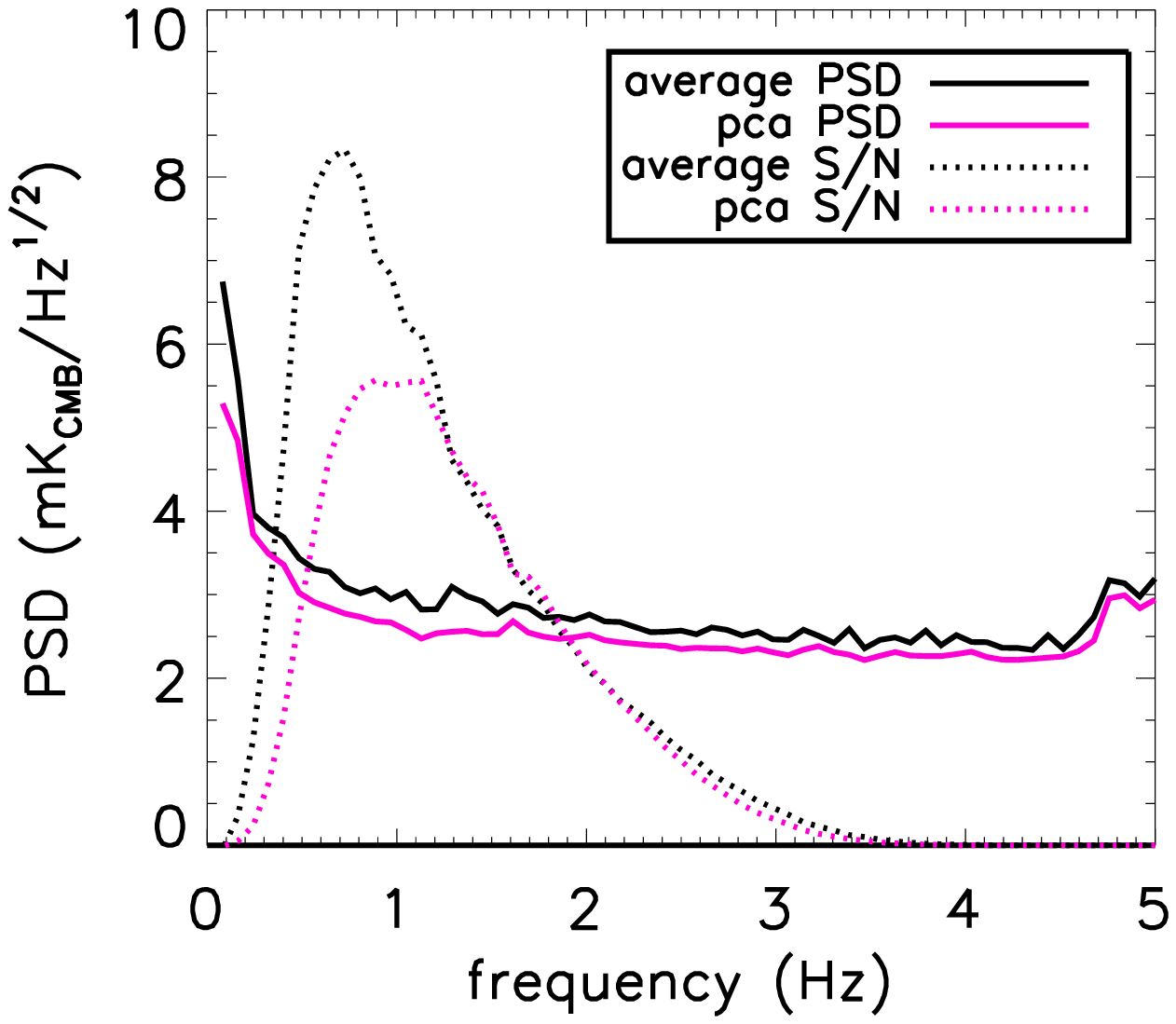}{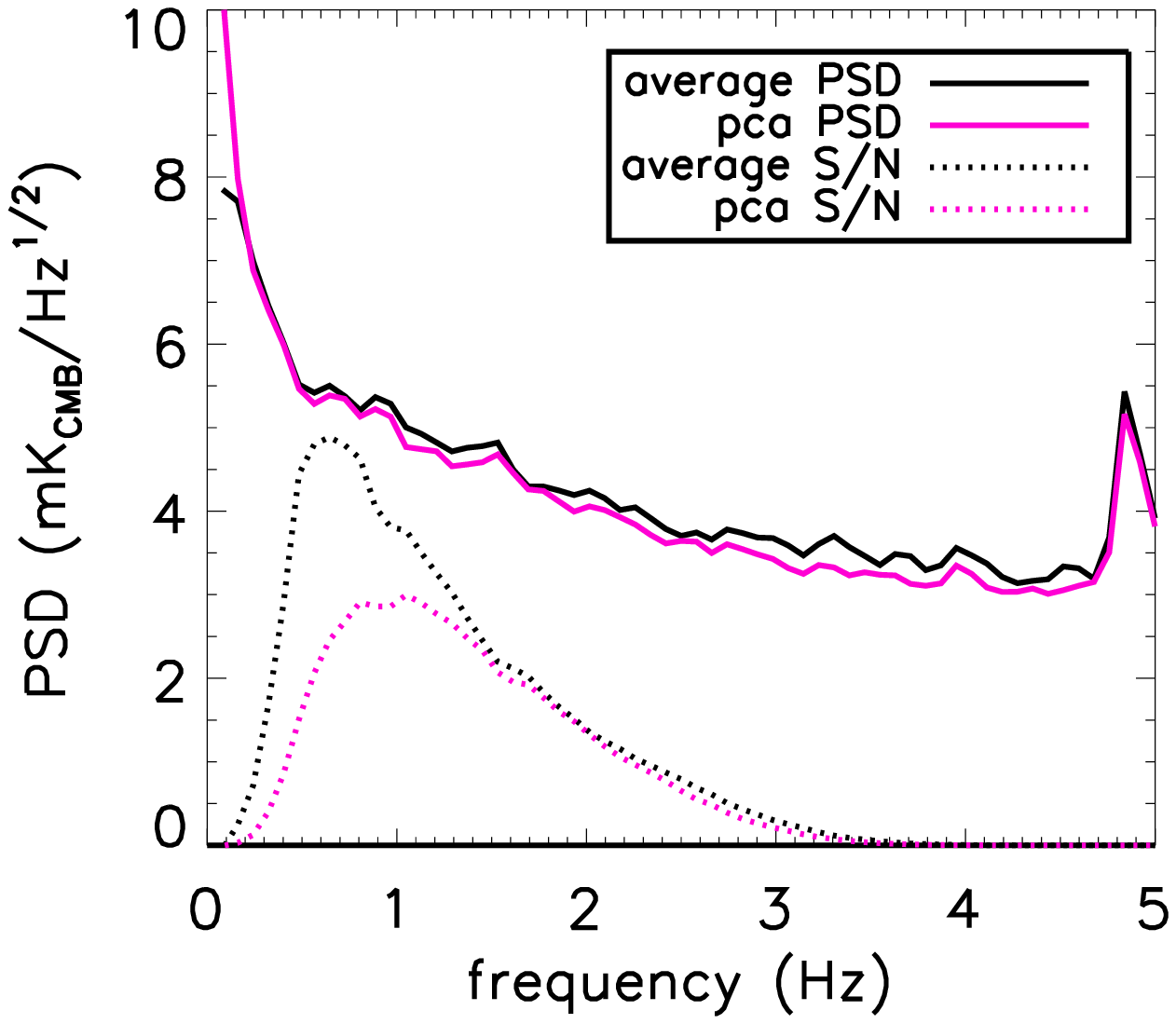}
  \caption{Plots of 143~GHz time-stream PSDs averaged over all
    scans and all bolometers for a single observation
    for average subtraction and PCA subtraction.
    The left plot shows data from an observation made in relatively
    good weather, and the right plot shows an observation made
    in relatively poor weather.
    Overplotted as dotted lines is the S/N 
    for each subtraction 
    method (in arbitrary units), 
    calculated by dividing the window function for a CMB shaped signal
    by the noise PSD.
    Note that the S/N is significantly higher for
    average subtraction compared to PCA 
    subtraction at low frequencies because average subtraction
    attenuates much less signal.}
  \label{fig:avg_skysub_2}
\end{figure}

    Typically, adaptive PCA only removes one or two eigenvectors
    from the 143~GHz data.
%    At high frequencies, adaptive PCA works slightly
%    better than the quadratic method described
%    in Section~\ref{sec:planar_skysub},
%    but there is very little astronomical signal at these frequencies
%    because of attenuation by the beam.
    In good weather, adaptive PCA produces slightly better time-stream
    noise PSDs than average subtraction, while average subtraction
    produces slightly better noise PSDs in bad weather.
%    In good weather, the results from adaptive PCA
%    are comparable to the results from quadratic
%    subtraction, but adaptive PCA performs much worse
%    than quadratic subtraction in bad weather.
    See Figure~\ref{fig:avg_skysub_2}.
%    Additionally, adaptive PCA attenuates more of the 
%    astronomical signal than quadratic 
%    subtraction,\footnote{
%      For a dim point-like source, adaptive PCA removes 
%      approximately 19\% of the flux, while quadratic
%      subtraction only removes about 12\% of the flux.
%      For reference, average subtraction removes
%      around 2\% of the flux and planar subtraction
%      removes approximately 6\% of the flux.}
%    so 
    However, adaptive PCA attenuates much more signal than
    average subtraction at low frequencies,
    which means that average subtraction produces
    a better post-subtraction S/N compared to
    adaptive PCA subtraction in all conditions.
    Therefore, 
    adaptive PCA was never the optimal subtraction algorithm
    for our analysis of blankfield data.
    Note that for observations of bright sources an
    iterative map-making technique can be used
    to recover a substantial amount
    of the signal that is lost in the process of subtracting
    the atmospheric noise \citep{enoch06}.
    Such flux recovery may change which
    subtraction algorithm that is optimal for
    a given observation.

  \subsection{Prospects for Improving Atmospheric Noise Subtraction}

    Although none of our subtraction algorithms allow us
    to reach BLIP limited performance with Bolocam
    below $\simeq 0.5$~Hz,
    this does not mean that BLIP performance
    is impossible from Mauna Kea.
    SuZIE I.5 was able to achieve instrument-limited performance\footnote{
      For reference, SuZIE I.5's BLIP limit was a factor of 
      $\simeq 3$ below the instrument noise limit
      at 100~mHz and a factor of $\simeq 6$ below
      the instrument noise limit at 10~mHz.}
    down to 10~mHz
    at 150~GHz at the CSO by subtracting a combination of spatial
    and spectral common-mode signals~\citep{mauskopf97_2}.
    The initial subtraction of the spatial common mode signal was
    obtained by differencing detectors separated by 
    $\simeq 4$~arcmin and removed the atmospheric noise
    to within a factor of two of the instrument noise level
    below a couple hundred mHz.
    In addition, SuZIE I.5 had three observing bands 
    (143, 217, and 269~GHz) per spatial
    pixel, which allowed determination of 
    the correlated signal over a range of frequencies.
    The remaining atmospheric noise at low frequency
    was removed down to the instrument noise level
    by subtracting this spectral common-mode signal.

    SuZIE II was able to employ a similar subtraction method,
    using observing bands at 143, 221, and 355~GHz for each
    spatial pixel~\citep{benson04_2}.
    Additionally, SuZIE II had a much lower instrument
    noise level at 150~GHz compared to SuZIE I.5, 
    within 50\% of the BLIP limit.
    Similar to Bolocam, SuZIE II reached the instrument
    noise level at frequencies above a couple hundred mHz
    by subtracting a spatial common mode signal.
    However, by subtracting the spectral common mode
    signal, SuZIE II achieved instrument noise limited 
    performance below 100~mHz, and was within a factor
    of 1.5 of the instrument noise limit at 10~mHz.
    Therefore, spectral subtraction of the atmospheric
    noise does provide a method to achieve
    nearly BLIP performance from the CSO.
    The MKIDCam CSO facility camera, due to be deployed
    in 2010,
    will make use of these lessons;
    it will have 576 pixels each sensing 4 colors, thus
    providing the ability to perform both spatial
    and spectral subtraction of the atmospheric noise
    \citep{glenn08}.

    Additionally, scanning the telescope more quickly
    can increase the amount of astronomical signal band
    that is free from atmospheric noise.
    As long as the telescope scan speed is slower than
    the angular wind speed of the turbulent layer, the
    atmospheric noise power spectrum will remain
    unchanged in the time-stream data as the telescope
    scan speed is increased.
    For Bolocam at the CSO, this means that the atmospheric
    noise will remain below $\simeq 0.5$~Hz for scan speeds
    below the average angular wind speed of $\simeq 30$~arcmin/sec.
    Increasing the scan speed for Bolocam observations
    from $2-4$~arcmin/sec to $30$~arcmin/sec 
    would increase the
    half-width of the beam profile from $\simeq 1-2$~Hz to
    $\simeq 10-20$~Hz, significantly increasing the 
    amount of astronomical signal band that is at 
    frequencies above the atmospheric noise.
    Unfortunately, we are not able to collect Bolocam data
    at these fast scan speeds because
    it is impossible/inefficient to scan the
    CSO telescope faster than a few arcmin/sec.
    See footnote~\ref{fn:CSO}.

\section{Residual Time-Stream Correlations}
  \label{sec:resid_corr}

  \subsection{Adjacent Bolometer Correlations}
    \label{sec:bolo_corr}

    There is a large excess correlation, above what is
    predicted by the K-T model of the atmosphere, 
    between the time-streams
    of adjacent bolometers for 143~GHz Bolocam observations.
    This excess correlation appears mainly at low frequencies
    in the time-stream data ($f \le 1$~Hz), and can be seen
    in the data in both of the following ways:
%    These excess correlations appear in the data as:
    1) a residual offset between the correlation value for
    adjacent bolometers and the K-T model 
    (see Figure~\ref{fig:corr_plots}) and as
    2) a non-zero fractional correlation between adjacent bolometers after
    subtracting most of the atmospheric noise 
    (see Figures~\ref{fig:cross_psd_neighbor}
    and \ref{fig:cross_psd_neighbor_2} and Table~\ref{tab:correlations}).
    On average, this excess correlation between adjacent bolometers is
    $\simeq 2$~mK$^2$ for $f \le 1$~Hz.
    However, the amount of excess correlation depends on the amplitude
    of the atmospheric fluctuations;
    when the observations are sorted by the value of $B_{\nu}^2$, the
    average excess correlation in the lowest quartile is $\lesssim 1$~mK$^2$
    and the average excess correlation in the highest
    quartile is $\gtrsim 5$~mK$^2$.
    See Table~\ref{tab:correlations}.
%    Additionally, most of the excess correlation between adjacent
%    bolometers is at low frequencies
%    in the time-stream data ($f \le 1$~Hz).
    Since the amplitude of this excess correlation depends on the value
    of $B_{\nu}^2$,\footnote{
      At 143~GHz the total optical load is almost independent of 
      atmospheric conditions because most of the load is not
      sourced by the atmosphere.
      See Section~\ref{sec:inst}.
      Therefore, there will only be a very weak correlation between
      $B_{\nu}^2$ and the amount of photon noise.}
    and since
    it has a rising spectrum at low frequency,\footnote{
      The Bolocam electronics noise is white down to $\lesssim 10$~mHz,
      so the only noise in the time-stream data with a rising 
      spectrum at low frequency is the atmospheric noise.}
    the source of this correlated noise appears to
    be atmospheric fluctuations.
 
    In addition to the excess correlation between adjacent bolometers,
    there is also excess noise in the bolometer time-streams
    at low frequencies.
    After accounting for the electronics noise, photon noise, and 
    atmospheric noise, there is an excess
    of $\simeq 4$~mK$^2$ for $f \le 1$~Hz.
    This excess noise increases when the value of 
    $B_{\nu}^2$ increases, so it also appears to 
    be sourced by the atmosphere,
    and thus we interpret it as excess correlation at zero
    spacing that should be considered together with the
    excess correlation between adjacent bolometers.
    Given this $\simeq 4$~mK$^2$ of excess low-frequency time-stream noise,
%2009/10/30
    we speculate that 
    the $\simeq 2$~mK$^2$ of excess correlated noise between adjacent
    detectors is explained by the fact that adjacent detectors
    are separated by less than the smallest
    possible size of a spatial mode of the electromagnetic field
    (EM) field that propagates through the optical system and 
    arrives at the focal plane.\footnote{
      This fact is a consequence of the spatial coherence of the
      EM field from classical electromagnetism.
      It is interesting to note that the same effect
      holds for photon noise in addition to atmospheric
      noise, since pixels separated 
      by $\lesssim (f/\#)\lambda$ form an intensity interferometer
      of the kind first discussed by 
      \citet{hanbury56, hanbury57,hanbury58}.
      Therefore, atmospheric noise and photon noise
      (both the shot noise
      and wave noise terms) will be correlated for
      pixels separated by $\lesssim (f/\#)\lambda$.
      We discuss this correlated photon noise below.}
%
%    In general, for 143~GHz observations made in bad weather,
%    a large number of bolometer pairs have
%    highly correlated time-streams after an atmospheric
%    template is removed from the data.
%    Additionally, 
%    the pairs with very large correlations in bad weather are almost always
%    adjacent bolometers.
%    See Figure~\ref{fig:cross_psd_neighbor_2}.
%    This adjacent bolometer
%    correlation is present because 
%    the smallest possible size 
%    of a spatial mode of the
%    electromagnetic (EM) field 
%    that propagates through the optical system and arrives
%    at the focal plane is approximately
%    the FWHM of the telescope's
%    diffraction spot size at the focal plane;
%    any power incident on the telescope, such as atmospheric noise,
%    thus appears as a highly correlated signal between detectors
%    that are separated by less than the diffraction
%    spot size, or 
%    $\lesssim (f/\#)\lambda$.\footnote{
%      This fact is a consequence of the spatial coherence of the
%      EM field from classical electromagnetism.
%      It is interesting to note that the same effect likely
%      holds for photon noise in addition to atmospheric
%      noise, since pixels separated 
%      by $\lesssim (f/\#)\lambda$ form an intensity interferometer
%      of the kind first discussed by 
%      \citet{hanbury56, hanbury57,hanbury58}.
%      Therefore, atmospheric noise and photon noise
%      (both the shot noise
%      and wave noise terms) will be correlated for
%      pixels separated by $\lesssim (f/\#)\lambda$.}
    The 143~GHz Bolocam optics provide a detector spacing
    of $0.7(f/\#)\lambda$, compared to the diffraction spot
    size of $\simeq (f/\#)\lambda$, which means there
    will be significant correlations in the signal recorded by
    adjacent detectors.
    Using the optical properties of the telescope and Bolocam optics, along 
    with the geometry of the focal plane, we calculated the amount
    of correlation between adjacent bolometers for a 
    beam-filling source (like the atmosphere).
    The result is that approximately 50\% of the 143~GHz power received by
    adjacent bolometers is completely correlated, which is what
    we observe in this excess low-frequency noise.

    Although this excess noise appears to be caused by
    atmospheric fluctuations, we do not have an adequate
    model to explain its source.
%2009/10/30
    The excess noise appears in single detector time-streams
    (along with adjacent detectors for the reasons argued above),
    which means it must be localized to a single beam.
    Additionally, since the noise appears 
    at low frequencies in the time-streams,
    it must be sourced by fluctuations larger than
    $2-4$~arcmin.\footnote{
%    the 40 arcsec separation between adjacent detectors.
	Since the telescope scan speed is $2-4$~arcmin/sec,
	noise appearing below 1~Hz must be sourced by modes
	larger than $2-4$~arcmin.}
    But, the Bolocam beams for adjacent pixels
    are only separated by 40~arcsec;
    dozens of pixels are separated by less than $2-4$~arcmin.
    Therefore, fluctuations with an angular size of $2-4$~arcmin
    will cause correlations between a large number of 
    Bolocam detectors, not just adjacent ones.
%    do not separate by $\gtrsim 2-4$~arcmin
%    until they are
%    extremely high in the atmosphere ($\gtrsim 200$~km), and
%    it seems unlikely that there is a significant amount of 
%    water vapor at these heights.
%    However, note 
    An alternate explanation is motivated by the fact 
    that the median amount of excess low-frequency noise
    ($\simeq 4$~mK$^2$)
    is much less than the total amount 
    of atmospheric noise in the	
    Bolocam data below 1~Hz ($\simeq 240$~mK$^2$).
    Therefore, this excess noise could be explained by
    atmospheric fluctuations at a reasonable height if there is
    an optical non-ideality that couples
    $1-2$\% of the beam to the atmosphere in a manner
    that is uncorrelated across the array,
    excluding the adjacent bolometer correlations
    discussed above.
% could
%    explain this excess noise.
%
%    so a non-ideality at the $1-2$\% level could explain
%    the excess noise.
%     (\emph{e.g.}, far sidelobes of the 
%    beam coupling to the atmosphere).

    This excess correlated noise is difficult to remove because it is only 
    correlated among bolometers that are close
    to each other on the focal plane.
    We have attempted to remove this noise by constructing
    localized templates using the data from a bolometer
    and the $\le 6$ bolometers that are adjacent to 
    it on the focal plane.
    We have removed these localized templates from the 
    data both before and after applying our atmospheric
    noise removal algorithm to the data.
    Unfortunately, subtracting these templates 
    from the data resulted in
    an unacceptable amount of signal attenuation, and not
    all of the locally correlated noise was removed.

    Additionally, as a consequence of the Bolocam detector spacing,
    we expect the atmospheric photon noise
    will also be $\simeq 50$\% correlated
    between adjacent detectors.
    Since the photon noise has a white spectrum, these correlations
    will have a larger effect at
    high frequencies in the time-stream data
    where there is almost no contamination from atmospheric noise.
    For Bolocam, 
    the median white noise of 5~mK$^2$/Hz is composed of 
    2.5~mK$^2$/Hz of detector plus electronics noise and 2.5~mK$^2$/Hz
    of photon noise.
    At frequencies above 2.5 Hz, well above the sky noise, the median 
    correlation between adjacent bolometer time-streams is 5\%,
    which means 
    the median correlated noise is $5 \times 0.05 = 0.3$~mK$^2$/Hz.
%2009/10/30
    As mentioned above,
    EM-field overlap between adjacent pixels implies that 50\%
    of the photon noise should be correlated, yielding an expectation
    of 1.3~mK$^2$/Hz of correlated white noise,
    roughly 4 times the observed value.

    We speculate that 
    this deficit of correlation in the photon noise
    is explained by the fact that
    high-angle scattering to warm surfaces in the relay optics
    is the dominant source of optical loading.\footnote{
      Physical optics 
      calculations with ZEMAX indicate that overillumination 
      of the relay optics is 
      negligible, and optical tests with a cold source indicate very high 
      angle scattering, not mirror spillover,
      produces most of our observed optical load.} 
%      Propagating the pixels' beams outward in time-reversed fashion, 
%      scattering does not necessarily result in the EM-field overlap 
%      between nearby pixels that would be expected from mirror 
%      overillumination, and hence one should abandon the expectation 
%      of significant correlation in the photon noise.}
    Such scattering does not necessarily preserve the correlation
    of the EM-field between adjacent pixels in the 
    way that it is preserved for the transmitted beam.
    The EM-field correlations between adjacent pixels are 
    only guaranteed to be preserved for the 10\% of our
    optical loading that is received from the atmosphere 
    via the transmitted beam.
    However, we caution that we
    have no positive evidence supporting this scattering
    hypothesis for the observed deficit of correlated photon
    noise between adjacent detectors.

%    optical loading from the relay optics box appears to be due to 
%    scattering, not overillumination of the optics.
 
%    However, less than 10\% of the photon noise
%    is due to the atmosphere.
%    Rather, most is sourced by the warm
%    relay optics.
%    Therefore, correlated atmospheric photon noise will not
%    produce a significant amount of correlation between 
%    adjacent bolometers.
%    At frequencies above 2.5~Hz the median correlation between
%    adjacent bolometer time-streams is 5\%, which means that
%    the median amount of correlated noise between adjacent
%    detectors is only $\simeq 5 \times 0.05 = 0.3$~mK$^2$/Hz.
%    Therefore, below 1~Hz there will only be
%    $\simeq 0.3$~mK$^2$ of correlated white noise.
    
%2009/10/30
    Finally, our hypothesis of EM-field overlap between adjacent
    detectors implies that the atmospheric noise will also be
    $\simeq 50$\% correlated between adjacent detectors as
    a result of our spacing.
    However, since most of the fluctuation
    power in the atmosphere is at large scales, the
    atmospheric noise in these detectors
    is already highly correlated.
    Therefore, the excess adjacent bolometer correlations
    will only appear in the atmospheric noise that the
    K-T model predicts will be uncorrelated
%    the only noise that will be affected by
%    the adjacent detector correlations is the atmospheric
%    noise that the K-T model predicts will not be
%    correlated for adjacent bolometers 
    (\emph{i.e.},
    the difference between the K-T model prediction
    for adjacent bolometers and bolometers with 
    zero separation).
    The median amount of noise predicted by the K-T
    model to be uncorrelated between adjacent bolometers
    is $\simeq 0.2$~mK$^2$, which means there
    will be $\simeq 0.1$~mK$^2$ of correlated noise
    between adjacent bolometers that is not
    predicted by the K-T model.
%    Since almost all of the atmospheric noise is below
%    1~Hz, 
    This means that the atmospheric
    noise will only cause an excess correlated noise signal of
    $\simeq 0.1$~mK$^2$ between adjacent detectors.

    In summary, there is an excess noise that appears at 
    low frequencies in the Bolocam time-stream data.
    The amount of excess noise depends on the amplitude
    of the atmospheric fluctuations, and it is approximately
    $\simeq 50$\% correlated between adjacent detectors.
%2009/10/30
    We hypothesize that this correlation 
    is due to the EM-field overlap engendered by the geometry of the 
    optical system and the physical separation between
    adjacent detectors.
%    We emphasize 
    The available evidence suggests that this excess noise
    is due to the atmosphere, but we emphasize that we do not have 
    a physical model to explain it, nor do we
    have direct evidence for our EM-field overlap hypothesis.
%    Additionally, small, but non-negligible, 
%    amounts of photon noise
%    and atmospheric noise also appear as 
%    residual correlated noise between adjacent
%    detectors due to their close spacing.

%    We note that, to our knowledge, this excess sub-$(f/\#)\lambda$ 
%    correlation has  
%    not been appreciated before in the context of
%    mm/submm camera design.  
%    Given the above physical model for these  
%    excess correlations, as well as a model such as the one we have  
%    obtained for the atmospheric conditions at a given site and observing  
%    band (distribution of $B_{\nu}^2$, 
%    effective turbulent layer height $h$, angular wind  
%    speed $w$, and assumed K-T power law 
%    $b = 11/3$), it is straightforward to  
%    include this effect in a calculation of the optimal pixel size,  
%    spacing, and detector count for any given application.  

  \subsection{Sensitivity Losses Due to Residual Atmospheric Noise and
    Adjacent Bolometer Correlations}

    Ideally, the noise in our data would be uncorrelated
    between bolometers and have a white spectrum.
    This is approximately what we would expect if instrumental
    or photon
    noise was the dominant source of unwanted signal
    in our data time-streams.
    However, our data contains a significant amount of 
    noise with a rising spectrum at low frequency.
    Some of this noise is due to residual 
    atmospheric noise, and some is due to the excess 
    low frequency noise described in Section~\ref{sec:bolo_corr}.
%    However, at low temporal frequencies our data is 
%    significantly contaminated by
%    atmospheric noise, even after subtracting most of the
%    atmospheric noise using the
%    algorithms in Section~\ref{sec:atm_noise_rem}.
    As mentioned in Section~\ref{sec:bolo_corr}, the excess
    low frequency noise (along with some residual 
    atmospheric noise and photon noise) is 
    highly correlated among adjacent bolometers.
    Additionally, there are correlations between all
    bolometer pairs on the focal plane due to the
    residual atmospheric noise.
    Finally, the atmospheric template used in our subtraction
    algorithms is constructed as a superposition of all
    the bolometer time-streams, so removing this template
    from each bolometer time-stream will cause it to 
    be slightly correlated with every other bolometer 
    time-stream.
%    Additionally, there are correlations between the bolometer
%    time-streams for two reasons related to the 
%    atmospheric noise.
%    First, the unremoved atmospheric noise produces
%    correlations between the 143~GHz bolometer time-streams
%    of adjacent bolometers for the reasons discussed in 
%    Section~\ref{sec:bolo_corr}.
%    Second, the atmospheric noise indirectly creates correlations
%    among the bolometer time-streams due to our removal
%    algorithms.
%    This is because the atmospheric template is a superposition of 
%    all the bolometer data, so a small amount of signal from
%    each bolometer is subtracted from the time-stream of every
%    other bolometer when the template is subtracted from
%    the data.

\begin{figure}
  \epsscale{1}
  \plotone{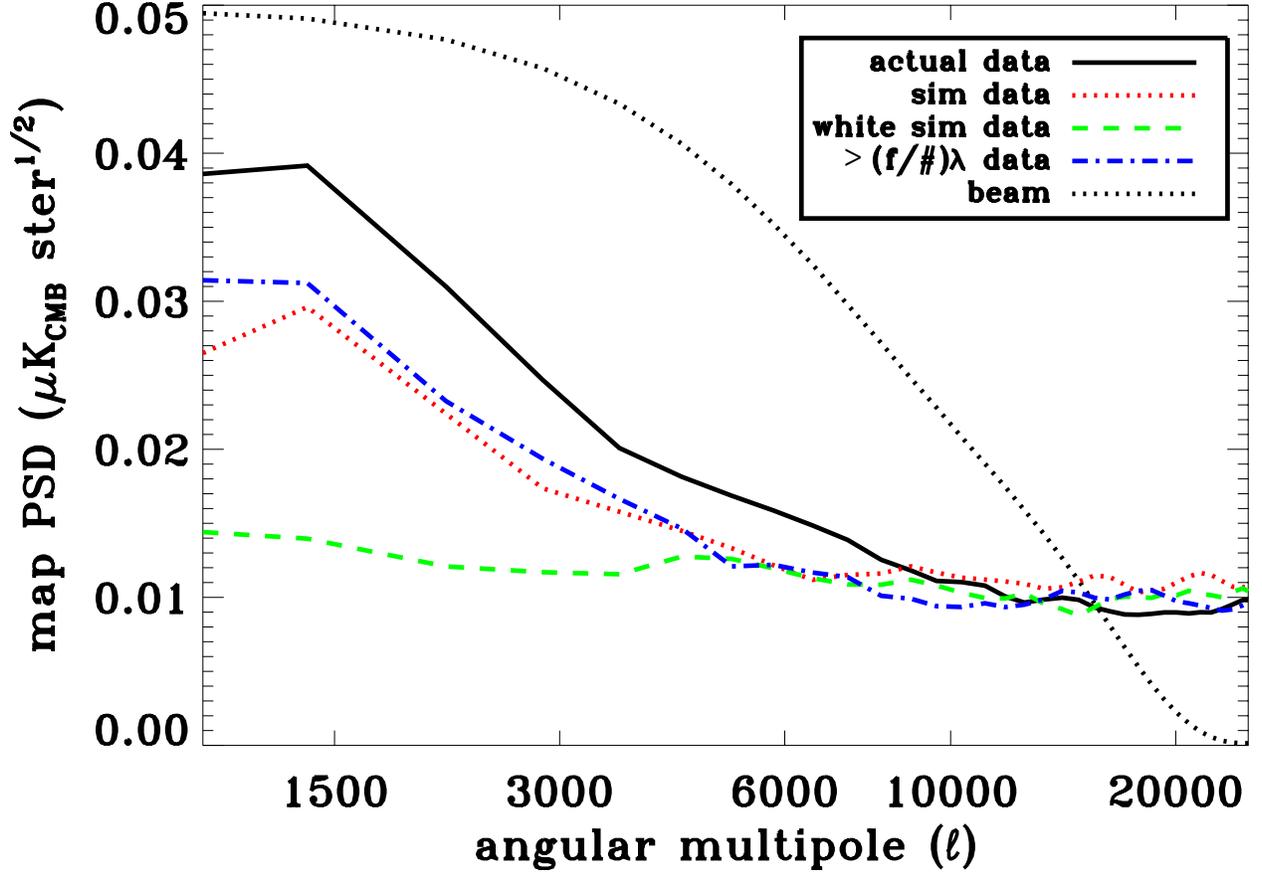}
  \caption{
    Map PSDs for actual and simulated time-streams.
    The solid black line shows the map PSD for 
    all of the 143~GHz Lynx data.
    The red dotted line shows the map PSD for simulated
    data generated using the noise spectrum
    of our actual time-streams, except
    that the simulated data are uncorrelated
    between detectors.
    The green dashed line shows the map PSD for
    uncorrelated simulated data that have a flat
    frequency spectrum and is based on the white
    noise level of our actual data.
    The blue dot-dashed line shows the map PSD for
    a map made from our actual data, after masking out
    some detectors so that the spacing between all
    detectors is $\gtrsim 1.3(f/\#)\lambda$.
    This reduces the number of detectors from 115 to
    36, but it discards the highly correlated 
    data between
    adjacent detector pairs.
    Note that this spectrum has been multiplied by
    $\sqrt{36/115}$ to account for the change in the
    number of detectors.
    Since this PSD overlaps with the uncorrelated simulated
    PSD, we can conclude that most of the correlations
    between detector time-streams
    are among adjacent detector pairs,
    and these residual correlations have a significant impact
    on the noise of the resulting maps.}
  \label{fig:sim_noise}
\end{figure}

    To understand how these non-idealities affect our data,
    we have generated two sets of simulated data.
    A different simulated data set was generated for each
    detector for each $\simeq 10$-minute-long observation,
    based on the measured PSD of each bolometer for 
    each observation.
    One simulated data
    set contains randomly generated data with the same
    noise PSD as our actual data, except the simulated
    data is completely uncorrelated between bolometers.
    The second set was generated using a flat
    noise spectrum (\emph{i.e.}, white noise), based on the white
    noise level observed in our actual data at high frequency.
    This simulated data set provides a best-case scenario
    for Bolocam.
    For each simulation we generated data corresponding to all
    of the 143~GHz
    observations of the Lynx science field, and the results are shown
    in Figure~\ref{fig:sim_noise}.
    Additionally, we made a map from our actual data after masking
    off 79 of the 115 detectors.
    This data set includes 36 detectors, all of which are separated 
    by $\gtrsim 1.3(f/\#)\lambda$, allowing us to test if the time-stream
    correlations are isolated to adjacent bolometer pairs.
    The results from this data set are also shown in 
    Figure~\ref{fig:sim_noise}.

    At high spatial frequency ($\ell \gtrsim 10000$), 
    the simulated data sets produce
    noise levels that are similar to our actual data,
    which implies that the correlations between
    detectors occur at low frequency and are caused by
    the atmospheric noise.
    However, both simulated data sets have a much
    lower noise level than our actual data at low
    spatial frequencies.
    To quantify the difference between the simulated data sets
    and our actual data set, we have estimated the 
    uncertainty in determining the amplitude of a flat CMB
    power spectrum (see \cite{sayers09} for details of the calculation).
    Additionally, we estimated the uncertainty in
    determining the amplitude of a flat CMB power spectrum
    for the data set that contains our actual data for
    36 detectors.
    This uncertainty was multiplied by 36/115 to account
    for the degradation caused by masking off 79 detectors.
    The results are shown in Table~\ref{tab:sim_noise}.
    The simulated data indicate that our uncertainty
    on the amplitude of a flat CMB power spectrum would be
    improved by a factor of $\simeq 1.6$ if the detector
    time-streams were uncorrelated, and by another factor
    of $\simeq 1.7$ if the time-streams had a white spectrum
    instead of a rising spectrum at low frequency
    due to the residual atmospheric noise.

    Additionally, after correcting for the loss of 79 detectors,
    the data set with 36 detectors produces a similar result
    to the simulated data set based on our actual noise spectra.
    This indicates that the correlations between
    time-streams of non-adjacent bolometers are
    negligible. 
    The implication is that, if we had used larger horns
    (in $(f/\#)\lambda$) while maintaining the same number
    of detectors, we would have improved our sensitivity
    in $\mu$K$^2_{CMB}$
    by a factor of 1.6.
    By going to larger horns, we
    would also have had a larger FOV, which would have had
    both positive (\emph{e.g.}, sensitivity to larger
    scales) and negative (\emph{e.g.}, less uniform map 
    coverage) effects on our data.\footnote{
      Additionally, there would be less correlation in the 
      atmospheric noise signal over a larger FOV.
      However, given how well the K-T model describes the
      correlations as a function of separation
      in our data (see Figure~\ref{fig:corr_plots}),
      the correlation over an 8 arcmin
      subregion of the FOV would be approximately equal to
      what we observed.
      Therefore, similar atmospheric noise removal could
      be obtained by performing the atmospheric noise
      subtraction algorithms on subregions of the larger
      FOV and/or subtracting higher-order polynomials.} 
    It seems likely that these negative effects would have been 
    small compared to the large gain in sensitivity we would
    have 
    obtained by eliminating the excess correlations between 
    adjacent bolometer time-streams.
    Another implication is that, at fixed detector count, 
    it is more advantageous from the atmospheric noise 
    point-of-view to use $\gtrsim(f/\#)\lambda$ pixel
    spacing and increase 
    the FOV than it is to hold the FOV fixed and sample it more 
    finely with $\lesssim(f/\#)\lambda$ pixel spacing. 
    Increasing the Bolocam FOV was not possible by the time 
    this effect was observed, 
    but this lesson is being applied for MKIDCam.
      
\section{Conclusions}

  We have studied the atmospheric noise above Mauna Kea 
  at millimeter wavelengths from the CSO using Bolocam.
  Under all observing conditions, the data time-streams
  are dominated by atmospheric noise at frequencies below 
  $\simeq 0.5$~Hz.
  The data are consistent with
  a K-T turbulence model for a thin wind-driven
  screen, and the median
  amplitude of the fluctuations is 280~mK$^2$ rad$^{-5/3}$
  at 143~GHz and 4000~mK$^2$ rad$^{-5/3}$ at 268~GHz.
  Based on a comparison to the ACBAR data in~\cite{Bussmann05},
  we conclude that these atmospheric noise fluctuation amplitudes
  are a factor of $\simeq 80$ larger than they
  would be at the South Pole for identical observing bands.
  This large difference in atmospheric noise amplitudes is
  due 
  primarily to the South Pole being a much drier site than
  Mauna Kea, with a small factor of $\simeq 2$ arising
  from the fact that 
  the fractional fluctuations in the column depth of water
  vapor are a factor of $\simeq \sqrt{2}$ lower at the South Pole.
  Based on our atmospheric modeling, we developed several algorithms
  to remove atmospheric noise, and the best results 
  were achieved when we described the fluctuations using 
  a low-order polynomial in detector position over the 
  8~arcmin focal plane.
  However, even with these algorithms, we were not able to obtain
  BLIP performance at 
  frequencies below $\simeq 0.5$~Hz in any observing 
  conditions. 
  Therefore, we conclude that 
  BLIP performance 
  is not possible from Mauna Kea below $\simeq 0.5$~Hz
  for broadband $\simeq 1-2$ mm receivers
  with subtraction of a spatial atmospheric template on scales
  of several arcmin.
  We also observed an excess low-frequency noise that
  is highly correlated between detectors separated by
  $\lesssim (f/\#)\lambda$; 
  this noise appears to be caused by atmospheric fluctuations,
  but we do not have an adeqaute model to explain its source.
  We hypothesize that the correlations arise from
%  We also observed excess correlations in the atmospheric noise 
%  between pixels separated by $\lesssim (f/\#)\lambda$, which 
%  are 
%  explained by 
  the classical coherence of 
  the EM field across a distance of $\simeq (f/\#)\lambda$ 
  on the focal plane. 
%  We also observed excess correlations in the atmospheric noise 
%  between pixels separated by $\lesssim (f/\#)\lambda$, which 
%  are explained by the classical coherence of 
%  the EM field across a distance of $\simeq (f/\#)\lambda$ 
%  on the focal plane. 

\section{Acknowledgements}

  We acknowledge the assistance of: Minhee Yun and Anthony D. Turner of
  NASA's Jet Propulsion Laboratory, who fabricated the Bolocam science
  array; Toshiro Hatake of the JPL electronic packaging group, who
  wirebonded the array; Marty Gould of Zen Machine and Ricardo Paniagua
  and the Caltech PMA/GPS Instrument Shop, who fabricated much of the
  Bolocam hardware; Carole Tucker of Cardiff University, who tested
  metal-mesh reflective filters used in Bolocam; Ben Knowles of
  the University of Colorado, who contributed to the software pipeline,
  the day crew and Hilo
  staff of the Caltech Submillimeter Observatory, who provided
  invaluable assistance during commissioning and data-taking for this
  survey data set; 
  high school teacher Tobias Jacoby and high school students
  Jonathon Graff, Gloria Lee, and Dalton Sargent,
  who helped as summer research assistants;
  and Kathy Deniston, who provided effective
  administrative support at Caltech.
  Bolocam was constructed and
  commissioned using funds from NSF/AST-9618798, NSF/AST-0098737,
  NSF/AST-9980846, NSF/AST-0229008, and NSF/AST-0206158.  
  J. Sayers and G. Laurent
  were partially supported by
  NASA Graduate Student Research Fellowships, 
  J. Sayers was partially supported by a NASA Postdoctoral
  Program Fellowship,
  J. Aguirre was partially supported by a Jansky
  Postdoctoral Fellowship, and
  S. Golwala
  was partially supported by a R.~A.~Millikan Postdoctoral Fellowship
  at Caltech.
  The research described in this paper was carried out at the 
  Jet Propulsion Laboratory, California Institute of Technology,
  under a contract with the National Aeronautics and Space
  Administration.

{\it Facilities:} \facility{CSO}.

\appendix
\section{Appendix Material}
\label{sec:appen}

  In order to account for the time lags and advances
  between bolometer time-streams that are
  described by the K-T thin-screen model,
  we in general have to shift the time-streams by 
  a fractional number of samples.
  For example, if a given bolometer time-stream
  is advanced by $\Delta t_b$ seconds,
  then we will account for this advance by shifting
  the time-stream according to
  \begin{equation}
    d_n' = \left( 1 - \left|\frac{\Delta t_b}{\Delta t} 
    \right| \right) d_n + 
    \left|\frac{\Delta t_b}{\Delta t} \right| 
    d_{n+\Delta t_b/\Delta t}
    \label{eqn:DAS_shift}
  \end{equation}
  where $d_n'$ is the interpolated data time-stream, 
  $d_n$ is the original data time-stream,
  $\Delta t$ is the time between
  samples, and $n$ is the sample number.
  Note that we have assumed that 
  $\Delta t_b < \Delta t$, since shifts by
  integer multiples of $\Delta t$ are trivial.
  Alternatively, this shift can be performed in frequency space
  by applying
  \begin{equation}
    S_m = \left( 1 - \left| \frac{\Delta t_b}{\Delta t} \right| + 
    \left| \frac{\Delta t_b}{\Delta t} \right|
    e^{-\mathrm{sign}(\Delta t_b) i 2 \pi f_m \Delta t} \right)
  \end{equation}
  to the Fourier transform of the time-stream data,
  where $f_m$ is frequency
  in Hz and $m$ is the frequency-space index.
  $S_m$ acts like a filter, and, for all non-zero frequencies,
  $|S_m| < 1$.
  Therefore, to preserve the noise properties of our
  data, we divide the Fourier transform of the
  shifted time-stream by $|S_m|$.
  In summary, we shift the time-stream data according to
  Equation~\ref{eqn:DAS_shift}, then correct for
  the filtering effects of this shift in frequency-space
  by dividing by $|S_m|$.

\clearpage

\clearpage

\clearpage

\clearpage

\clearpage

\clearpage

\clearpage

\clearpage

\clearpage

\clearpage

\clearpage

\clearpage

\clearpage

\clearpage

\clearpage

\clearpage

\clearpage
\begin{deluxetable}{ccccc} 
  \tablewidth{0pt}
  \tablecaption{Atmospheric Noise Amplitude ($B_{\nu}^2$)} 
  \tablehead{\colhead{instrument} & \colhead{frequency} &
    \colhead{quartile 1} & \colhead{quartile 2} & 
    \colhead{quartile 3}} 
  \startdata
     Bolocam & 143 GHz & 100 mK$^2$ rad$^{-5/3}$	& 280 mK$^2$ rad$^{-5/3}$
     & 980 mK$^2$ rad$^{-5/3}$ \\	    
     ACBAR & 151 GHz & 3.7 mK$^2$ rad$^{-5/3}$	& 10 mK$^2$ rad$^{-5/3}$
     & 37 mK$^2$ rad$^{-5/3}$ \\	    
     \multicolumn{2}{c}{Bolocam/ACBAR} & 27 & 28 & 26 \\ \hline
     Bolocam & 268 GHz & 1100 mK$^2$ rad$^{-5/3}$ & 4000 mK$^2$ rad$^{-5/3}$
     & 14000 mK$^2$ rad$^{-5/3}$ \\	    
     ACBAR & 282 GHz & 28 mK$^2$ rad$^{-5/3}$	& 74 mK$^2$ rad$^{-5/3}$
     & 230 mK$^2$ rad$^{-5/3}$ \\	    
     \multicolumn{2}{c}{Bolocam/ACBAR} & 39 & 54 & 61 \\
  \enddata
  \tablecomments{The observed quartile values of $B_{\nu}^2$ for the
    two Bolocam observing
    bands from Mauna Kea and two of the ACBAR observing bands
    from the South Pole.
    The ratio of $B_{\nu}^2$ for the two instruments is given
    for each of the bands ($\simeq 150$ and $\simeq 275$ GHz.)}
	\label{tab:noise_amp}
\end{deluxetable}

\clearpage
\begin{deluxetable}{cccccc}
  \tablewidth{0pt}
  \tablecaption{143 GHz data, $f \le 1$~Hz}
  \tablehead{\colhead{} & \colhead{bin 1} & \colhead{bin 2} & 
     \colhead{bin 3} & \colhead{bin 4} & \colhead{all data}}
  \startdata
     $B_{\nu}^2$ (mK$^2$rad$^{-5/3}$) 
       & $46 \pm 2$ & $170 \pm 10$ & $580 \pm 20$ 
       & $4000 \pm 400$ & $280 \pm 60$ \\
     raw atmosphere (mK$^2$) 
       & $77 \pm 11$ & $131 \pm 22$ & $310 \pm 60$ 
       & $1060 \pm 150$ & $240 \pm 50$ \\
     adj. corr. noise (mK$^2$) 
       & $0.8 \pm 0.1$ & $1.1 \pm 0.1$ & $1.9 \pm 0.2$ 
       & $5.8 \pm 0.3$ & $1.9 \pm 0.1$ \\
     adj. corr. fraction 
       & $0.39 \pm 0.04$ & $0.52 \pm 0.04$ & $0.42 \pm 0.07$ 
       & $0.53 \pm 0.03$ & $0.46 \pm 0.03$ \\
%     corr. fraction & 27\% & 35\% & 43\% & 84\% & 45\% \\
  \enddata
  \tablecomments{Description of the excess low-frequency noise
    that appears in the 143~GHz time-stream data
    and is likely sourced by the atmosphere.
    The first four columns give the 
    median value, and uncertainty on the median value,
    of the data when they are
    binned as a function of the
    amplitude of the atmospheric fluctuations, $B_{\nu}^2$.
    The final column gives the median values,
    and uncertainties on the median values, for the full 
    data set.
    From top to bottom the rows give the value of 
    $B_{\nu}^2$; the raw atmospheric noise 
    below 1~Hz prior to subtraction;
    the excess correlated noise between adjacent
    detectors below 1~Hz after accounting for
    residual atmospheric noise, and correlated
    photon/white noise;
    and the correlation fraction between adjacent detectors
    below 1~Hz after accounting for the correlations
    expected from residual atmospheric noise and
    photon/white noise.
    The excess noise rises at low frequency and 
    increases as a function of $B_{\nu}^2$,
    indicating that it is sourced by the atmosphere.
    Additionally, the excess noise is $\simeq 50$\%
    correlated between adjacent detectors; we hypothesize that this 
    correlation is a consequence of 
    the $0.7(f/\#)\lambda$ spacing between these detectors.}
  \label{tab:correlations}
\end{deluxetable}

\clearpage
\begin{deluxetable}{ccc} 
  \tablewidth{0pt}
  \tablecaption{Lynx data} 
  \tablehead{\colhead{data type} & \colhead{data spectrum} &
    \colhead{CMB amplitude uncertainty}} 
  \startdata
	    actual data & actual data & 270 $\mu$K$_{CMB}^2$ \\
	    simulated & actual data & 170 $\mu$K$_{CMB}^2$ \\
	    simulated & white & 100 $\mu$K$_{CMB}^2$ \\
	    actual data, $> (f/\#)\lambda$ & actual data & 
	    170 (550) $\mu$K$_{CMB}^2$ \\
  \enddata
  \tablecomments{The estimated uncertainty on measuring the amplitude
	  of a flat CMB power spectrum 
	  for all of the 143~GHz Lynx observations.
	  The four data sets include:
	  our actual data, simulated data using our actual time-stream
	  noise spectra, simulated data using our actual time-stream
	  white noise level, and
	  our actual data after masking off 79 of our 115 detectors so that
	  the spacing between all detectors is 
	  $\gtrsim 1.3(f/\#)\lambda$.
	  For the two simulated data sets the bolometer time-streams
	  are uncorrelated.
	  The results for the second and fourth data sets are similar,
	  after accounting for the reduction in detector number
	  in the fourth set, indicating that the majority of the
	  correlations between our detector time-streams are
	  between adjacent detector pairs.
	  The results show that 
	  our sensitivity to a CMB amplitude is reduced by a factor
	  of $\simeq 1.6$ due to these correlations, and by 
	  another factor of $\simeq 1.7$ due to the 
	  residual atmospheric noise in our data at low frequencies.}
	\label{tab:sim_noise}
\end{deluxetable}


\begin{thebibliography}{}
  \bibitem[Aguirre et al.(2009)]{aguirre09}
    Aguirre, J. E. et al, in preparation
  \bibitem[Archibald et al.(2002)]{archibald02}
    Archibald, E. N. et al., 2002, \mnras, 336, 1
  \bibitem[Benson(2004)]{benson04_2}
    Benson, B. A., 2004, PhD Thesis, Stanford
  \bibitem[Borys et al.(1999)]{borys98}
    Borys, C. et al., 1999, \mnras, 308, 527
  \bibitem[Bussmann et al.(2005)]{Bussmann05}
    Bussmann, R. S., Holzapfel, W. L., and Kuo, C. L.,
    2005, \apj, 622, 1343
  \bibitem[Chamberlin(2004)]{chamberlin99}
    Chamberlin, R. A., 2004, PASAu, 21, 264
  \bibitem[Church(1995)]{church95}
    Church, S. E., 1995, \mnras, 272, 551
  \bibitem[Conway et al.(1965)]{conway63}
    Conway, R. G. et al., 1965, \mnras, 131, 159
  \bibitem[Dobbs et al.(2006)]{dobbs06}
    Dobbs, M. et al., 2006, New Astr. Rev., 50, 960
  \bibitem[Dowell et al.(2003)]{dowell03}
    Dowell, C. D. et al., 2003, \procspie, 4855, 73
  \bibitem[Enoch et al.(2006)]{enoch06}
    Enoch, M. et al., 2006, \apj, 638, 293
  \bibitem[Glenn et al.(1998)]{glenn98} 
    Glenn, J. et al., 1998, \procspie, 3357, 326
  \bibitem[Glenn et al.(2003)]{glenn03} 
    Glenn, J. et al., 2003, \procspie, 4855, 30
  \bibitem[Glenn et al.(2008)]{glenn08}
    Glenn, J. et al., 2008, \procspie, 7020, 10
  \bibitem[Haig et al.(2004)]{haig04}
    Haig, D. J. et al., 2004, \procspie, 5498, 78
  \bibitem[Hanbury Brown and Twiss(1956)]{hanbury56}
    Hanbury Brown, R. and Twiss, R. Q., 
    1956, \nat, 178, 1046
  \bibitem[Hanbury Brown and Twiss(1957)]{hanbury57}
    Hanbury Brown, R. and Twiss, R. Q., 
    1957, Proc. Roy. Soc. Lon. A, 242, 300
  \bibitem[Hanbury Brown and Twiss(1958)]{hanbury58}
    Hanbury Brown, R. and Twiss, R. Q., 
    1958, Proc. Roy. Soc. Lon. A, 243, 291
  \bibitem[Holland et al.(1999)]{holland99}
    Holland, W. S. et al., 1999, \mnras, 
    303, 659
  \bibitem[Jenness et al.(1998)]{jenness98}
    Jenness, T. et al., 1998, \procspie, 3357, 638
  \bibitem[Kreysa et al.(1998)]{kreysa98}
    Kreysa, E. et al., 1998, \procspie, 3357, 319
  \bibitem[Kreysa et al.(2003)]{kreysa03}
    Kreysa, E. et al., 2003, \procspie, 4855, 41
  \bibitem[Kolmogorov(1941)]{kolmogorov41}
    Kolmogorov, A. N., 1941, ANSSSR, 30, 301
  \bibitem[Kosowsky(2003)]{kosowsky03}
    Kosowsky, A., 2003, New Astr. Rev., 47, 939
  \bibitem[Kovacs(2008)]{kovacs08}
    Kovacs, A., 2008 preprint (astro-ph/0805.3928)
  \bibitem[Lane(1998)]{lane98}
    Lane, A. P., 1998, Astro Antarctica, 141, 289
  \bibitem[Laurent et al.(2005)]{laurent05} 
    Laurent, G. T. et al., 2005, \apj, 623, 742
  \bibitem[Lay and Halverson(2000)]{Lay00}
    Lay, O. P. and Halverson, N. W., 2000, \apj, 543, 787
  \bibitem[Masson(1994)]{masson94}
    Masson, C. R., 1994, IAU Colloq 140, 59, 87
  \bibitem[Murtagh and Heck(1987)]{murtagh87}
    Murtagh, F. and Heck, A., 1987,
    \emph{Multivariate Data Analysis}, Kluwer Academic Publishers, 
    Boston
  \bibitem[Mauskopf et al.(1997)]{mauskopf97}
    Mauskopf, P. D. et al.,
    1997, \ao, 36, 765
  \bibitem[Mauskopf(1997)]{mauskopf97_2}
    Mauskopf, P. D., 1997, PhD Thesis, University of California
    at Berkeley
  \bibitem[Pardo et al.(2001a)]{pardo01}
    Pardo, J. R., Cernicharo, J., and Serabyn, E., 
    2001a, ITAP, 49, 1683
  \bibitem[Pardo et al.(2001b)]{pardo01_2}
    Pardo, J. R., Serabyn, E., and Cernicharo, J., 
    2001b, JQSRT, 68, 419
  \bibitem[Pardo et al.(2005)]{pardo05}
    Pardo, J. R. et al., 2005, JQRST, 96, 537
  \bibitem[Peterson et al.(2003)]{peterson03}
    Peterson, J. B. et al., 2003, 115, 383
    Publ Astron Soc Pac
  \bibitem[Radford and Chamberlin(2000)]{radford00}
    Radford, S. J. and Chamberlin, R. A., 2000, ALMA memo 334
  \bibitem[Reichertz et al.(2001)]{reichertz01}
    Reichertz, L. A. et al., 2001, A\&A, 379, 735
  \bibitem[Ruhl et al.(2004)]{ruhl04}
    Ruhl, J. et al., 2004, \procspie, 5498, 11
  \bibitem[Sayers(2007)]{golwala08}
    Sayers, J., 2007, PhD Thesis, Caltech
  \bibitem[Sayers et al.(2009)]{sayers09}
    Sayers, J. et al., 2009, \apj, 690, 1597
%  \bibitem[Smith et al.(2001)]{smith01}
%    Smith, G. J., et al., 2001, Int. J. Infrared Milli., 22, 661
  \bibitem[Stark et al.(2001)]{stark01}
    Stark, A. A., et al., 2001, Publ Astron Soc Pac, 113, 567
  \bibitem[Tatarskii(1961)]{tatarskii61}
    Tatarskii, V. I., 1961, \emph{Wave Propagation in a Turbulent Medium},
    McGraw-Hill, New York
  \bibitem[Taylor(1938)]{taylor38}
    Taylor, G. I., 1938, Proc. R. Soc. Lond. A, 164, 476
  \bibitem[Weferling et al.(2002)]{weferling01}
    Weferling, B. et al., 2002, A\&A, 383, 1088
  \bibitem[Wright(1996)]{wright96}
    Wright, M. C. H., 1996, Publ Astron Soc Pac, 108, 520
\end{thebibliography}
\end{document}